\documentclass[sts]{imsart}
 
\RequirePackage[OT1]{fontenc}
\usepackage{amsthm,amsmath}
\usepackage[square, semicolon]{natbib}
\RequirePackage[colorlinks,citecolor=blue,urlcolor=blue]{hyperref}

\usepackage{amsthm}
\usepackage{subfigure} 
\usepackage{multirow}
\usepackage{tikz}
\usepackage{amssymb,amstext,amsmath}
\usepackage{bbm}
\usepackage{bm}
\usepackage{tikz} 
\usepackage{xcolor}
\colorlet{BLUE}{blue}
\colorlet{RED}{red}
\usepackage{pdflscape}

\startlocaldefs
\numberwithin{equation}{section}
\theoremstyle{plain}
\newtheorem{theorem}{Theorem}[section]
\newtheorem{remark}{Remark}[section]
\newtheorem{proposition}{Proposition}[section]
\definecolor{orange}{RGB}{255,127,0}
\endlocaldefs

\newcommand{\x}{{\bf x}}
\newcommand{\y}{{\bf y}}
\newcommand{\z}{{\bf z}}
\newcommand{\X}{{\bf X}}

\newcommand{\hI}{{\hat{I}}}

\newcommand{\Ra}{\texttt{R1}}
\newcommand{\Rb}{\texttt{R2}}
\newcommand{\Rc}{\texttt{R3}}
\newcommand{\Na}{\texttt{N1}}
\newcommand{\Nb}{\texttt{N2}}
\newcommand{\Nc}{\texttt{N3}}

\newcommand{\normalized}{\tilde}

\newcommand{\E}{\mathrm{E}}
\newcommand{\Var}{\mathrm{Var}}

\newcommand{\gauss}{\mathcal{N}}

\newcommand{\RThreeProposals} {
 \begin{tikzpicture}[scale=1, transform shape]
 \draw [rounded corners=.2cm] (-5,-1.5) rectangle (9.6,4.6);
 \draw (-5,3.5) -- (9.6,3.5);
 \draw[dash dot] (-2,2.5) -- (9.6,2.5);
  \draw (-5,1.5) -- (9.6,1.5);
  \draw (-2,0.5) -- (9.6,0.5);
  \draw (-2,-0.5) -- (9.6,-0.5);
  \draw (-1,-1.5) -- (-1,4.6);
  \draw (-2,-1.5) -- (-2,3.5);

\node [anchor=west] at (-4.5, 4) {Available proposals};

\node [anchor=west] at (-4.5, 2.5) {Sampling};
\node [anchor=west] at (-1.8, 3) {$j_n$};
\node [anchor=west] at (-1.8, 2) {$\x_n$};
\node [anchor=west] at (-1.8, 1) {\texttt{R1}};
\node [anchor=west] at (-1.8, 0) {\texttt{R2}};
\node [anchor=west] at (-1.8, -1) {\texttt{R3}};

\node [anchor=west] at (-4.9, 0.3) {Weighting options};
\node [anchor=west] at (-4.4, -0.4) {$w_n = \frac{\pi(\x)}{\varphi_n(\x)}$};

\draw [rounded corners=.2cm] (0,3.7) rectangle (1.6,4.3);
\draw [anchor=center,fill=cyan](0.3,4) circle (0.18);
\node at (0.3,4) {$1$};
\draw [anchor=center,fill=green](0.8,4) circle (0.18);
\node at (0.8,4) {$2$};
\draw [anchor=center,fill=orange](1.3,4) circle (0.18);
\node at (1.3,4) {$3$};

\draw [rounded corners=.2cm] (3.5,3.7) rectangle (5.1,4.3);
\draw [anchor=center,fill=cyan](3.8,4) circle (0.18);
\node at (3.8,4) {$1$};
\draw [anchor=center,fill=green](4.3,4) circle (0.18);
\node at (4.3,4) {$2$};
\draw [anchor=center,fill=orange](4.8,4) circle (0.18);
\node at (4.8,4) {$3$};

\draw [rounded corners=.2cm] (7,3.7) rectangle (8.6,4.3);
\draw [anchor=center,fill=cyan](7.3,4) circle (0.18);
\node at (7.3,4) {$1$};
\draw [anchor=center,fill=green](7.8,4) circle (0.18);
\node at (7.8,4) {$2$};
\draw [anchor=center,fill=orange](8.3,4) circle (0.18);
\node at (8.3,4) {$3$};

\draw [anchor=center,fill=orange](0.8,3) circle (0.18);
\node at (0.8,3) {$3$};
\draw [anchor=center,fill=orange](4.3,3) circle (0.18);
\node at (4.3,3) {$3$};
\draw [anchor=center,fill=cyan](7.8,3) circle (0.18);
\node at (7.8,3) {$1$};

\node [anchor=center] at (0.8,2) {$\x_1 \sim $ \textcolor{orange}{$q_3$}};
\node [anchor=center] at  (4.3,2) {$\x_2 \sim $  \textcolor{orange}{$q_3$}};
\node [anchor=center] at  (7.8,2) {$\x_3\sim $ \textcolor{cyan}{$q_1$}};

\node [anchor=center] at (0.8,1) {$\frac{\pi(\x_1)}{\textcolor{orange}{q_3}(\x_1)}$};
\node [anchor=center] at  (4.3,1){$\frac{\pi(\x_2)}{\textcolor{orange}{q_3}(\x_2)}$};
\node [anchor=center] at  (7.8,1) {$\frac{\pi(\x_3)}{\textcolor{cyan}{q_1}(\x_3)}$};

\node [anchor=center] at (0.8,0) {$\frac{\pi(\x_1)}{\frac{1}{3}\left(\textcolor{orange}{q_3}(\x_1)+\textcolor{orange}{q_3}(\x_1)+\textcolor{cyan}{q_1}(\x_1)\right)}$};
\node [anchor=center] at  (4.3,0){$\frac{\pi(\x_2)}{\frac{1}{3}\left(\textcolor{orange}{q_3}(\x_2)+\textcolor{orange}{q_3}(\x_2)+\textcolor{cyan}{q_1}(\x_2)\right)}$};
\node [anchor=center] at  (7.8,0) {$\frac{\pi(\x_3)}{\frac{1}{3}\left(\textcolor{orange}{q_3}(\x_3)+\textcolor{orange}{q_3}(\x_3)+\textcolor{cyan}{q_1}(\x_3)\right)}$};

\node [anchor=center] at (0.8,-1) {$\frac{\pi(\x_1)}{\frac{1}{3}\left(\textcolor{cyan}{q_1}(\x_1)+\textcolor{green}{q_2}(\x_1)+\textcolor{orange}{q_3}(\x_1)\right)}$};
\node [anchor=center] at  (4.3,-1){$\frac{\pi(\x_2)}{\frac{1}{3}\left(\textcolor{cyan}{q_1}(\x_2)+\textcolor{green}{q_2}(\x_2)+\textcolor{orange}{q_3}(\x_2)\right)}$};
\node [anchor=center] at  (7.8,-1) {$\frac{\pi(\x_3)}{\frac{1}{3}\left(\textcolor{cyan}{q_1}(\x_3)+\textcolor{green}{q_2}(\x_3)+\textcolor{orange}{q_3}(\x_3)\right)}$};

\end{tikzpicture}
}

\newcommand{\NacThreeProposals} {
 \begin{tikzpicture}[scale=1, transform shape]
 \draw [rounded corners=.2cm] (-5,-0.5) rectangle (9.6,4.6);
 \draw (-5,3.5) -- (9.6,3.5);
 \draw[dash dot] (-2,2.5) -- (9.6,2.5);
  \draw (-5,1.5) -- (9.6,1.5);
  \draw (-2,0.5) -- (9.6,0.5);

  \draw (-1,-0.5) -- (-1,4.6);
  \draw (-2,-0.5) -- (-2,3.5);

\node [anchor=west] at (-4.5, 4) {Available proposals};

\node [anchor=west] at (-4.5, 2.5) {Sampling};
\node [anchor=west] at (-1.8, 3) {$j_n$};
\node [anchor=west] at (-1.8, 2) {$\x_n$};
\node [anchor=west] at (-1.8, 1) {\texttt{N1}};
\node [anchor=west] at (-1.8, 0) {\texttt{N3}};

\node [anchor=west] at (-4.9, 0.9) {Weighting options};
\node [anchor=west] at (-4.4, 0.2) {$w_n = \frac{\pi(\x)}{\varphi_n(\x)}$};

\draw [rounded corners=.2cm] (0,3.7) rectangle (1.6,4.3);

\draw [anchor=center,fill=cyan](0.8,4) circle (0.18);
\node at (0.8,4) {$1$};

\draw [rounded corners=.2cm] (3.5,3.7) rectangle (5.1,4.3);

\draw [anchor=center,fill=green](4.3,4) circle (0.18);
\node at (4.3,4) {$2$};

\draw [rounded corners=.2cm] (7,3.7) rectangle (8.6,4.3);

\draw [anchor=center,fill=orange](7.8,4) circle (0.18);
\node at (7.8,4) {$3$};

\draw [anchor=center,fill=cyan](0.8,3) circle (0.18);
\node at (0.8,3) {$1$};
\draw [anchor=center,fill=green](4.3,3) circle (0.18);
\node at (4.3,3) {$2$};
\draw [anchor=center,fill=orange](7.8,3) circle (0.18);
\node at (7.8,3) {$3$};

\node [anchor=center] at (0.8,2) {$\x_1 \sim $ \textcolor{cyan}{$q_1$}};
\node [anchor=center] at  (4.3,2) {$\x_2 \sim $  \textcolor{green}{$q_2$}};
\node [anchor=center] at  (7.8,2) {$\x_3\sim $ \textcolor{orange}{$q_3$}};

\node [anchor=center] at (0.8,1) {$\frac{\pi(\x_1)}{\textcolor{cyan}{q_1}(\x_1)}$};
\node [anchor=center] at  (4.3,1){$\frac{\pi(\x_2)}{\textcolor{green}{q_2}(\x_2)}$};
\node [anchor=center] at  (7.8,1) {$\frac{\pi(\x_3)}{\textcolor{orange}{q_1}(\x_3)}$};

\node [anchor=center] at (0.8,0) {$\frac{\pi(\x_1)}{\frac{1}{3}\left(\textcolor{cyan}{q_1}(\x_1)+\textcolor{green}{q_2}(\x_1)+\textcolor{orange}{q_3}(\x_1)\right)}$};
\node [anchor=center] at  (4.3,0){$\frac{\pi(\x_2)}{\frac{1}{3}\left(\textcolor{cyan}{q_1}(\x_2)+\textcolor{green}{q_2}(\x_2)+\textcolor{orange}{q_3}(\x_2)\right)}$};
\node [anchor=center] at  (7.8,0) {$\frac{\pi(\x_3)}{\frac{1}{3}\left(\textcolor{cyan}{q_1}(\x_3)+\textcolor{green}{q_2}(\x_3)+\textcolor{orange}{q_3}(\x_3)\right)}$};

\end{tikzpicture}
}

\newcommand{\NbThreeProposals} {
 \begin{tikzpicture}[scale=1, transform shape]
 \draw [rounded corners=.2cm] (-5,0.5) rectangle (9.6,4.6);
 \draw (-5,3.5) -- (9.6,3.5);
 \draw[dash dot] (-2,2.5) -- (9.6,2.5);
  \draw (-5,1.5) -- (9.6,1.5);

  \draw (-1,0.5) -- (-1,4.6);
  \draw (-2,0.5) -- (-2,3.5);

\node [anchor=west] at (-4.5, 4) {Available proposals};

\node [anchor=west] at (-4.5, 2.5) {Sampling};
\node [anchor=west] at (-1.8, 3) {$j_n$};
\node [anchor=west] at (-1.8, 2) {$\x_n$};
\node [anchor=west] at (-1.8, 1) {\texttt{N2}};

\node [anchor=west] at (-4.4, 1.2) {Weighting};
\node [anchor=west] at (-4.4, 0.8) {$w_n = \frac{\pi(\x)}{\varphi_n(\x)}$};

\draw [rounded corners=.2cm] (0,3.7) rectangle (1.6,4.3);
\draw [anchor=center,fill=cyan](0.3,4) circle (0.18);
\node at (0.3,4) {$1$};
\draw [anchor=center,fill=green](0.8,4) circle (0.18);
\node at (0.8,4) {$2$};
\draw [anchor=center,fill=orange](1.3,4) circle (0.18);
\node at (1.3,4) {$3$};

\draw [rounded corners=.2cm] (3.5,3.7) rectangle (5.1,4.3);
\draw [anchor=center,fill=cyan](3.8,4) circle (0.18);
\node at (3.8,4) {$1$};
\draw [anchor=center,fill=green](4.3,4) circle (0.18);
\node at (4.3,4) {$2$};

\draw [rounded corners=.2cm] (7,3.7) rectangle (8.6,4.3);

\draw [anchor=center,fill=green](7.8,4) circle (0.18);
\node at (7.8,4) {$2$};

\draw [anchor=center,fill=orange](0.8,3) circle (0.18);
\node at (0.8,3) {$3$};
\draw [anchor=center,fill=cyan](4.3,3) circle (0.18);
\node at (4.3,3) {$1$};
\draw [anchor=center,fill=green](7.8,3) circle (0.18);
\node at (7.8,3) {$2$};

\node [anchor=center] at (0.8,2) {$\x_1 \sim $ \textcolor{orange}{$q_3$}};
\node [anchor=center] at  (4.3,2) {$\x_2 \sim $  \textcolor{cyan}{$q_1$}};
\node [anchor=center] at  (7.8,2) {$\x_3\sim $ \textcolor{green}{$q_2$}};

 \node [anchor=center] at (0.8,1) {$\frac{\pi(\x_1)}{\frac{1}{3}\left(\textcolor{cyan}{q_1}(\x_1)+\textcolor{green}{q_2}(\x_1)+\textcolor{orange}{q_3}(\x_1)\right)}$};
 \node [anchor=center] at  (4.3,1){$\frac{\pi(\x_2)}{\frac{1}{2}\left(\textcolor{cyan}{q_1}(\x_2)+\textcolor{green}{q_2}(\x_2) \right)}$};
 \node [anchor=center] at  (7.8,1) {$\frac{\pi(\x_3)}{\textcolor{green}{q_2}(\x_3)}$};

\end{tikzpicture}
}

\begin{document}

\begin{frontmatter}
\title{Generalized Multiple Importance Sampling}
\runtitle{Generalized Multiple Importance Sampling}

\begin{aug}
\author{\fnms{V\'ictor} \snm{Elvira}$^1$\thanksref{t1}\ead[label=e1]{victor.elvira@ed.ac.uk}},
\author{\fnms{Luca} \snm{Martino}$^2$\ead[label=e2]{luca.martino@uv.es}},
\author{\fnms{David} \snm{Luengo}$^3$\ead[label=e3]{david.luengo@upm.es}},
\and
\author{\fnms{M\'onica} \snm{F. Bugallo}$^4$\ead[label=e4]{monica.bugallo@stonybrook.edu}
}

\thankstext{t1}{Contact author for the manuscript \printead{e1}}
\affiliation{$^1$University of Edinburgh (United Kingdom), $^2$Universidad Rey Juan Carlos (Spain), $^3$Universidad Polit\'ecnica de Madrid (Spain),  $^4$Stony Brook University (USA)}

\end{aug}

\begin{abstract}
Importance sampling (IS) methods are broadly used to approximate posterior distributions or their moments. In the standard IS approach, samples are drawn from a single proposal distribution and weighted adequately. However, since the performance in IS depends on the mismatch between the targeted and the proposal distributions, several proposal densities are often employed for the generation of samples. Under this multiple importance sampling (MIS) scenario, extensive literature has addressed the selection and adaptation of the proposal distributions, interpreting the sampling and weighting steps in different ways. In this paper, we establish a novel general framework with sampling and weighting procedures when more than one proposal is available. The new framework encompasses most relevant MIS schemes in the literature, and novel valid schemes appear naturally. All the MIS schemes are compared and ranked in terms of the variance of the associated estimators. Finally, we provide illustrative examples revealing  that, even with a good choice of the proposal densities, a careful interpretation of the sampling and weighting procedures can make a significant difference in the performance of the method.
\end{abstract}


\end{frontmatter}

%%%%%%%%%%%%%%%%%%%%%%%%%%%%%%%%%%%%%%%%%
%%%%%%%%%%%%%%%%%%%%%%%%%%%%%%%%%%%%%%%%%
\section{Introduction}
%%%%%%%%%%%%%%%%%%%%%%%%%%%%%%%%%%%%%%%%%
%%%%%%%%%%%%%%%%%%%%%%%%%%%%%%%%%%%%%%%%%

Importance sampling (IS) is a well-known Monte Carlo technique that can be applied to compute integrals involving target probability density functions (pdfs) \citep{Robert04,Liu04b}. The standard IS technique draws samples from a single proposal pdf and assigns them weights based on the ratio between the target and the proposal pdfs, both evaluated at the sample value. 
The choice of a suitable proposal pdf is crucial for obtaining a good approximation of the target pdf using the IS method. Indeed, although the validity of this approach is guaranteed under mild assumptions, the variance of the estimator depends on the discrepancy between the shape of the proposal and the target \citep{Robert04,Liu04b}. 

Several advanced strategies have been proposed in the literature to design more robust IS schemes \citep[Chapter 2]{Liu04b}, \citep[Chapter 9]{Owen13}, \citep{Liang02}. A powerful approach is based on using a population of different proposal pdfs. 
This approach is often referred to as {\it multiple} importance sampling (MIS) and several possible implementations have been proposed {depending on the specific assumptions of the problem, e.g. the knowledge of the normalizing constants, prior information of the proposals, etc} \citep{Veach95,Hesterberg95,Owen00,tan2004likelihood,he2014optimal,elvira2015efficient}. In general, MIS strategies provide more robust algorithms, since they avoid entrusting the performance of the method to a single proposal. Moreover, many algorithms have been proposed in order to conveniently adapt the set of proposals in MIS \citep{Cappe04,martino2017layered,elvira2017improving}. 

When a set of proposal pdfs is available, the way in which the samples can be drawn and weighted is not unique, unlike the case of using a single proposal. Indeed, different MIS algorithms in the literature (both adaptive and non-adaptive) have implicitly and independently interpreted the sampling and weighting procedures in different ways \citep{Owen00,Cappe04,Cappe08,elvira2015efficient,APIS15,CORNUET12,bugallo2017adaptive}. 
Namely, there are several possible combinations of {sampling} and {weighting} schemes, when a set of proposal pdfs is available, which lead to valid MIS approximations of the target pdf. However, these different possibilities can largely differ in terms of performance of the corresponding estimators.
   
In this paper, we introduce a unified framework for MIS schemes, providing a general theoretical description of the possible sampling and weighting procedures when a set of proposal pdfs is used to produce an IS approximation. Within this unified context, it is possible to interpret that all the MIS algorithms draw samples from an equally-weighted mixture distribution obtained from the set of available proposal pdfs. 
Three different sampling approaches and five different functions to calculate the weights of the generated samples are proposed and discussed. Moreover, we state two basic rules for possibly devising new valid sampling and weighting strategies within the proposed framework. All the analyzed combinations of sampling/weighting provide consistent estimates of the parameters of interest.

The proposed generalized framework includes all of the existing MIS methodologies that we are aware of (applied within different algorithms, e.g. in \citep{elvira2015efficient,Cappe04,CORNUET12,APIS15,martino2017layered,elvira2017improving}) and allows the design of novel techniques (here we propose three new schemes, but more can be introduced). An exhaustive theoretical analysis is provided by introducing general expressions for sampling and weighting in this generalized MIS context, and by proving that they yield consistent estimators. Furthermore, we compare the performance of the different MIS schemes (the proposed and existing ones) in terms of the variance of the estimators.

The rest of this paper is organized as follows. In Section \ref{sec_problem}, we describe the problem and we revisit the standard IS methodology. In Section \ref{sec_sampling}, we discuss the sampling procedure in MIS, propose three new sampling strategies, and
%we 
analyze some distributions of interest. In Section \ref{sec_weighting}, we propose five different weighting functions, some of them completely new, and 
%we 
show their validity. The different combinations of sampling/weighting strategies are analyzed in Section \ref{sec_MIS_schemes}, establishing the connections with
% the 
existent MIS schemes, 
%in the literature, 
and describing three novel MIS schemes.
In Section \ref{sec_variance}, we analyze the performance of the different MIS schemes in terms of the variance of the estimators.
{Then, Section \ref{sec:applyingMIS} discusses some relevant aspects about the application of the proposed MIS schemes, including their use in adaptive settings.
Finally, Section \ref{sec_results} presents some descriptive numerical examples where the different MIS schemes are simulated, and Section \ref{sec_conclusions} contains some concluding remarks. A running example is introduced in Section \ref{sec_sampling} and continued in Section \ref{sec_weighting}, Section \ref{sec_MIS_schemes}, Section \ref{sec_variance} and Section \ref{sec_results} in order to clarify the flow of the paper. In addition, we perform numerical simulations on the running example, where the proposal pdfs are intentionally well chosen, to evidence the significant effects produced by the different interpretations of the sampling and weighting schemes.

%%%%%%%%%%%%%%%%%%%%%%%%%%%%%%%%%%%%%%%%%
%%%%%%%%%%%%%%%%%%%%%%%%%%%%%%%%%%%%%%%%%
\section{Problem Statement and Background}
\label{sec_problem}
%%%%%%%%%%%%%%%%%%%%%%%%%%%%%%%%%%%%%%%%%
%%%%%%%%%%%%%%%%%%%%%%%%%%%%%%%%%%%%%%%%%

Let us consider a system characterized by a vector of $d_x$ unknown parameters, $\x\in \mathbb{R}^{d_x}$, and a set of $d_y$ observed data, ${\bf y}\in \mathbb{R}^{d_y}$.\footnote{Vectors are denoted by bold-faced letters, e.g., $\x$, while regular-faced letters are used for scalars, e.g., $x$.} A general objective is to extract the complete information about the latent state, $\x$, given the observations, ${\bf y}$, by means of  studying  the posterior distribution defined as
\begin{equation}
	\normalized{\pi}(\x| {\bf y})
		= \frac{\ell({\bf y}|\x) h(\x)}{Z({\bf y})} \propto \pi(\x|{\bf y})=\ell({\bf y}|\x) h(\x),
\label{eq_posterior}
\end{equation}
where $\ell({\bf y}|\x)$ is the likelihood function, $h(\x)$ is the prior pdf, and $Z({\bf y})$ is the normalization factor.\footnote{In the sequel, to simplify the notation, the dependence on ${\bf y}$ is removed, e.g., $Z \equiv Z(\y)$.} 
The objective is to approximate the pdf of interest (referred to as target pdf) by Monte Carlo-based sampling \citep{kong2003theory,Robert04,Liu04b,Owen13}. The resulting approximation of 
$\normalized \pi(\x)$ will be denoted as ${\hat \pi}(\x)$ and will be attained using IS techniques.
 
%%%%%%%%%%%%%%%%%
\subsection{Standard importance sampling}
%%%%%%%%%%%%%%%%%

IS is a general Monte Carlo technique for the approximation of a pdf of interest by a random measure
composed of samples and weights \citep{Robert04}. In its original formulation, a set of $N$ samples, $\{\x_n\}_{n=1}^N$, is drawn from a {single} proposal pdf, $q(\x)$, with heavier tails than those of the target pdf, $\pi(\x)$.
A particular sample, ${\x_n}$, is assigned an importance weight given by
 \begin{equation} 
	w_n= \frac{\pi(\x_n)}{{q(\x_n)}}, \quad n=1,\ldots,N,
\label{is_weights_static}
\end{equation} 
which represents the ratio between the target pdf, {$\pi$}, and the proposal pdf, {$q$}, both evaluated at $\x_n$. 
The samples and weights form the random measure $\chi=\{\x_n,w_n\}_{n=1}^N$ that approximates the measure of the target pdf as
\begin{equation}
	{\hat \pi}_{\text{IS}} (\x)= \frac{1}{N\hat{Z}}  \sum_{n=1}^N w_n \delta_{\x_n}(\x),
\label{approx_pi}
\end{equation}
where  $\delta_{\x_n}(\x)$ is the unit delta measure 
concentrated at $\x_n$ and $\hat{Z}=\frac{1}{N}\sum_{j=1}^N w_j$ is an unbiased estimator of $Z=\int \pi(\x) d\x$ \citep{Robert04}. Fig. \ref{fig_approx} (a) displays an example of a target pdf and a proposal pdf, as well as the samples and
weights that form a random measure approximating the posterior. Note that, unlike Markov Chain Monte Carlo (MCMC) methods, all the generated samples are used to build the estimators, e.g., there is no burn-in period.

\begin{figure*}[htp]
%\centering
\subfigure[Single proposal pdf (standard IS).]{
\includegraphics[scale=0.24]{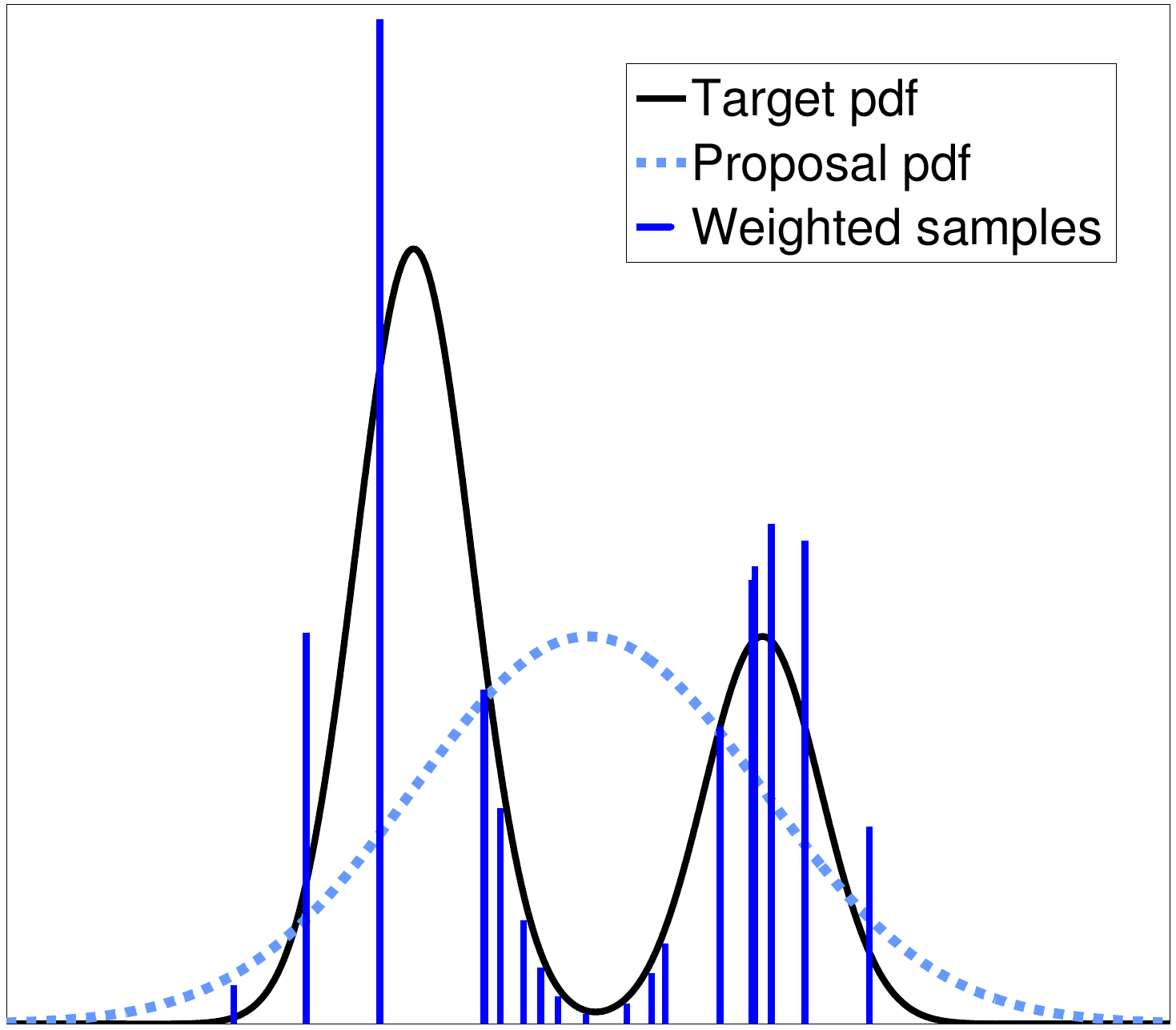}
} 
%\centering
\subfigure[Two proposal pdfs (MIS).]{
\includegraphics[scale=0.24]{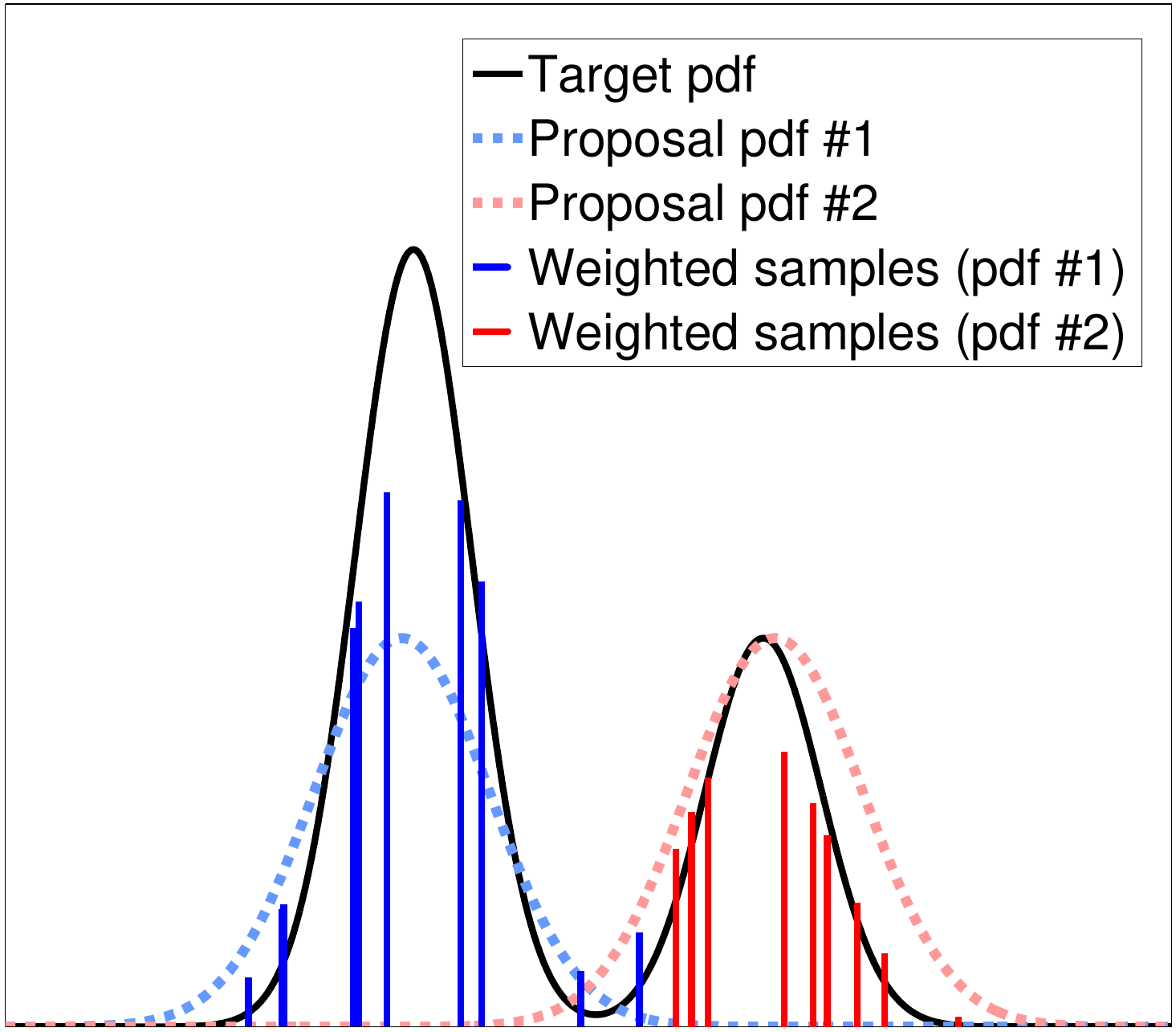}
}\qquad
\caption{Approximation of the target pdf, $\pi(\x)$, by the random measure $\chi$.}
\label{fig_approx}
\end{figure*}

\subsection{Estimators in importance sampling}
%
%The weighting step is used to evaluate the adequacy of the samples generated from the proposal pdfs with respect to the general objective of approximating the target $\pi(\x)$. 
Let us consider the integral $I = \int g(\x) \normalized \pi(\x) d\x$, where $g$ is any integrable function w.r.t. $\normalized \pi(\x)$. When $Z$ is known, an unbiased IS estimator of $I$ is given by 
\begin{equation}
\hat{I} = \frac{1}{NZ} \sum_{n=1}^N w_n  g(\x_n).
%\label{eq_IS_estimator_unnorm}
\label{eq_IS_estimator_unnorm}
\end{equation}
%
%where $w_n$ is the importance weight of the $n$-th sample, $\x_n$, and $Z=\int \normalized \pi(\x) d\x$ is the normalizing constant. %, i.e. a specific function evaluated in $\x_n$. 
Otherwise, if the target distribution is only known up to the normalizing constant, $Z$, one can use the self-normalized estimator 
\begin{equation}
	\tilde{I} = \frac{1}{N\hat{Z}}  \sum_{n=1}^N w_n g(\x_n),
\label{eq_IS_estimator_norm}
\end{equation}
where $Z$ is approximated by the estimate
\begin{eqnarray}
\hat{Z} = \frac{1}{N} \sum_{n=1}^N w_n.
\label{eq_Z_estimator}
\end{eqnarray}
Under some mild assumptions regarding the tails of the proposal and target distributions, $Z$ is an unbiased and consistent estimator of $Z$, and $\tilde{I}$ is a consistent estimator of $I$ \citep{Robert04}. Furthermore, the variance of $\hat{I}$ and $\tilde{I}$ is directly related to the discrepancy between $\normalized\pi(\x)|g(\x)|$ and $q(\x)$ \citep{Robert04, kahn1953methods}. For a general $g$, a common strategy is decreasing the mismatch between the proposal $q(\x)$ and the target $\normalized \pi(\x)$. A very common strategy consists in using several proposal pdfs. 

%%%%%%%%%%%%%%%%%%%%%%%%%%%%%%%%%%%%%%%%%
%%%%%%%%%%%%%%%%%%%%%%%%%%%%%%%%%%%%%%%%%
\section{Sampling in Multiple Importance Sampling}
\label{sec_sampling}
%%%%%%%%%%%%%%%%%%%%%%%%%%%%%%%%%%%%%%%%%
%%%%%%%%%%%%%%%%%%%%%%%%%%%%%%%%%%%%%%%%%
MIS schemes consider a set of $N$ proposal pdfs,
$\{q_{{n}}(\x)\}_{n=1}^N \equiv \{q_1(\x),\ldots,q_N(\x) \}$, 
and proceed by drawing $M$ samples, $\{\x_m\}_{m=1}^M$ (where $M\neq N$, in general) and properly weighting them. As a visual example, Fig. \ref{fig_approx} (b) displays a target pdf and two proposal pdfs, as well as the samples and
weights that form a random measure approximating the posterior. 

It is in the way that the sampling and the weighting are performed that different variants of MIS  can be devised. In this section, we focus on the generation of samples $\{\x_m\}_{m=1}^M$.  For clarity in the explanations and the theoretical proofs, we always consider 
$M=N$, i.e., the number of samples to be generated coincides with the number of  proposal pdfs. All the considerations can be automatically extended to the case with $M=kN$ samples, with $k\in \mathbb{N}^+$. The sampling and weighting procedures that we propose in the following can be easily applied to each block of $N$ samples. Then, the estimators described in the previous section would use all the $M=kN$ samples. In this work, we consider that we have no prior information about the adequacy of the proposals. Hence, all the proposals will be equally used for sampling purposes (see more details in Section \ref{sec:samplingMixture}).
The use of the complete set of $N$ proposal pdfs {with no prior information about them} can {also represent  a} single equally weighted mixture proposal,
\begin{equation}
\psi(\x)\equiv\frac{1}{N}\sum_{n=1}^N q_n(\x).
\label{eq_mixture}
\end{equation}
Unequal weights could also be considered in the mixture. In \citep{he2014optimal}, the weights can be optimized to minimize the variance for a certain integrand.

%%%%%%%%%%%%%%%%%
\subsection{Sampling from the set of proposal pdfs}
\label{sec:samplingMixture}
%%%%%%%%%%%%%%%%%

Let us consider a generic mechanism for the simulation of $N$ samples from the set of $N$ proposals. 
Starting with $n=1$:
\begin{enumerate}
\item Choose an index $j_n\in\{1,\ldots, N\}$, which corresponds to the selection of the proposal pdf $q_{j_n}$. % using some suitable approach. This corresponds to the selection of a proposal pdf, $q_{j_n}$.
\item Generate a sample $\x_n$ from the selected proposal pdf, i.e., $\x_n\sim q_{j_n}(\x_n)$.
\item Set $n=n+1$ and go to step 1.
\end{enumerate}
Note that, in step 1, the probabilities associated to each possible value of $j_n$ are not specified yet. The graphical model corresponding to this sampling scheme is shown in Fig. \ref{fig_sampling}.
\begin{figure}[htb]
\centering
{\includegraphics[scale=0.6]{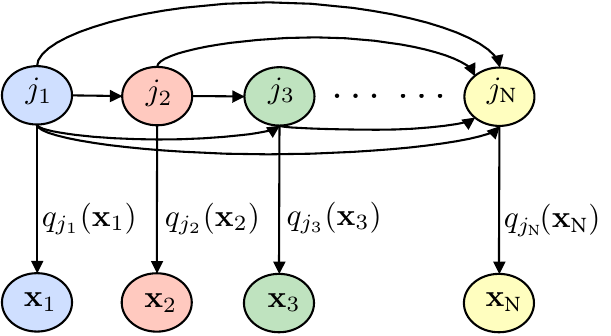} } 
\qquad
\caption{Graphical model associated to the generic sampling scheme.}
\label{fig_sampling}
\end{figure}

Therefore, obtaining the set of samples  $\{\x_n\}_{n=1}^N\equiv \{\x_1,...,\x_N\}$ is in general a two step sequential procedure. First,  the $n$-th index $j_n$ is drawn according to some conditional pdf, $P(j_n|j_{1:n-1})$, where $j_{1:n-1}\equiv \{j_1,\ldots, j_{n-1}\}$ is the sequence of the previously generated indexes.\footnote{We use a simplified argument-wise notation, where $p(\x_n)$ denotes the pdf of the continuous random variable (r.v.) $\X_n$, while $P(j_n)$ denotes the probability mass function (pmf) of the discrete r.v. $J_n$. Also, $p(\x_n,j_n)$ denotes the joint pdf and $p(\x_n|j_n)$ is the conditional pdf of $\X_n$ given $J_n=j_n$. If the argument of $p(\cdot)$ is different from $\x_n$, then it denotes the evaluation of the pdf as a function, e.g., $p(\z|j_n)$ denotes the pdf $p(\x_n|j_n)$ evaluated at $\x_n=\z$.}
Then, the $n$-th sample is drawn from the selected proposal pdf as $\x_n \sim p(\x_n|j_n)$. The joint probability distribution of the current sample and all the indexes used to generate the  samples from $1$ to $n$ is
 \begin{eqnarray}
 \label{jointPDF}
p(\x_n,j_{1:n})&=&P(j_{1:n-1})P(j_n|j_{1:n-1}) p(\x_n|j_n)\nonumber \\
&=&P(j_{1})\left[\prod_{i=2}^n P(j_i|j_{1:i-1})\right] q_{j_n}(\x_n),
\end{eqnarray}
where $p(\x_n|j_n)=q_{j_n}(\x_n)$ is the $n$-th selected proposal pdf, $q_{j_n}(\x_n)$.
%

%%%%%%%%%%%%%%%%%
\subsection{Selection of the proposal pdfs}
%%%%%%%%%%%%%%%%%

In the sequel, we describe three mechanisms for obtaining the sequence of indexes, $j_{1:N}$. 
All the mechanisms share the property that 
 \begin{equation}
\frac{1}{N}\sum_{n=1}^N P({J_n}=k)=\frac{1}{N}, \quad \forall k\in \{1,\ldots,N\},
\end{equation}
i.e., all the indexes have the same (marginal) probability of being selected.

\begin{itemize}

\item[$\mathcal{S}_1$:]{\it Random index selection with replacement}:
The $N$ indexes are independently drawn from the set $\{1,\ldots,N\}$ with equal probability.
Thus, we have  
\begin{equation}
P(j_n|j_{1:n-1}) = P(j_n)=\frac{1}{N}.
\label{eq_index_conditional_1}
\end{equation}
With this type of index sampling, there may be more than one sample drawn from some proposal, and there may be proposal pdfs that are not used at all.

%\subsubsection{Random index selection without replacement}
\item[$\mathcal{S}_2$:]{\it Random index selection without replacement}:
The indexes are uniformly and sequentially drawn from different sets as $j_1\in \mathcal{I}_1=\{1,\ldots,N\}$, ... , $j_n\in\mathcal{I}_n=\{1,\ldots,N\} \setminus\{j_{1:n-1}\}$, i.e., removing the proposals previously used. Hence, the conditional probability mass function (pmf) of the $n$-th index given the previous ones is now
\begin{gather}
\label{EqMonica}
P({J_n}=k|j_{1:n-1})=\left\{
\begin{split}
&\frac{1}{N-n+1}   \quad  \mbox{ if } k\in \mathcal{I}_n, \\
&0   \quad\quad\quad\quad\quad \mbox{ } \mbox{ if } k\notin \mathcal{I}_n, 
\end{split}
\right.
\end{gather}
where $|\mathcal{I}_n|=N-n+1$. Note that the marginal pmf of the $j$-th index is still given by \eqref{eq_index_conditional_1}.\footnote{There are $N!$ equiprobable configurations (permutations) of the sequence $\{j_1,\ldots,j_N \}$, and in $(N-1)!$ the $k$-th index is drawn at the $n$-th position $\forall k,n = 1,\ldots,N$. Therefore, $P(J_n=k) = \frac{(N-1)!}{N!}=\frac{1}{N}$ $\forall k,n = 1,\ldots,N$.}
However, exactly one sample is drawn from each of the proposal pdfs by following this strategy.

%\subsubsection{Deterministic index selection without replacement} 
\item[$\mathcal{S}_3$:]{\it Deterministic index selection without replacement}:
This sampling is a particular case of sampling $\mathcal{S}_2$, where a fixed deterministic sequence of indexes is drawn. For instance, and without loss of generality: $j_1=1, j_2=2, \ldots, j_n=n, \ldots, j_N=N$. Therefore, $\x_n\sim q_{j_n}({\x_n})=q_n(\x_n)$, and the conditional pmf of the $n$-th index given the $n-1$ previous ones becomes  
\begin{equation}
P(j_n|j_{1:n-1}) = P(j_n) = \mathbbm{1}_{j_n=n},
\label{eq_index_conditional_3}
\end{equation} 
where $ \mathbbm{1}$ denotes the indicator function.
%This procedure could be interpreted as a particular case of the previous index selection scheme without replacement. 
%
Again, each of the $N$ proposal pdfs is used to generate exactly one sample of the set $\{\x_n \}_{n=1}^N$. 
This index selection procedure has been used by several MIS algorithms (e.g., in \citep{CORNUET12,elvira2017improving}), and it is also implicitly used in some particle filters (PFs), such as the bootstrap PF \citep{Gordon93}.

\end{itemize}

The connexions of the sampling mechanisms with some resampling schemes are discussed in Appendix \ref{appendix_resampling}.

%%%%%%%%%%%%%%%%% 
\subsection{Running example}
\label{running_example_sampling}
%%%%%%%%%%%%%%%%% 

Let us consider $N=3$ Gaussian proposal pdfs $q_1(x)=\mathcal{N}(x;\mu_1,\sigma_1^2)$,  $q_2(x)=\mathcal{N}(x;\mu_2,\sigma_2^2)$ and $q_3(x)=\mathcal{N}(x;\mu_3,\sigma_3^2)$ with predefined means and variances. In $\mathcal{S}_1$, a possible realization of the indexes is the sequence $\{j_1,j_2,j_3 \} = \{3,3,1\}$. Therefore, in this situation, $\x_1 \sim q_3$, $\x_2 \sim q_3$, and $\x_3 \sim q_1$. In  $\mathcal{S}_2$, the realization could {result from} the permutation $\{j_1,j_2,j_3 \} = \{3,1,2\}$. In  $\mathcal{S}_3$, the sequence is deterministically {obtained as} $\{j_1,j_2,j_3 \} = \{1,2,3\}$.

%%%%%%%%%%%%%%%%%
\subsection{Distributions of interest of the $n$-th sample, $\x_n$} % Distributions of $\x_n$
\label{DistInXn}
%%%%%%%%%%%%%%%%%

In the following, we discuss some important distributions related to the set of samples drawn. These distributions are of utmost importance to understand the different methods for weighting the samples discussed in the following section.

Note that the distribution of the $n$-th sample given all the knowledge of the process up to that point is $p(\x_n|j_{1:n-1},\x_{1:n-1})=p(\x_n|j_{1:n-1})$. In $\mathcal{S}_1$, this distribution corresponds to $p(\x_n|j_{1:n-1}) =\psi({\x_n})$. We recall that $\psi$ is the mixture of proposals defined in Eq. \eqref{eq_mixture}. In $\mathcal{S}_2$, we have $p(\x_n|j_{1:n-1}) = \frac{1}{|\mathcal{I}_n|} \sum_{\forall k\in \mathcal{I}_n} q_k(\x)$. Finally, under $\mathcal{S}_3$, $p(\x_n|j_{1:n-1}) = q_n(\x_n)$.
Once the $n$-th index $j_n$ has been selected, the $n$-th sample, $\x_n$, is distributed as $p(\x_n|j_n) = q_{j_n}(\x_n)$ in any sampling method within the proposed framework. The marginal distribution of this $n$-th sample, $\x_n$, is then given by
\begin{equation}
\label{Eqmarginal}
p(\x_n) = \sum_{k=1}^{N}q_k(\x_n)P(J_n=k),%\rfootnote{Consider removing the ``$j_n = k$.}
\end{equation}
where we have used the fact that $p(\x_n|J_n=k) = q_k(\x_n)$, and the marginal distribution, $P(J_n=k)$, depends on the sampling method. When randomly selecting the indexes ($\mathcal{S}_1$ or $\mathcal{S}_2$), $P(J_n=k)=\frac{1}{N}, \forall n,k$, and thus $p(\x_n) = \frac{1}{N} \sum_{k=1}^Nq_k(\x_n) = \psi(\x_n)$. In the case of the deterministic index selection ($\mathcal{S}_3$), $P(J_n=k) = \mathbbm{1}_{k=n}$, and thus $p(\x_n) = q_n(\x_n)$, i.e., the distribution of the r.v. $\X_n$ is the $n$-th proposal pdf, and not the whole mixture. % as in the other two sampling schemes with random index selection.

%%%%%%%%%%%%%%%%%
\subsection{Distributions of interest beyond $\x_n$}
\label{sec_BeyondXn}
%%%%%%%%%%%%%%%%%

The traditional IS approach focuses just on the distribution of the r.v. $\X_n$. In MIS, we are also interested in the statistical properties of the set of samples, regardless of their index $n$, since the $N$ samples are used jointly in the estimators, regardless their order of appearance. Hence, we introduce a generic r.v.,
\begin{equation}
\X=\X_n \quad \mbox{ with } \quad  n \sim \mathcal{U}\{1,2,\ldots,N\},
\label{eq_pdf_x}
\end{equation}
where $\mathcal{U}\{1,2,\ldots,N\}$ is the discrete uniform distribution on the set $\{1,2,\ldots,N\}$. The density of $\X$ is then given by
\begin{equation}
\label{eq_events_union}
f(\x) = \frac{1}{N} \sum_{n=1}^N p_{\x_n}(\x) = \psi(\x),
\end{equation}
where $p_{\x_n}(\x)$ denotes the marginal pdf of $\X_n$, given by Eq. \eqref{Eqmarginal}, evaluated at $\x$, and $\psi(\x)$ is the mixture pdf.\footnote{For the sake of clarity, in Eq. \eqref{eq_events_union} we have used the notation $p_{\x_n}(\x)$, instead of $p(\x)$ as in Eq. \eqref{Eqmarginal} and the rest of the paper, to denote the marginal pdf of $\X_n$ evaluated at $\x$.}
Moreover, one can also obtain the conditional pdf of $\X$ given the sequence of indexes as
\begin{equation}
\label{eq_exotic_mixture}
f(\x|j_{1:N}) = \frac{1}{N} \sum_{k=1}^N p_{\x_k}(\x|j_{1:N}) = \frac{1}{N} \sum_{k=1}^N q_{j_k}(\x).
\end{equation}
Note that, in this case, $f(\x|j_{1:N}) = \psi(\x)$ for the schemes without replacement at the index selection ($\mathcal{S}_2$ and $\mathcal{S}_3$), but $f(\x|j_{1:N}) =\frac{1}{N}\sum_{n=1}^N q_{j_n}(\x)$ for the case with replacement ($\mathcal{S}_1$), i.e., some proposal pdfs may not appear while others may appear repeated.

\begin{remark} (Sampling): \label{Rem1} In the proposed framework, we consider valid, any sequential sampling scheme for generating the set $\{\X_1,\ldots,\X_N\}$ such that the pdf of the r.v. $\X$ defined in Eq. \eqref{eq_pdf_x} is given by $\psi(\x)$. 
Further considerations about the r.v. $\X$ and connections with variance reduction methods \citep{Robert04,Owen13} are given in Appendix \ref{sec_further_obs}. \label{remark_sampling} \label{remark_sampling}\end{remark}

Table \ref{table_distributions} summarizes all the distributions of interest. Note that, the pdf of the r.v. $\X$ is always the mixture $\psi(\x)$, but different sampling procedures yield different conditional and marginal distributions that will be exploited to justify different strategies for calculation of the importance weights in the next section.
Finally, the last row of the table shows the joint distribution $p(\x_{1:N})$ of the variables $\X_1,\ldots,\X_N$, i.e., $p(\x_{1:N})=\prod_{n=1}^N\psi({\x_n})$ and $p(\x_{1:N})=\prod_{n=1}^Nq_n({\x_n})$ for $\mathcal{S}_1$ and $\mathcal{S}_3$, respectively. For $\mathcal{S}_2$,
\begin{equation}
\label{eq_joint_monica}
p(\x_{1:N}) = \psi(\x_1) \prod_{n=2}^N \frac{1}{|\mathcal{I}_{n}|} \sum_{\ell \in \mathcal{I}_{n}} q_{\ell}(\x_n),
\end{equation}
with $\mathcal{I}_n=\{1,\ldots,N\} \setminus\{j_{1:n-1}\}$.

%%%%%%%%%%%%%%%%%%%%%%%%%%%%%%%%%%%%%%%%%
%%%%%%%%%%%%%%%%%%%%%%%%%%%%%%%%%%%%%%%%%
\section{Weighting in Multiple Importance Sampling}
\label{sec_weighting}
%%%%%%%%%%%%%%%%%%%%%%%%%%%%%%%%%%%%%%%%%
%%%%%%%%%%%%%%%%%%%%%%%%%%%%%%%%%%%%%%%%%

Our approach is based on analyzing which weighting functions yield \emph{proper} MIS estimators. 
We consider that the set of weighting functions $\{w_n\}_{n=1}^N$ is proper if
\begin{eqnarray}
\frac{E_{p(\x_{1:N},j_{1:N})}\left[\frac{1}{N} \sum_{n=1}^N w_n  g(\x_n)\right]}{E_{p(\x_{1:N},j_{1:N})}\left[\frac{1}{N} \sum_{n=1}^N w_n \right]} =  E_{\pi}[g(\x)].
\end{eqnarray}
This is equivalent to imposing the restriction
\begin{eqnarray}
\frac{E_{p(\x_{1:N},j_{1:N})}\left[Z \hat I\right]}{E_{p(\x_{1:N},j_{1:N})}\left[\hat Z \right]} = I,
\end{eqnarray}
which is fulfilled if $\E[\hat I]=I$ and $\E[\hat Z]=Z$. 
In order to narrow down the set of all possible proper functions, we impose the weight function to have the (deterministic) structure $w_n = \frac{\pi(\x_n)}{\varphi_{\mathcal{P}_n}(\x_n)}$,
where $\varphi_{\mathcal{P}_n}$
is a generic function parametrized by a set of parameters $\mathcal{P}_n \subseteq \{j_1,...,j_N \}$ (further details are given below). 
Note that $\varphi_{\mathcal{P}_n}$ plays the role of the \emph{interpreted} proposal from which $\x_n$ is drawn.
It is on this interpretation of what the proposal pdf used for the generation of the sample is that different weighting strategies can be devised.\footnote{In an even more generalized framework, $w_n$ could hypothetically depend on more than one sample of the set $\x_{1:N}$ if one could properly design the function $\varphi_n$ that yields {valid} estimators.} 
{The expectation of the generic estimator $\hat{I}$ of Eq. \eqref{eq_IS_estimator_unnorm} {can be} computed as}
{\small
\begin{eqnarray}
\E[\hat I] &=&  \frac{1}{ZN}\sum_{n=1}^N \sum_{j_{1:N}} \int \frac{\pi(\x_n)g(\x_n)}{\varphi_{\mathcal{P}_n}(\x_n)} P(j_{1:N}) p(\x_n|j_n) d\x_n,
\label{general_estimator}
\end{eqnarray}
}
{where we use the joint distribution of indexes and samples {from} Eq. \eqref{jointPDF}.}
\begin{remark} (Weighting): \label{Rem2} In the proposed framework, we consider valid any weighting scheme (i.e., any function $\varphi_{\mathcal{P}_n}$ at the denominator of the weight) that yields $\E[\hat I] \equiv I$  in Eq. \eqref{general_estimator}. \end{remark}

%%%%%%%%%%%%%%%%%%%
\subsection{Weighting functions}
%%%%%%%%%%%%%%%%%%%
\label{sec_wfunctions}

Here we present several possible functions $\varphi_{\mathcal{P}_n}$, that yield an unbiased estimator of ${I}$ according to Eq. \eqref{general_estimator}. The different choices for {$\varphi_{\mathcal{P}_n}$}, used in the denominator of the weight $w_n = \frac{\pi(\x_n)}{\varphi_{\mathcal{P}_n}(\x_n)}$, come naturally from the sampling densities discussed in Section \ref{sec_sampling}. More precisely, they correspond to the five different functions in Table \ref{table_distributions} related to the distributions of the generated samples. From now on, $p(\cdot)$ and $f(\cdot)$, which correspond to the pdfs of $\X_n$ and $\X$ respectively, are used as functions and the argument represents a functional evaluation.
\begin{enumerate}
\item[$\mathcal{W}_1$:] %First option: ``Natural'' or ``iterative'' weights, 
$\varphi_{\mathcal{P}_n}(\x_n) =  \varphi_{j_{1:n-1}}(\x_n) = p(\x_n|j_{1:n-1})$

Since the sampling process is sequential, this option is of particular interest. It interprets the proposal pdf as the conditional density of $\x_n$ given all the previous proposal indexes of the sampling process.
\item[$\mathcal{W}_2$:] %\item Second option,
{$\varphi_{\mathcal{P}_n}(\x_n) = \varphi_{j_n}(\x_n)=  p(\x_n|j_n) =  q_{j_n}(\x_n)$}

It interprets that if the index $j_n$ is known, $\varphi_{\mathcal{P}_n}$ is the proposal $q_{j_n}$.
\item[$\mathcal{W}_3$:] %\item Third option, (usual MIS weights), 
{$\varphi_{\mathcal{P}_n}(\x_n) = p(\x_n)$}

It interprets that $\x_n$ is a realization of the marginal $p(\x_n)$. This is probably the most ``natural'' option (as it does not assume any further knowledge in the generation of $\x_n$) and is a usual choice for the calculation of the weights in some of the existing MIS schemes (see Section \ref{sec_MIS_schemes}).
\item[$\mathcal{W}_4$:] %\item Third option, (usual MIS weights), 
{$\varphi_{\mathcal{P}_n}(\x_n) = \varphi_{j_{1:N}}(\x_n) = f(\x_n|j_{1:N}) = \frac{1}{N}\sum_{k=1}^N q_{j_k}(\x_n) $}

This interpretation makes use of the distribution of the r.v. $\X$ conditioned on the whole set of indexes (defined in Section \ref{sec_BeyondXn}). 

\item[$\mathcal{W}_5$:] {$\varphi_{\mathcal{P}_n}(\x_n) =\varphi(\x_n) = f(\x_n) = \frac{1}{N}\sum_{k=1}^N q_k(\x_n) $}

This option considers that all the $\x_n$ are realizations of the r.v. $\X$ defined in Section \ref{sec_BeyondXn} (see Appendix \ref{sec_further_obs} for a thorough discussion of this interpretation). 

\end{enumerate}

Table \ref{table_weights_options} summarizes the discussed functions {$\varphi_{\mathcal{P}_n}$}. Although some of the selected functions {$\varphi_{\mathcal{P}_n}$} may seem more natural than others, all of them yield valid estimators. The proofs can be found in Appendix \ref{appendix_unbiasedness}.  %that can be used to evaluate the denominator for the weight calculation, $w_n = \frac{\pi(\x_n)}{\varphi_{\mathcal{P}_n}(\x_n)}$. 
Other proper weighting functions are described in Section \ref{sec_partial}.

\begin{table*}[!t]
\caption{Summary of the different generic functions $\varphi_{\mathcal{P}_n}$. The distributions depend on the specific sampling scheme used for drawing the samples as shown in Table \ref{table_sampling_weights}. 
}
%\vspace{-0.5cm}
\begin{center}
\begin{tabular}{|c|c|c|c|c|c|}
\hline
\multirow{2}{*}{$\varphi_{\mathcal{P}_n}$}   & $\mathcal{W}_1$  &  $\mathcal{W}_2$  & $\mathcal{W}_3$  & $\mathcal{W}_4$ & $\mathcal{W}_5$ \\
%&Opt. 1& Opt. 2 &Opt. 3 &Opt. 4 & Opt. 5 \\
%\hline
%\hline
& $p(\x_n|j_{1:n-1})$ &$p(\x_n|j_n)$ & $p(\x_n)$ & $f(\x|j_{1:N})$ & $f(\x)$  \\ 
\hline \hline
$w_n = \frac{\pi(\x_n)}{\varphi_{\mathcal{P}_n}(\x_n)}$ & $\frac{\pi(\x_n)}{p(\x_n|j_{1:n-1})}$ & $\frac{\pi(\x_n)}{p(\x_n|j_n)}$ & $\frac{\pi(\x_n)}{p(\x_n)}$ & $\frac{\pi(\x_n)}{f(\x_n|j_{1:N})}$ & $\frac{\pi(\x_n)}{f(\x_n)}$  \\ 
\hline
\end{tabular}
\end{center}
\label{table_weights_options}
\end{table*}

%%%%%%%%%%%%%%%%% 
\subsection{Connection with Liu-properness of single IS}
\label{sec_weight_interpretation}

We consider the definition of properness by Liu \cite[Section 2.5]{Liu04b} and we extend (or relax) it to the MIS scenario. Namely, Liu-properness in standard IS states that a weighted sample $\{\x_n,w_n\}$ drawn from a single proposal $q$ is proper if, for any square integrable function $g$,
\begin{eqnarray}
\frac{E_q[g(\x)w(\x)]}{E_q[\pi(\x)]} =  E_{\pi}[g(\x)],
\label{eq_liu_properness}
\end{eqnarray}
i.e., $w$ can be in any form as long as the condition of Eq. \eqref{eq_liu_properness} is fulfilled. Note that, for a deterministic weight assignment, the only proper weights are the ones considered by the standard IS approach. 
Note also that the MIS properness is a relaxation of the one proposed by Liu, i.e., any Liu-proper weighting scheme is also proper a according to our definition, but not vice versa.

%%%%%%%%%%%%%%%%% 
\subsection{Running example}
\label{running_example_weighting}
%%%%%%%%%%%%%%%%% 

Here we follow the running example of Section \ref{running_example_sampling}. For instance, let us consider the sampling method $\mathcal{S}_1$ and {let the realization of the indexes be} the sequence $\{j_1,j_2,j_3 \} = \{3,3,1\}$. Under the weighting scheme  $\mathcal{W}_2$, the weights would be computed as $w_1 = \frac{\pi(\x_1)}{q_3(\x_1)}$, $w_2 = \frac{\pi(\x_2)}{q_3(\x_2)}$, and $w_3 = \frac{\pi(\x_3)}{q_1(\x_3)}$. However, under $\mathcal{W}_4$, $w_1 = \frac{\pi(\x_1)}{\frac{1}{3} \left(q_1(\x_1) + 2 q_3(\x_1) \right)}$, $w_2 = \frac{\pi(\x_2)}{\frac{1}{3} \left(q_1(\x_2) + 2 q_3(\x_2 \right)}$, and $w_3 = \frac{\pi(\x_3)}{\frac{1}{3} \left(q_1(\x_3) + 2 q_3(\x_3) \right)}$. Note that all weighing schemes require the same number of target evaluations (which are usually more expensive) but different numbers of proposal evaluations. 

%%%%%%%%%%%%%%%%%%%%%%%%%%%%%%%%%%%%%%%%%
%%%%%%%%%%%%%%%%%%%%%%%%%%%%%%%%%%%%%%%%%
\section{Multiple Importance Sampling Schemes}
\label{sec_MIS_schemes}
%%%%%%%%%%%%%%%%%%%%%%%%%%%%%%%%%%%%%%%%%
%%%%%%%%%%%%%%%%%%%%%%%%%%%%%%%%%%%%%%%%%

In this section, we describe the different possible combinations of the three sampling strategies considered in {Section \ref{sec_sampling}} and the five weighting functions devised in Section \ref{sec_weighting}. Once combined, the fifteen possibilities only lead to six unique MIS methods. Three of the methods are associated to the sampling scheme with replacement ($\mathcal{S}_1$), while the other three methods correspond to the sampling schemes without replacement ($\mathcal{S}_2$ and $\mathcal{S}_3$). 
Table \ref{table_sampling_weights} summarizes the possible combinations of sampling/weighting and indicates the resulting MIS method within brackets. The six MIS methods are labeled either by an \texttt{R} (sampling with \textit{replacement}) or with an \texttt{N} (sampling with \textit{no} replacement). We remark that these schemes are examples of proper MIS techniques fulfilling Remarks \ref{Rem1} and \ref{Rem2}.

%%%%%%%%%%%%%%%%%
\subsection{MIS schemes with replacement}
%%%%%%%%%%%%%%%%%

In all {\tt R} schemes, the $n$-th sample is drawn with replacement (i.e., $\mathcal{S}_1$) from the whole mixture $\psi$:

\begin{description}
\item[{\tt [R1]}:] \emph{Sampling with replacement, $\mathcal{S}_1$, and weight denominator $\mathcal{W}_2$:}\\ 
For the weight calculation of the $n$-th sample, only the proposal selected for generating the sample is evaluated in the denominator. 
\item[{\tt [R2]}:] \emph{Sampling with replacement, $\mathcal{S}_1$, and weight denominator $\mathcal{W}_4$:}\\
With the $N$ selected indexes $j_n$, for  $n=1,...,N$, one forms a mixture comprising all the corresponding proposal pdfs. The weight calculation of the $n$-th sample considers this {\it a posteriori} mixture evaluated at the $n$-th sample in the denominator, i.e., some proposals might be used more than once while other proposals might not be used.
\item[{\tt [R3]}:] \emph{Sampling with replacement, $\mathcal{S}_1$, and weight denominator $\mathcal{W}_1$, $\mathcal{W}_3$, or $\mathcal{W}_5$:} \\
For the weight calculation of the $n$-th sample, the denominator applies the value of the $n$-th sample to the whole mixture {$\psi$} composed of the set of initial proposal pdfs (i.e., the function in the denominator of the weight does not depend on the sampling process). This is the approach followed by the so called mixture PMC method \citep{Cappe08}.
\end{description}

%%%%%%%%%%%%%%%%%%%%%%%%%
\subsection{MIS schemes without replacement}
%%%%%%%%%%%%%%%%%%%%%%%%

In all {\tt N} schemes, exactly one sample is generated from each proposal pdf. This corresponds to having a sampling strategy without replacement.
\begin{description}
\item [{\tt [N1]}:] \emph{Sampling without replacement (random or deterministic), $\mathcal{S}_2$ or $\mathcal{S}_3$, and weight denominator $\mathcal{W}_2$ (for $\mathcal{S}_2$) or $\mathcal{W}_1$, $\mathcal{W}_2$, or $\mathcal{W}_3$ (for $\mathcal{S}_3$):}\\
For calculating the denominator of the $n$-th weight, the specific proposal used for the generation of the sample is used. This is the approach frequently used in particle filtering \citep{Gordon93} and in the standard PMC method \citep{Cappe04}.
\item [{\tt [N2]}:] \emph{Sampling without replacement (random), $\mathcal{S}_2$,  and weight denominator $\mathcal{W}_1$:} \\
This MIS implementation draws one sample from each proposal, but the order matters (it must be random) since the calculation of the $n$-th weight uses for the evaluation of the denominator the mixture pdf formed by the proposal pdfs that were still available at the generation of the $n$-th sample.
\item  [{\tt [N3]}:] \emph{Sampling without replacement (random or deterministic), $\mathcal{S}_2$ or $\mathcal{S}_3$, and weight denominator $\mathcal{W}_3$, $\mathcal{W}_4$, or $\mathcal{W}_5$ (for $\mathcal{S}_2$), or $\mathcal{W}_4$ or $\mathcal{W}_5$ (for $\mathcal{S}_3$):}\\
In the calculation of the $n$-th weight, one uses for the denominator the whole mixture. This is the approach, for instance, of \citep{APIS15,CORNUET12}. As shown in Section \ref{sec_variance}, this scheme has several benefits over the others.
\end{description}

Table \ref{table_mis_schemes_summary} summarizes the six resulting MIS schemes and their references in literature,  indicating {the} sampling procedure and weighting function {that are applied} to obtain the $n$-th weighted sample $\x_n$. We consider \texttt{N1} and  \texttt{N3} associated to $\mathcal{S}_3$ (they can also be obtained with $\mathcal{S}_2$) since it is simpler than $\mathcal{S}_2$. All the different algorithms in the literature (as far as we know) correspond to one of the MIS schemes described above (see Section \ref{sec_adaptiv_mis}). Moreover, several new valid schemes have also appeared naturally ($\Ra$, $\Rb$, and $\Nb$), and new ones can be proposed within this framework.

\begin{table}[!t]
\caption{Summary of the sampling procedure and the weighting function of each MIS scheme.}
%\vspace{-0.5cm}
\begin{center}
\begin{tabular}{|l|c|c|c|}
\hline
MIS scheme & Sampling & $w(\x_n)$ & Used in \\ 
\hline \hline
$\Ra$ & $\mathcal{S}_1$ &   $\frac{\pi(\x_n)}{q_{j_n}(\x_n)}$ & Novel scheme \\ 
\hline
$\Rb$&$\mathcal{S}_1$ &  $\frac{\pi(\x_n)}{\frac{1}{N}\sum_{k=1}^N q_{j_k}(\x_n)}$ & Novel scheme\\ 
\hline
$\Rc$&$\mathcal{S}_1$ &  $\frac{\pi(\x_n)}{\psi(\x_n)}$ & \citep{Cappe08} \\ 
\hline
\hline
$\Na$ & $\mathcal{S}_3$ &   $\frac{\pi(\x_n)}{q_{n}(\x_n)}$ & \citep{Cappe04} \\ 
\hline
$\Nb$&$\mathcal{S}_2$ &  $\frac{\pi(\x_n)}{\frac{1}{|\mathcal{I}_n|} \sum_{\forall k\in \mathcal{I}_n} q_k(\x_n)}$ & Novel scheme\\ 
\hline
$\Nc$&$\mathcal{S}_3$ &  $\frac{\pi(\x_n)}{\psi(\x_n)}$ &  \citep{APIS15,CORNUET12} \\ 
\hline 
\end{tabular}
\end{center}
\label{table_mis_schemes_summary}
\end{table}%

%%%%%%%%%%%%%%%%% 
\subsection{Running example}
\label{running_example_schemes}
%%%%%%%%%%%%%%%%% 

Let us consider the example {from} Section \ref{running_example_sampling} where the realizations of the sequence of indexes for the sampling schemes  $\mathcal{S}_1$, $\mathcal{S}_2$, and $\mathcal{S}_3$ are respectively $\{j_1,j_2,j_3 \} = \{3,3,1\}$, $\{j_1,j_2,j_3 \} = \{3,1,2\}$, and $\{j_1,j_2,j_3 \} = \{1,2,3\}$. Figure \ref{fig_schemes}(a) shows the three first schemes of Table  \ref{table_mis_schemes_summary} related to the sampling with replacement, $\mathcal{S}_1$. The figure shows a possible realization of all MIS schemes with $M=N=3$ samples and pdfs. For the $n$-th sample, we show the set of available proposals, the index $j_n$ of the proposal pdf that was actually selected to draw the sample, the function $\varphi_n$, and the importance weight. Similarly, Figs. \ref{fig_schemes}(b)-(c) depict the three schemes of Table  \ref{table_mis_schemes_summary} related to the sampling without replacement, $\mathcal{S}_2$ and $\mathcal{S}_3$, where exactly one sample is drawn from each available proposal.

\begin{figure*}[htp]
\centering
\subfigure[Schemes \texttt{R1}, \texttt{R2}, and \texttt{R3}]{
\RThreeProposals
} 
\centering
\subfigure[Scheme \texttt{N2}]{
\NbThreeProposals
}\qquad
\subfigure[Schemes \texttt{N1} and \texttt{N3}]{
\NacThreeProposals
} 
\caption{(a) Example of a realization of the indexes selection ($N=3$) with the sampling procedure $\mathcal{S}_1$ (with replacement), and all weighting possibilities, yielding the MIS schemes \texttt{R1}, \texttt{R2}, and \texttt{R3}. (b) Example of a realization of the indexes selection ($N=3$) with the sampling procedure $\mathcal{S}_2$ (without replacement) yielding the MIS scheme \texttt{N2}. (c) Sampling procedure $\mathcal{S}_3$ (deterministic index selection), and all weighting possibilities, yielding the MIS schemes \texttt{N1} and \texttt{N3}.}
\label{fig_schemes}
\end{figure*}

%%%%%%%%%%%%%%%%%%%%%%%%%%%%%%%%%%%%%%%%%
%%%%%%%%%%%%%%%%%%%%%%%%%%%%%%%%%%%%%%%%%
\section{Variance Analysis of the Schemes} 
\label{sec_variance}
%%%%%%%%%%%%%%%%%%%%%%%%%%%%%%%%%%%%%%%%%
%%%%%%%%%%%%%%%%%%%%%%%%%%%%%%%%%%%%%%%%%
%
Although the six different MIS schemes that appear in Section \ref{sec_MIS_schemes} yield the estimator $\hat{I}$ of Eq. \eqref{eq_IS_estimator_unnorm} unbiased (see Appendix \ref{appendix_unbiasedness}), the performance of each of the possible obtained estimators can be dramatically different. 
In this section, we provide an exhaustive variance analysis of the MIS schemes presented in the previous section. 
The details of the derivations are in Appendix \ref{sec_appendix_var_t1}. The estimators of the three methods with \emph{replacement} present the following variances:
{\small
\begin{eqnarray}
&&\Var ( \hI_{\Ra} ) =   \frac{1}{Z^2N^2}  \sum_{k = 1}^N \int  \frac{\pi^2(\x)g^2(\x)}{q_k(\x)} d\x  - \frac{I^2}{N},
\label{eq_clean_variance_r1}
\end{eqnarray}
}
{\small
\begin{eqnarray}
	&&\Var ( \hI_{\Rb} )  =  \frac{1}{Z^2N}  \frac{1}{N^N} \sum_{j_{1:N}} \int \frac{\pi^2(\x)g^2(\x)}{f(\x|j_{1:N})}d\x\nonumber \\
 && -  \frac{1}{Z^2N^2} \frac{1}{N^N}  \sum_{j_{1:N}}   \sum_{n=1}^N  \left(\int \frac{\pi(\x_n)g(\x_n)}{f(\x_n|j_{1:N})} q_{j_n}(\x_n)d\x_n \right)^2,
\end{eqnarray}
}
and
{\small
\begin{eqnarray}
	\Var ( \hI_{\Rc} )  &=& \frac{1}{Z^2N} \int  \frac{ {\pi}^2(\x)g^2(\x)}{\psi(\x)}d\x - \frac{I^2}{N}.
\label{eq_var_r3}
\end{eqnarray}
}
On the other hand, the variances associated to the estimators of the three methods with \emph{no} replacement are
{\small
\begin{eqnarray}
	\Var ( \hI_{\Na} ) &=& \frac{1}{Z^2N^2} \sum_{n=1}^N  \int  \frac{ {\pi}^2(\x_n)g^2(\x_n)}{q_n(\x_n)}d\x_n - \frac{I^2}{N},
\end{eqnarray}
}
{\small
\begin{eqnarray}
&& \Var ( \hI_{\Nb} ) = \left[ \frac{1}{Z^2N^2}\sum_{n=1}^N  \sum_{j_{1:n-1}} \int  \frac{\pi^2(\x_n)g^2(\x_n)}{p(\x_n|j_{1:n-1})}P(j_{1:n-1}) d\x_n \right] \nonumber \\
&& -    \left[ \frac{1}{Z^2N^2} \sum_{n=1}^N \sum_{j_{1:n}} \left(\int \frac{\pi(\x_n)g(\x_n)}{p(\x_n|j_{1:n-1})} q_{j_n}d\x_n \right)^2 \right] P(j_{1:n}),\nonumber \\
\end{eqnarray} 
}
\noindent and
{\small
\begin{eqnarray}
	\Var ( \hI_{\Nc} ) &=& \frac{1}{Z^2N}\int \frac{ {\pi}^2(\x)g^2(\x)}{\psi(\x)}d\x  - \frac{1}{Z^2N^2}\sum_{n=1}^N \left(\int \frac{{\pi}(\x)g(\x)}{\psi(\x)} q_n(\x) d\x \right)^2.
    \label{eq_clean_variance_n3}
\end{eqnarray}
}

One of the goals of this paper is to provide the practitioner with solid theoretical results about the superiority of some specific MIS schemes. In the following, we state two theorems that relate the variance of the estimator with these six methods, establishing a hierarchy among them. Note that obtaining an IS estimator with finite variance essentially amounts to having a proposal with heavier tails than the target. See \citep{Robert04,Geweke89} for sufficient conditions that guarantee this finite variance.

\begin{theorem}
\label{theorem_1}
%For {any set of proposal densities} $\{ q_n(\x) \}_{n=1}^N$, any target distribution $\pi(\x)$, and any square integrable function $g$,% it can be proved that
For any target distribution $\pi(\x)$, any square integrable function $g$, and any set of proposal densities $\{ q_n(\x) \}_{n=1}^N$ such that the variance of the corresponding MIS estimators is finite,
{%\small
\begin{equation}
\Var(\hI_{\Ra}) =  \Var(\hI_{\Na}) \geq  \Var(\hI_{\Rc}) \geq \Var(\hI_{\Nc}) \nonumber
\label{eq_whole_inequality}
\end{equation}
}
\end{theorem}
\noindent\emph{\textbf{Proof:}} See Appendix \ref{sec_appendix_var_t1}. \hfill \qed

 \begin{theorem}
 \label{theorem_2}
%For any set of proposal densities $\{ q_n(\x) \}_{n=1}^2$, any target distribution $\pi(\x)$, and any square integrable function $g$, % it can be proved that
For any target distribution $\pi(\x)$, any square integrable function $g$, and any set of proposal densities $\{ q_n(\x) \}_{n=1}^2$ such that the variance of the corresponding MIS estimators is finite,
{%\small
\begin{equation}
\Var(\hI_{\Ra})  = \Var(\hI_{\Na}) \geq \Var(\hI_{\Rb}) =  \Var(\hI_{\Nb}) \geq \Var(\hI_{\Nc})
\end{equation}
}
\end{theorem}
\noindent\emph{\textbf{Proof:}} See Appendix \ref{sec_appendix_var_t2}. \hfill \qed

First, let us note that the scheme \texttt{N3} outperforms (in terms of the variance) any other MIS scheme in the literature that we are aware of. Moreover, for $N=2$, it also outperforms the other novel schemes $\texttt{R2}$ and $\texttt{N2}$. While the MIS schemes \texttt{R2} and \texttt{N2} do not appear in Theorem \ref{theorem_1}, we hypothesize that the conclusions of Theorem  \ref{theorem_2} might be extended to $N>2$. The intuitive reason is that, regardless of $N$, both methods partially reduce the variance of the estimators by placing more than one proposal at the denominator of some or all the weights. A possible interpretation of the superiority of $\Nc$ is that it uses the whole mixture at the denominator of each weight, thus providing an exchange of information between all the proposals. This exchange of information is essential in multimodal scenarios, where the whole set of proposals, seen as a mixture, should mimic the whole target, but each proposal should adapt locally to the target. Since the variance of the IS weight depends on the mismatch of the target (numerator) w.r.t. the proposal (denominator), the use of the whole mixture in the denominator reduces the variance of the weight in general, and therefore, also the variance of the estimator (see the variance analysis in Appendix \ref{sec_appendix_variance}). The scheme $\Nc$ goes a step further w.r.t. $\Rc$, drawing deterministically one sample from each mixand of $\psi(\x)$, which can be seen as drawing $N$ samples from the mixture $\psi(\x)$ with a modified version of stratified sampling, a well-known variance reduction technique (see Appendix \ref{sec_further_obs} and \cite[Section 9.12]{Owen13}), which is also related to the residual resampling.

The variance analysis of the self-normalized estimator $\tilde{I}$ in Eq. \eqref{eq_IS_estimator_norm} implies a ratio of dependent r.v.'s, and therefore, it cannot be performed without resorting to an approximation, e.g., by means of a Taylor expansion as it is performed in \citep{Kong92,Kong94,Owen13}. In this case, the bias of $\tilde{I}$ is usually considered negligible compared to the variance for large $N$. With this approximation, the variance depends on the variances of the numerator (which is a scaled version of $\hat{I}$), the variance of $\hat Z$, and the covariance of both. Therefore, the variance results that we have proved above for $\hat{I}$ and $\hat Z$, cannot be directly extrapolated for $\tilde{I}$. However, it is reasonable to assume that methods that reduce the variance of $\hat{I}$ and $\hat Z$, in general will also reduce the variance of $\tilde{I}$. In Section \ref{sec_results}, this hypothesis is reinforced by means of numerical simulations. Therefore, \texttt{N3} should always be used whenever possible (it requires extra proposal evaluations). See a detailed discussion in Section \ref{sec:applyingMIS}.

%%%%%%%%%%%%%%%%% 
\subsection{Running example: Exact variances of the MIS estimators of $Z$}
\label{running_example_variances}
%%%%%%%%%%%%%%%%% 

{Here we focus on computing the} exact variances of estimators related to the running example. {We simplify the case study} to $N=2$ proposals, {for the sake for conciseness in the proofs}. The proposal pdfs are then $q_1(x)=\mathcal{N}(x;\mu_1,\sigma^2)$ and $q_2(x)=\mathcal{N}(x;\mu,\sigma^2)$ with means $\mu_1=-3$ and $\mu_2=3$, and variance $\sigma^2=1$. {We} consider a normalized bimodal target pdf $\pi(x)=\frac{1}{2} \mathcal{N}(x;\nu_1,c_1^2) + \frac{1}{2} \mathcal{N}(x;\nu_2,c_2^2)$ {and} set $\nu_1 = \mu_1$, $\nu_2 = \mu_2$, and $c_1^2 = c_2^2 =  \sigma^2$. Then, both proposal pdfs can be seen as a whole mixture that exactly replicates the target, i.e., $\pi(x) = \frac{1}{2} q_1(x) + \frac{1}{2} q_2(x)$. This is the desired situation pursued by an {AIS} algorithm: {Each} proposal is centered at each target mode, and {the} scale parameters perfectly match the scale of the modes. The goal {consists in} estimating the normalizing constant with the six schemes described in Section \ref{sec_MIS_schemes}. We use the $\hat Z$ estimator of Eq. \eqref{eq_Z_estimator} and the estimator $\hat I$ of \eqref{eq_IS_estimator_unnorm} when $g=x$, both with $N=2$ samples. The closed-form variance expressions of the six schemes are presented in the following: 

The {variances} of the estimators of the normalizing constant (true value $Z=\int \pi(x)dx=1$) are given by
\begin{eqnarray}
\Var(\hat Z_{\Ra}) = \Var(\hat Z_{\Na})   &=& \frac{3 + \exp{\left( \frac{4 \mu^2}{\sigma^2}\right)}}{8} -\frac{1}{2} = \frac{\exp{(36)} -1}{8} \approx 5.4 \cdot 10^{14}, \nonumber \\
\Var(\hat Z_{\Rb}) = \Var(\hat Z_{\Nb})  &=& \frac{3 + \exp{\left( \frac{4 \mu^2}{\sigma^2}\right)}}{16} -\frac{1}{4}= \frac{\exp{(36)}-1}{16} \approx 2.7 \cdot 10^{14}, \nonumber
\end{eqnarray}
and
\begin{eqnarray}
\Var(\hat Z_{\Rc}) = \Var(\hat Z_{\Nc}) &=& 0. \nonumber
\end{eqnarray}

The {variances} of the estimators of the target mean (true value $I=\int x \pi(x)dx=0$) are given by
\begin{eqnarray}
\Var(\hat I_{\Ra})  = \Var(\hat I_{\Na}) &=& \frac{3(\sigma^2+\mu^2)}{8} + \frac{\sigma^2 + 9 \mu^2}{8} \exp{\left( \frac{4\mu^2}{\sigma^2} \right) }  \nonumber \\
&=&\frac{30}{8} + \frac{82}{8}\exp{(36)} \approx 4.42 \cdot 10^{16}, \nonumber \\
\Var(\hat I_{\Rb}) = \Var(\hat I_{\Nb})  &=& \frac{3(\sigma^2+\mu^2)}{16} + \frac{\sigma^2 + 9 \mu^2}{16} \exp{\left( \frac{4\mu^2}{\sigma^2} \right) } + \frac{\sigma^2}{4}   \nonumber \\
&=&\frac{30}{16} + \frac{82}{16}\exp{(36)} + \frac{1}{4}\approx 2.21 \cdot 10^{16}, \nonumber \\
\Var(\hat I_{\Rc}) &=& \frac{\sigma^2 + \mu^2 }{2}  = 5, \nonumber
\end{eqnarray}
and
\begin{eqnarray}
\Var(\hat I_{\Nc}) &=& \frac{\sigma^2}{2} = \frac{1}{2} \nonumber
\end{eqnarray}

The derivations can be found in Appendix \ref{sec_ex1_a}. We observe that, for a very simple bimodal scenario where the proposals are perfectly placed in the target modes, the schemes $\Rc$ and $\Nc$ present a good performance while the other schemes do not work.

%%%%%%%%%%%%%%%%%%%%%%%%%%%%%%%%%%%%%%%%%
%%%%%%%%%%%%%%%%%%%%%%%%%%%%%%%%%%%%%%%%%
\section{Applying the MIS Schemes}
\label{sec:applyingMIS}
%%%%%%%%%%%%%%%%%%%%%%%%%%%%%%%%%%%%%%%%%
%%%%%%%%%%%%%%%%%%%%%%%%%%%%%%%%%%%%%%%%%

%%%%%%%%%%%%%%%%% 
\subsection{Computational complexity}
%%%%%%%%%%%%%%%%%

Table \ref{table_comp_cost} compares the total number of target and proposal evaluations in each MIS scheme. First, note that the estimators of any MIS scheme within the proposed general framework perform $N$ target evaluations in total. However, depending on the function $\varphi_{\mathcal{P}_n}$ used by each specific scheme at the weight denominator, a different number of proposal evaluations is performed. We see that $\Rc$ and $\Nc$ always require the largest number of proposal evaluations. In $\Rb$, the number of proposal evaluations is variable: although each weight evaluates $N$ proposals, some proposals may be repeated, whereas others may not be used.

In many relevant scenarios, the cost of evaluating the proposal densities is negligible compared to the cost of evaluating the target function. In this scenario, the MIS scheme $\Nc$ should always be chosen, since it yields a lower variance with a negligible increase in computational cost. For instance, this is the case in the \emph{Big Data} Bayesian framework, where the target function is a posterior distribution with a large amount of data in the likelihood function.  However, in some other scenarios, e.g. when the number of proposals $N$ is too large and/or the target evaluations are not very expensive, limiting the number of proposal evaluations can result in a better cost-performance trade off.

Unlike most MCMC methods, several strategies of parallelization can be applied in IS-based techniques. In the adaptive context, the adaptation of all proposals usually depends on the performance of all previous proposals, and therefore the adaptivity is the bottleneck of the {parallelization}. The {six} schemes proposed in this paper can be {parallelized} to some {extent}.  Once all the proposals are available, the schemes $\Ra$, $\Na$, $\Rc$, and $\Nc$ can draw and weight the $N$ samples in parallel, which represents a large advantage w.r.t. MCMC methods. In the schemes $\Rb$ and $\Nb$, the samples can be drawn independently, but the denominator of the weight cannot be computed in a parallel way. However, since the target evaluation in the numerator of the weights is fully parallelizable, the drawback of these schemes can be considered negligible for a small/medium number of proposals.

%%%%%%%%%%%%%%%%%
\subsection{\emph{A priori} partition approach}
\label{sec_partial}
%%%%%%%%%%%%%%%%%
The extra computational cost of some MIS schemes occurs because each sample must be evaluated in more than one proposal $q_n$, or even in all of the available proposals (e.g. the MIS scheme $\Nc$). In order to limit the number of proposal evaluations, let us first define a partition of the set of the indexes of all proposals, $\{1,\ldots,N\}$, into $P$ disjoint subsets of $L$ elements (indexes), $\mathcal{J}_p$ with $p=1,\ldots,P$, s.t.
\begin{equation}
	\mathcal{J}_1 \cup \mathcal{J}_2\cup \ldots \cup \mathcal{J}_P= \{1,\ldots,N\},
	\label{eq_partition}
\end{equation}
where $\mathcal{J}_k \cap \mathcal{J}_q = \emptyset$ for all $k,q=1,\ldots,P$ and $k\neq q$.\footnote{Note that, for the sake of simplifying the notation, we assume that all $P$ subsets have the same number of elements. However this is not necessary, and it is straightforward to extend the conclusions of this section to the case where each subset has different number of elements.}
Therefore, each subset, $\mathcal{J}_p = \{j_{p,1}, j_{p,2},\ldots, j_{p,L}\}$, contains $L$ indexes, $j_{p,\ell}\in \{1,\ldots,N\}$ for $\ell=1,\ldots,L$ and $p=1,\ldots, P$.

After this \emph{a priori} partition, one could apply any MIS scheme in each (partial) subset of proposals, and then perform a suitable convex combination of the partial estimators. This general strategy is inspired by a specific scheme, partial deterministic mixture MIS (p-DM-MIS), which was recently proposed in \citep{elvira2015efficient}. That work applies the idea of the partitions just for the MIS scheme $\Nc$, denoted there as full deterministic mixture MIS (f-DM-MIS). The sampling procedure is then $\mathcal{S}_3$, i.e., exactly  one sample is drawn from each proposal. The weight of each sample in p-DM-MIS, instead of evaluating the whole set of proposals (as in $\Nc$), evaluates only the proposals within the subset that the generating proposal belongs to. Mathematically, the weights of the samples corresponding to the $p$-th mixture are computed as
\begin{equation}
	w_n = \frac{\pi({\bf x}_{n})}{\psi_{p}({\bf x}_{n})} 
		= \frac{\pi({\bf x}_{n})}{\frac{1}{L}\sum_{j\in\mathcal{J}_p}q_{{j}}({\bf x}_{n})}, \quad n \in\mathcal{J}_p.
\label{p_dm_weights_static}
\end{equation}
Note that the number of proposal evaluations is $N \leq \frac{N^2}{P}\leq N^2$. Specifically, we have the particular cases $P=1$ and $P=N$ corresponding to the MIS schemes $\Nc$ (best performance) and $\Na$ (worst performance), respectively. In \citep{elvira2015efficient}, it is proved that for a specific partition with $P$ subsets of proposals, merging any pair of subsets decreases the variance of the estimator $\hat I$ of Eq. \eqref{eq_IS_estimator_unnorm}.

The previous idea can be applied to the other MIS schemes presented in Section \ref{sec_MIS_schemes} (not only $\Nc$). In particular, one can make an \emph{a priori} partition of the proposals as in Eq. \eqref{eq_partition}, and apply independently any different MIS scheme in each set. For instance, and based on some knowledge about the performance of the different proposals, one could make two disjoint sets of proposals, applying the MIS scheme $\Na$ in the first set, and the MIS scheme $\Nc$ in the second set. {Recently, a novel partition approach has been proposed in  \citep{elvira2016heretical}. In this case, the sets of proposals are performed \emph{a posteriori}, once the samples have been drawn. The variance of the estimators is reduced at the price of introducing a bias.}

\begin{table*}[!t]
\begin{center}
\begin{tabular}{|c||c|c|c|c|c|c|}
\hline
 MIS Scheme& \texttt{R1}&\texttt{N1}& \texttt{R2}&\texttt{N2}& \texttt{R3}& \texttt{N3}\\
\hline
\hline
Target Evaluations & $N$   &     $N$ &     $N$ &  $N$ &      $N$        &    $N$  \\
\cline{1-7}
Proposal Evaluations & $N$   &    $N$    &    $\leq N^2$  &     $N(N+1)/2$       &        $N^2$   &    $ N^2$  \\
\hline
\end{tabular}
\end{center}
\caption{Number of target and proposal evaluations for the different MIS schemes. Note that the number of proposal evaluations for R2 is a random variable with a range from $N$ to $N^2$.}
\label{table_comp_cost}
\end{table*}

%%%%%%%%%%%%%%%%%%%%%%%%%%%%%%%%%%%%
\subsection{Generalized Adaptive Multiple Importance Sampling}
\label{sec_adaptiv_mis}
%%%%%%%%%%%%%%%%%%%%%%%%%%%%%%%%%%%%

Adaptive importance sampling (AIS) methods iteratively update the parameters of the proposal pdfs using the information of the past samples 
(see a survey in \citep{bugallo2017adaptive})}. The sampling and weighting options, described in this work within a static framework for the sake of simplicity, can be straightforwardly applied in the adaptive context. 
%
%More specifically, 
Let us consider a set of proposal pdfs $\{q_{j,t}({{\bf x}})\}$, with $j=1,\ldots,J$ and $t=1,\ldots,T$, where the subscript $t$ indicates the iteration index of the adaptive algorithm, $T$ is the total number of adaptation steps, $J$ is the number of proposals per iteration, and $N=JT$ is the total number of proposal pdfs. A general adaptation procedure takes into account, at the $t$-th iteration, statistical information about the target pdf gathered in all of the previous iterations. Several algorithms have been proposed in the last decade \citep{Cappe08,CORNUET12,APIS15,martino2017layered,elvira2017improving}. 

The MIS schemes considered in Section \ref{sec_MIS_schemes} can be directly applied to the adaptive context. Moreover, the \emph{a priori} partition approach of Section \ref{sec_partial} can be very useful to limit the computational cost of the different MIS schemes when the number of iterations grows (and therefore also the total number of proposals).

\begin{figure}[!htb]
\centering 
\centerline{
 \includegraphics[width=8cm]{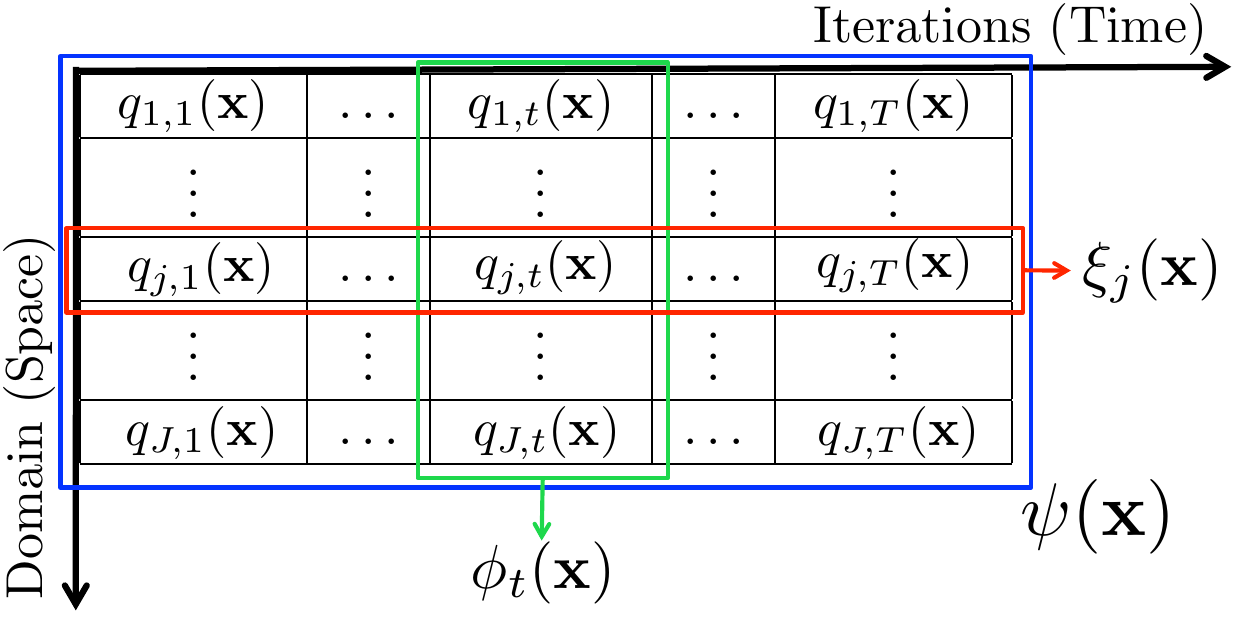}
  }
  \caption{Graphical representation of the $N=JT$ proposal pdfs used in the generalized adaptive MIS scheme, spread through the state space $\mathbb{R}^{d_x}$ ($j=1,\ldots,J$) and adapted over time ($t=1,\ldots,T$). 
  }
\label{figSpaceTime}
\end{figure}

Let us assume that, at the $t$-th iteration, one sample is drawn from each proposal $q_{j,t}$ (sampling $\mathcal{S}_3$), i.e., 
$$
{\x}_{j,t} \sim q_{j,t}(\x_{j,t}), 
$$
$j=1,\ldots,J$ and $t=1,\ldots,T$. Then, an importance weight $w_{j,t}$ is assigned to each sample ${\bf x}_{j,t}$. As described exhaustively in Section \ref{sec_weighting}, several strategies can be applied to build $w_{j,t}$ considering the different MIS approaches. Figure \ref{figSpaceTime} provides a graphical representation of this scenario, by showing both the spatial and temporal evolution of the $J=NT$ proposal pdfs. In a generic AIS algorithm, one weight 
\begin{equation}
\label{Gen_W_Eq}
w_{j,t}=\frac{\pi({\bf x}_{j,t})}{\varphi_{j,t}({\bf x}_{j,t})},
\end{equation}
is associated  to each sample ${\bf x}_{j,t}$.
In the MIS scheme $\Na$, the function employed in the denominator is  
\begin{equation}
  \label{Case1}
\varphi_{j,t}({\bf x})=q_{j,t}({\bf x}).
\end{equation}

In the following, we focus on the MIS scheme $\Nc$ in the adaptive framework, considering several choices of the partitioning of the set of proposals, since this scheme attains the best performance, as shown in Section \ref{sec_variance}. In the \emph{full} $\Nc$ scheme, the function $\varphi_{j,t}$ is 
\begin{equation}
  \label{CompleteProposal}
\varphi_{j,t}({\bf x})= \psi({\bf x})= \frac{1}{JT}\sum_{k=1}^J \sum_{r=1}^T q_{k,r}({\bf x}),
\end{equation} 
where $\psi(\x)$ is now the mixture of all the spatial and temporal proposal pdfs. 
{This case corresponds to the blue rectangle in Fig. \ref{figSpaceTime}. However, note that the computational complexity can become prohibitive as the product JT increases. Furthermore, two natural alternatives of partial $\Nc$ schemes appear in this scenario. The first one uses the following partial mixture 
 \begin{equation}
  \label{TemporalMixProposal}
 \varphi_{j,t}({\bf x})= \xi_{j}({\bf x})= \frac{1}{T}\sum_{r=1}^T q_{j,r}({\bf x}),
 \end{equation}
  with $j=1,\ldots,J$, as mixture-proposal pdf in the IS weight denominator, i.e. using the temporal evolution of the $j$-th single proposal $q_{j,t}$ at the weight denominator. In this case, there are $P=J$ mixtures, each one formed by $L=T$ components (red rectangle in Fig. \ref{figSpaceTime}). Another possibility is considering the mixture of all the $q_{j,t}$'s at the $t$-th iteration, i.e., 
 \begin{equation}
  \label{SpatialMixProposal}
  \varphi_{j,t}({\bf x})=\phi_{t}({\bf x})= \frac{1}{J}\sum_{k=1}^J q_{k,t}({\bf x}),  
\end{equation}  
with $t=1,\ldots,T$, so that  we have $P=T$ mixtures, each one formed by $L=J$ components (green rectangle in Fig. \ref{figSpaceTime}). The function $\varphi_{j,t}$ in Eq. \eqref{Case1} is used in the standard PMC scheme \citep{Cappe04}; Eq. \eqref{TemporalMixProposal}, in the particular case of $J=1$, has been considered in the \emph{adaptive multiple importance sampling} (AMIS) algorithm \citep{CORNUET12}. Note that the schemes that consider at the denominator of the weight the temporal sequence of adapted proposals can introduce a bias in the IS estimators (see \citep[Section 5]{CORNUET12} for more details). The choice in Eq. \eqref{SpatialMixProposal} has been applied in the \emph{adaptive population importance sampling} (APIS)  \citep{APIS15}, the \emph{layered adaptive importance sampling} (LAIS)  \citep{martino2017layered}, and the \emph{deterministic mixture population Monte Carlo} (DM-PMC) \citep{elvira2017improving} algorithms. In other techniques, such as mixture PMC (M-PMC) \citep{Douc07a,Douc07b,Cappe08},  %\citep{Douc07a,Douc07b} 
 a similar strategy is employed, but using sampling $\mathcal{S}_1$ in the mixture $\phi_{t}({\bf x})$, i.e., with the MIS scheme $\Rc$. 
}
%%%%%%%%%%%%%%%%
%%%%%%%%%%%%%%%% 

Table \ref{FantasticTable} summarizes all the possible cases discussed above.
The last row corresponds to a generic grouping strategy of the proposal pdfs $q_{j,t}$. As previously described, we can also divide the $N=JT$ proposals into $P=\frac{JT}{L}$ disjoint groups of $P$ mixtures with $L$ components. Namely, we denote the set of $L$ pairs of indexes corresponding to the $p$-th mixture ($p=1,\ldots, P$) as $\mathcal{J}_p=\{(k_{p,1},r_{p,1}), \dots,(k_{p,L},r_{p,L})\}$, where $k_{p,\ell}\in\{1,\ldots,J\}$,  $r_{\ell,p}\in\{1,\ldots,T\}$ (i.e., $|\mathcal{J}_p|=L$, with each element being a pair of indexes), and $\mathcal{J}_p \cap \mathcal{J}_q = \emptyset$ for any pair $p,q=1,\ldots,P,$ and $p\neq q$. In this scenario, we have
   \begin{equation}
  \label{PartialMixProposal2}
  \varphi_{j,t}({\bf x})= \frac{1}{L}\sum_{(k,r)\in \mathcal{J}_p} q_{k,r}({\bf x}),  \quad \mbox{with} \quad  (j,t) \in \mathcal{J}_p.  
\end{equation} 

Note that using $\psi({\bf x})$ and $\xi_j({\bf x})$ the computational cost per iteration increases as the total number of iterations $T$ grows.
Indeed, at the $t$-th iteration all the previous proposals $q_{j,1},\ldots,q_{j,t-1}$ (for all $j$) must be evaluated at all the new samples ${\bf x}_{j,t}$. 
Hence, algorithms based on these proposals quickly become unfeasible as the number of iterations grows.
On the other hand, using $\phi_t({\bf x})$ the computational cost  per iteration is controlled by $J$, remaining constant regardless of the number of adaptive steps performed.

%%%%%%%%%%%%%%
\subsection{Guidelines for applying MIS}
%%%%%%%%%%%%%%
The superiority of $\Nc$ is theoretically proved for the unnormalized estimator in Theorems \ref{theorem_1} and \ref{theorem_2}, and {practically} shown by means of several numerical simulations for the self-normalized estimator (see next section). However, the associated computational complexity is also increased w.r.t. the other MIS schemes in terms of proposal evaluations. If $N$ is small or the target evaluations are expensive (w.r.t. the cost of the proposal evaluations), $\Nc$ should be used. However, when the target evaluation is cheap and/or the number of proposals is large, the use of $\Nc$ increases notably the computational complexity. In this case, the novel schemes $\Rb$ or $\Nb$ seem to provide {very good results}, and their theoretical properties are superior to those of $\Na$ and $\Ra$. However, future studies will be required to characterize these novel schemes and investigate efficient parallelization techniques. We also recommend to combine adaptive schemes with the partition approach proposed in \citep{elvira2015efficient} and \citep{elvira2016heretical}, and summarized in Section  \ref{sec_partial}. Note that further investigation is also needed for efficiently constructing the partitions of the proposals that allow to reduce the computational complexity while retaining most of the variance reduction associated to the $\Nc$ scheme.

In the adaptive context, there is a big potential for the MIS schemes where all spatial and temporal proposals are used at the denominator of all weights (blue square in Fig. \ref{figSpaceTime}). However, the computational complexity for large number of proposals is prohibitive, and further theoretical analysis about the bias of the estimators is needed (see \cite[Section 5]{CORNUET12}). The adaptivity of MIS algorithms is essential in challenging high-dimensional setups. The $\Nc$ scheme has exhibited a very good performance when used within adaptive MIS algorithms due to two main reasons. First, the variance of the estimators at each iteration is reduced as proved in Theorems \ref{theorem_1} and \ref{theorem_2}, which explains part of the variance reduction attained in AMIS \citep{CORNUET12}, LAIS \citep{martino2017layered}, or  GAPIS \citep{elvira2015gradient}. Second, when the IS weights are used for adaptive purposes (e.g. in APIS \citep{APIS15} or DM-PMC \citep{elvira2017improving}), the use of the whole mixture of proposals in the denominator of the weights can be seen as a cooperative adaptive procedure (see \citep{elvira2017improving} for further details).

Finally, one of the strengths of {the} $\Nc$ scheme is its performance in multimodal scenarios, where $\Na$ should always be avoided. If $N$ is comparable to the number of modes, an adaptive $\Nc$ scheme  should be employed; the aforementioned cooperation in the proposals adaptation has an implicit repulsive behavior that promotes the adaptation to different modes. However, if $N$ is much larger, the adaptive algorithm may use $\Rb$ or $\Nb$ with potentially similar performance but less computational complexity.

%%%%%%%%%%%%%%%%%%%%%%%%%%%%%%%%%%%%%%%%%
%%%%%%%%%%%%%%%%%%%%%%%%%%%%%%%%%%%%%%%%%
\section{Numerical Examples}
\label{sec_results}
%%%%%%%%%%%%%%%%%%%%%%%%%%%%%%%%%%%%%%%%%
%%%%%%%%%%%%%%%%%%%%%%%%%%%%%%%%%%%%%%%%%

In the previous sections, we have provided several theoretical results for comparing different MIS schemes according to different quality measures, e.g., ranking them in terms of the variance of the corresponding estimators. In this section, we provide different numerical results in order to quantify numerically the gap among these methods. In the following, we show that even in the case where the different proposals are well  tuned (in the sense of a small or no mismatch with a multimodal target), the choices of the sampling and weighting procedures dramatically affect the performance of the MIS estimator.

\subsection{{Running exmple: Estimation of the target mean}}
\label{sec_ex2}
%%%%%%%%%%%%%%
{
Let us consider again the target pdf of the running example
$$\pi(\x)=\frac{1}{3} \mathcal{N}(x;\nu_1,c_1^2) + \frac{1}{3} \mathcal{N}(x;\nu_2,c_2^2) + \frac{1}{3} \mathcal{N}(x;\nu_3,c_3^2),$$
with means $\nu_1=-3$, $\nu_2=0$, and $\nu_3=3$, and variances $c_1^2=c_2^2=c_3^2=1$. As proposal functions we use $q_i(x)=\mathcal{N}(x;\mu_i,\sigma)$, with $\mu_i = \nu_i$ {and} $i=1,2,3$  and $\sigma^2 = 1$,} i.e., the proposal pdfs can be seen as a whole mixture that exactly replicates the target, i.e., $\pi(x) = \psi(x) = \frac{1}{3} q_1(x) + \frac{1}{3} q_2(x) + \frac{1}{3} q_3(x)$. 

\begin{figure}[!htb]
%\centering
\subfigure[MSE of $\hat I$ (unnormalized weights).]{
\includegraphics[width=0.48\textwidth]{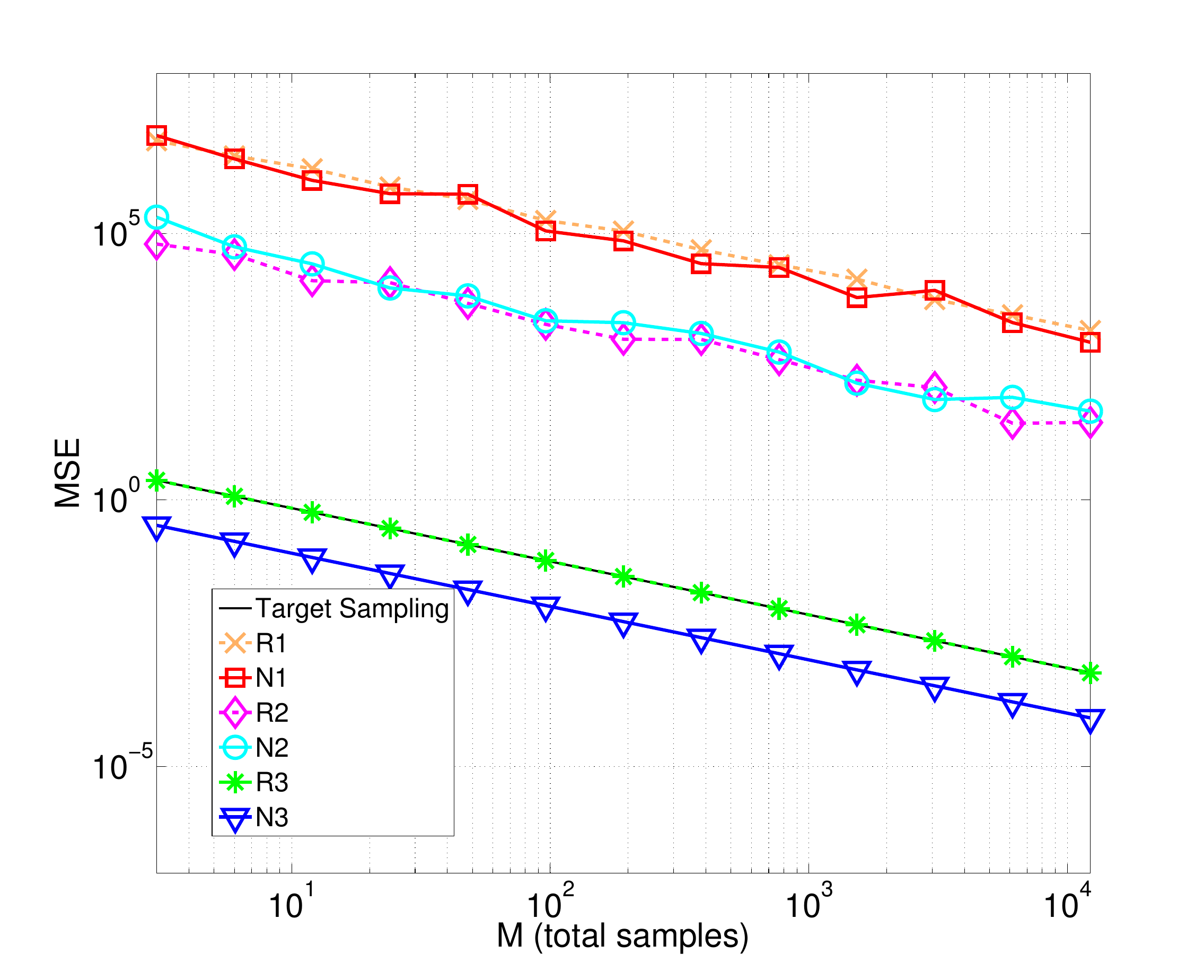}
}
\subfigure[MSE of $\tilde I$ (normalized weights).]{
\includegraphics[width=0.48\textwidth]{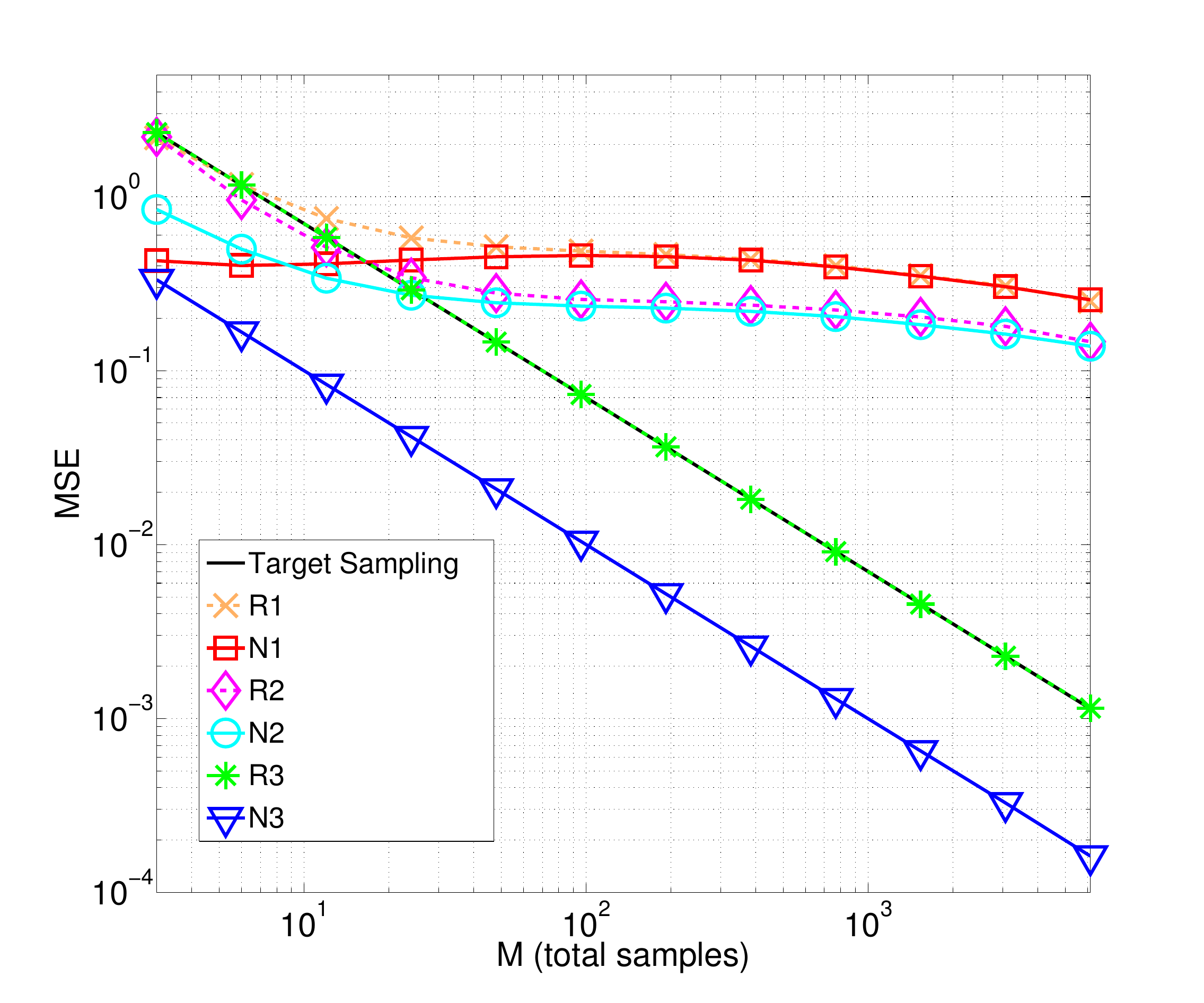}
}
\caption{(\textbf{Ex. of Section.} \ref{sec_ex2}) Performance of the estimators of the target mean for the different MIS schemes.}
\label{exMean_MSE_mean_unnorm}
\label{exMean_MSE_mean_norm}
\end{figure}

The goal is to estimate the mean of the target pdf with the six MIS schemes. Fig. \ref{exMean_MSE_mean_unnorm}(a) shows the MSE of the estimator $\hat I$ for all the methods w.r.t. the number of total samples (note that some schemes require that the total number of samples is multiple of $M=3$). The results have been averaged over $5\cdot 10^6$ runs. The solid black line shows the variance of the natural estimator, i.e. sampling directly from the target pdf (since this is possible in this easy example). Note that the method $\hat I_{\Rc}$ exactly replicates the performance of $\bar I$: this method samples from the mixture of Gaussians in the traditional way and the weights, due to the perfect match, are always $w=1$, i.e., $\hat I_{\Rc}$ and $\bar I$ are equivalent. We can see that $\hat I_{\Nc}$ is the best estimator in terms of variance, while $\hat I_{\Ra}$ and $\hat I_{\Na}$ present a high variance. Note that, surprisingly, $\hat I_{\Nc}$ has better performance than sampling from the target, i.e., estimator $\bar I$. This is because the sampling $\mathcal{S}_3$ can be seen as a sampling from the mixture of proposals $\psi(\x)$ (which coincides with the target in this example) with a variance reduction technique, as we discuss in Appendix \ref{sec_further_obs}. Note also that the inequality proved in Theorem \ref{theorem_1} holds since all methods are unbiased and therefore the MSE is due only to the variance. We can see that $\hat I_{\Rb}$ and $\hat I_{\Nb}$ also behave badly in terms of variance.  

Figure \ref{exMean_MSE_mean_norm}(b) shows the variance of the estimator $\tilde I$ of Eq. \eqref{eq_IS_estimator_norm} for all methods. First, note that the MSE of $\Rc$ and $\Nc$ is the same {as in Fig. \ref{exMean_MSE_mean_unnorm}{(b)}}, since the estimators $\hat I$ and $\tilde I$ are equivalent in this scenario (since they perfectly estimate the normalizing constant, i.e., $\hat Z = Z$). Note that the relations observed and proved for the different MIS schemes in terms of the variance of the estimator $\hat I$, are also kept here when we increase the number of samples. {The MSE curves are compared with the same number of samples $M$, i.e. the same number of target evaluations. {Note that each MIS scheme requires a different number of proposal evaluations per sample (see Table \ref{table_comp_cost}). However, a fair comparison is fully target dependent, and with few number of proposals we can consider that the computational complexity is similar in all schemes.}}

%%%%%%%%%%%%%%
\subsection{Applying the MIS schemes in adaptive IS (AIS)}
%%%%%%%%%%%%%%
{
We apply the different MIS schemes within an AIS context. In particular, we focus on the LAIS algorithm, recently proposed in \citep{martino2017layered}. The  method consists of an upper layer with a MCMC that draws samples from the target, while a lower layer uses those samples as location parameters (means) of some proposal pdfs for applying IS. In its basic version, $J$ Metropolis-Hastings chains independently {run} at the upper layer, and hence MIS is applied in the lower layer with $J$ proposals at each iteration. In the following, we implement the {six} adaptive MIS schemes in a spatial manner for two different target pdfs. For instance, the $\Nc$ scheme is implemented by sampling exactly one sample from each of the $J$ proposals at the $t$-th iteration, and applying at the denominator of the IS weight the whole mixture of $J$ proposals as in Eq. \eqref{SpatialMixProposal} (see the green square of Fig. \ref{figSpaceTime}).
%%%%%%%%%%%%%%
\subsubsection{{Mixture of bivariate Gaussians.}}
\label{sec_exp_mixt_Gaussians}
%%%%%%%%%%%%%%

{
Let us first consider a mixture of five bivariate Gaussians, 
\begin{equation}
\pi(\x)=\frac{1}{5}\sum_{i=1}^5 \mathcal{N}(\x;{\bm \nu}_i,{\bf \Sigma}_i), \quad \x\in \mathbb{R}^2,
\label{Target1b}
\end{equation}
where $\mathcal{N}(\x;{\bm \mu},{\bf C})$ denotes a Gaussian pdf with mean vector ${\bm \mu}$ and covariance matrix ${\bf C}$, ${\bf \nu}_1=[-10, -10]^{\top}$, ${\bm \nu}_2=[0, 16]^{\top}$, ${\bm \nu}_3=[13, 8]^{\top}$, ${\bm \nu}_4=[-9, 7]^{\top}$, ${\bm \nu}_5=[14, -14]^{\top}$, ${\bf \Sigma}_1=[2, \ 0.6; 0.6, \ 1]$, ${\bf \Sigma}_2=[2, \ -0.4; -0.4, \ 2]$, ${\bf \Sigma}_3=[2, \ 0.8; 0.8, \ 2]$, ${\bf \Sigma}_4=[3, \ 0; 0, \ 0.5]$, and ${\bf \Sigma}_5=[2, \ -0.1; -0.1, \ 2]$. We run the LAIS algorithm with $J=100$ spatial proposals that are adapted over $T=200$ iterations. The proposals of the upper and lower layers are isotropic Gaussians with $\sigma_{\text{upper}}=5$ and $\sigma_{\text{lower}}=2$, respectively. We also run the standard PMC algorithm of \citep{Cappe04}, computing at each iteration the weights according to $\Na$, which represents the standard PMC, and $\Nc$ which corresponds to the DM-PMC algorithm recently proposed in \citep{elvira2017improving}. The means of the proposals are randomly and uniformly initialized within the $[-4,4]\times[-4,4]$ square.  Table \ref{tabl_mixt_Gaussians} shows the MSE of the self-normalized estimator of the target mean, $\tilde I$, and the estimator of the normalizing constant (the true values are $\E[\X]=[1.6, 1.4]^{\top}$ and $Z=1$, respectively). The scheme $\Nc$ presents again the best performance in the adaptive setup, both in LAIS and PMC. Note that the novel schemes $\Rb$ and $\Nb$ show again a satisfactory performance.}
\begin{table*}[!t]
\begin{center}
\begin{tabular}{|c||c|c|c|c|c|c||c|c|}
\hline
 Alg. & \texttt{R1}-LAIS &\texttt{N1}-LAIS& \texttt{R2}-LAIS&\texttt{N2}-LAIS& \texttt{R3}-LAIS& \texttt{N3}-LAIS & \texttt{N1}-PMC & \texttt{N3}-PMC \\ 
\hline
\hline
$\Var(\hat Z)$ & 0.6471  &  0.6380   & 0.0004 &   0.0024  &  0.0005   & \textbf{0.0001} & 0.1528 & 0.0006 \\
\hline
$\Var(\tilde I)$  & 1.4509   & 2.0466 &   0.0335  &  0.0295   & 0.0423  &  \textbf{0.0088} & 0.3847 & 0.0363 \\
\hline
\end{tabular}
\end{center}
\caption{{\textbf{(Ex. of {Section \ref{sec_exp_mixt_Gaussians}})} {MSE of the LAIS and PMC algorithms with the different MIS schemes at the lower layer. $J=100$ proposals and $T=200$ iterations.}}}
\label{tabl_mixt_Gaussians}
\end{table*}

%%%%%%%%%%%%%%
\subsubsection{{Multidimensional banana-shaped distribution.}}
\label{sec_exp_banana}
%%%%%%%%%%%%%%

We consider the banana shape target example used in \citep{haario1999adaptive,Haario2001} which ``can be be calibrated to become extremely challenging'' \citep{CORNUET12}. The target is based on a $d_x$-dimensional multivariate Gaussian $\x \sim \mathcal{N}(\x;\textbf{0}_{d_x},{\bf \Sigma})$ with $\Sigma = \text{diag}(\sigma^2,1,...,1)$, where the second variable is nonlinearly transformed from $x_2$ to $x_2 - b(x_1^2 - \sigma^2)$. This transformation leads to a banana-shaped distribution with zero mean and uncorrelated components (note that the target dimension $d_x\geq 2$). 
 
We implement the MIS schemes within the LAIS algorithm as described in the previous example. We set $J=200$ proposals that are adapted over $T=1000$ iterations, and isotropic Gaussian proposals with $\sigma_{\text{upper}} = 0.2$ and $\sigma_{\text{lower}}=0.5$. The means of the proposals are randomly and uniformly initialized within the $[-4,4]\times[-5,5]$ square. In Fig. \ref{fig_banana}, we vary the dimension of the state space $d_x$ with, $2\leq d_x \leq 40$, and we show the MSE of the self-normalized estimator $\tilde I$ of the target mean. The results have been averaged over $300$ runs. We observe that $\Nc$ and $\Rc$ schemes provide a similar good performance as in previous examples, although if $N$ were smaller, $\Nc$ would clearly outperform $\Rc$. When the dimension increases, the performance of all schemes decays, but the same hierarchy in performance holds for all schemes. $\Nb$ presents a similar performance than $\Nc$ in high dimensions.

\begin{figure}[!htb]
\centering
\includegraphics[width=0.7\textwidth]{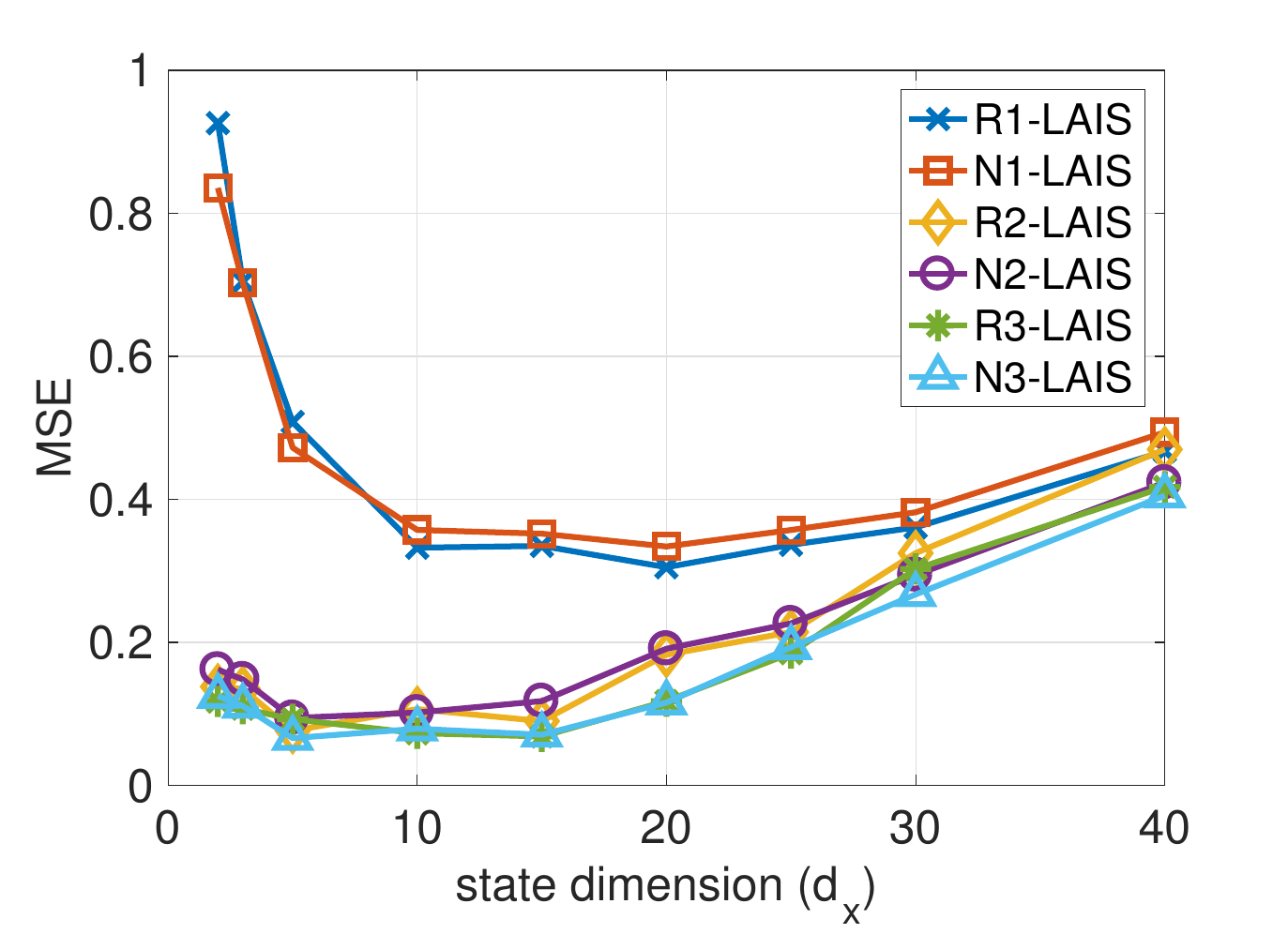}
\caption{{\textbf{(Ex. of {Section \ref{sec_exp_banana}})} LAIS algorithm with different MIS schemes in a multidimensional banana-shaped target. $J=100$ proposals and $T=200$ iterations.}}
\label{fig_banana}
\end{figure}

}
%%%%%%%%%%%%%%
\subsection{Discussion on the experimental results}
%%%%%%%%%%%%%%

First of all, note that the numerical simulations provided in this section corroborate the variance analysis of Section \ref{sec_variance}. More specifically, the hierarchy shown in Fig. \ref{exMean_MSE_mean_unnorm}, based on MSE of $\hat I$, corresponds to the hierarchy in terms of variance of $\hat I$ given in Theorems \ref{theorem_1} and \ref{theorem_2}. The same hierarchy is represented graphically in Fig. \ref{fig_ex3_mse_norm}. Furthermore, Fig. \ref{exMean_MSE_mean_norm}(b) depicts the MSE of the self-normalized estimator $\tilde I$: for large enough values of $M$ (so that a good approximation of $Z$ is attained), the MIS schemes are ordered exactly as in Fig. \ref{exMean_MSE_mean_unnorm} (as discussed in Section \ref{sec_variance}). 

The numerical experiments confirm that $\Nc$ provides the best performance. The scheme $\Rc$ also presents a good performance in most cases.
The performance of $\Ra$ and $\Na$ is, in general, much worse than the performance of the other schemes. Both schemes account at the weight denominator only for the proposal from which the sample is drawn, which in a multimodal scenario can be problematic. While $\Ra$ is a novel scheme that has naturally arisen in this work, and it probably has little interest from a practical point of view, $\Na$  has been applied in different adaptive MIS algorithms, such as the original version of PMC \citep{Cappe04}.

The novel schemes $\Rb$ and $\Nb$ have appeared in this new framework and  deserve a further analysis. The hierarchy theoretically proved for $N=2$ proposals in Theorem \ref{theorem_2}  still holds in the numerical examples for $N>2$, e.g. in Figs. \ref{exMean_MSE_mean_unnorm}(a) and \ref{exMean_MSE_mean_norm}(b). In some scenarios, for instance where there is a big number of proposals compared to the modes of the target, these schemes can attain most of the variance reduction of $\Na$ and $\Nc$ while reducing the number of proposal evaluations w.r.t. $\Nc$. In the example with AIS methods, both $\Rb$ and $\Nb$ present a very competitive performance w.r.t. to $\Nc$.

Finally, observe that in Fig. \ref{exMean_MSE_mean_norm}, when a small number of samples $M$ is employed, the schemes $\Na$, $\Nb$ and $\Nc$,  i.e., those with index selection without replacement ($\mathcal{S}_2$ and $\mathcal{S}_3$), behave better. This occurs because the variance associated to the index selection is reduced by guaranteeing that all proposal pdfs are always used.

%%%%%%%%%%%%%%%%%%%%%%%%%%%%%%%%%%%%%%%%%
%%%%%%%%%%%%%%%%%%%%%%%%%%%%%%%%%%%%%%%%%
\section{Conclusions}
\label{sec_conclusions}
%%%%%%%%%%%%%%%%%%%%%%%%%%%%%%%%%%%%%%%%%
%%%%%%%%%%%%%%%%%%%%%%%%%%%%%%%%%%%%%%%%%

In this work, we have introduced a unified framework for sampling and weighting in the context of multiple importance sampling (MIS). This {approach} extends the concept of a proper weighted sample, enabling the design of a wide range of sampling/weighting combinations. In particular, we have considered three specific sampling procedures and we have proposed five types of generic weighting functions (related to different conditional and marginal distributions which depend on the sampling scheme).  As a result of the combinations of sampling and weighting procedures, we have analyzed the six unique resulting schemes (three of them are not present in the literature to the best of our knowledge). We have provided a theoretical comparison of these schemes in terms of variance, establishing a ranking of the different methods in terms of performance and computational complexity. Moreover, we have discussed the application of the MIS schemes within adaptive procedures. In addition, we have provided the practitioner with several useful and easy-to-follow guidelines for applying the MIS schemes in different scenarios. We have analyzed the behavior of the MIS schemes in {three} different numerical examples which corroborate the previous theoretical analysis.

%%%%%%%%%%%%%%%%%%%%%%%%%%%%%%%%%%%%%%%%%
%%%%%%%%%%%%%%%%%%%%%%%%%%%%%%%%%%%%%%%%%
\section*{ACKNOWLEDGMENTS}
\label{sec_ack}
%%%%%%%%%%%%%%%%%%%%%%%%%%%%%%%%%%%%%%%%%
%%%%%%%%%%%%%%%%%%%%%%%%%%%%%%%%%%%%%%%%%
We thank the Editor, Associate Editor, and referees for their constructive comments that helped to improve the paper. V.E. acknowledges support from the \emph{Agence Nationale de la Recherche} of France under PISCES project (ANR-17-CE40-0031-01). M.F.B. thanks the support of the National Science Foundation (NSF) under Award CCF-1617986.

\bibliographystyle{unsrtnat}
%\bibliography{bibliografia}

\begin{thebibliography}{34}
\providecommand{\natexlab}[1]{#1}
\providecommand{\url}[1]{\texttt{#1}}
\expandafter\ifx\csname urlstyle\endcsname\relax
  \providecommand{\doi}[1]{doi: #1}\else
  \providecommand{\doi}{doi: \begingroup \urlstyle{rm}\Url}\fi

\bibitem[Robert and Casella(2004)]{Robert04}
C.~P. Robert and G.~Casella.
\newblock \emph{{M}onte {C}arlo Statistical Methods}.
\newblock Springer, 2004.

\bibitem[Liu(2004)]{Liu04b}
J.~S. Liu.
\newblock \emph{{M}onte {C}arlo Strategies in Scientific Computing}.
\newblock Springer, 2004.

\bibitem[Owen(2013)]{Owen13}
A.~Owen.
\newblock \emph{Monte Carlo theory, methods and examples}.
\newblock http://statweb.stanford.edu/$\sim$owen/mc/, 2013.

\bibitem[Liang(2002)]{Liang02}
F.~Liang.
\newblock Dynamically weighted importance sampling in {M}onte {C}arlo
  computation.
\newblock \emph{Journal of the American Statistical Association}, 97\penalty0
  (459):\penalty0 807--821, 2002.

\bibitem[Veach and Guibas(1995)]{Veach95}
E.~Veach and L.~Guibas.
\newblock Optimally combining sampling techniques for {M}onte {C}arlo
  rendering.
\newblock \emph{In SIGGRAPH 1995 Proceedings}, pages 419--428, 1995.

\bibitem[Hesterberg(1995)]{Hesterberg95}
T.~Hesterberg.
\newblock Weighted average importance sampling and defensive mixture
  distributions.
\newblock \emph{Technometrics}, 37\penalty0 (2):\penalty0 185--194, 1995.

\bibitem[Owen and Zhou(2000)]{Owen00}
A.~Owen and Y.~Zhou.
\newblock Safe and effective importance sampling.
\newblock \emph{Journal of the American Statistical Association}, 95\penalty0
  (449):\penalty0 135--143, 2000.

\bibitem[Tan(2004)]{tan2004likelihood}
Z.~Tan.
\newblock On a likelihood approach for monte carlo integration.
\newblock \emph{Journal of the American Statistical Association}, 99\penalty0
  (468):\penalty0 1027--1036, 2004.

\bibitem[He and Owen(2014)]{he2014optimal}
H.~Y. He and A.~B. Owen.
\newblock Optimal mixture weights in multiple importance sampling.
\newblock \emph{arXiv preprint (arXiv:1411.3954)}, Nov. 2014.

\bibitem[Elvira et~al.(2015{\natexlab{a}})Elvira, Martino, Luengo, and
  Bugallo]{elvira2015efficient}
V.~Elvira, L.~Martino, D.~Luengo, and M.~F. Bugallo.
\newblock Efficient multiple importance sampling estimators.
\newblock \emph{Signal Processing Letters, IEEE}, 22\penalty0 (10):\penalty0
  1757--1761, 2015{\natexlab{a}}.

\bibitem[Capp\'e et~al.(2004)Capp\'e, Guillin, Marin, and Robert]{Cappe04}
O.~Capp\'e, A.~Guillin, J.~M. Marin, and C.~P. Robert.
\newblock Population {M}onte {C}arlo.
\newblock \emph{Journal of Computational and Graphical Statistics}, 13\penalty0
  (4):\penalty0 907--929, 2004.

\bibitem[Martino et~al.(2017)Martino, Elvira, Luengo, and
  Corander]{martino2017layered}
L.~Martino, V.~Elvira, D.~Luengo, and J.~Corander.
\newblock Layered adaptive importance sampling.
\newblock \emph{Statistics and Computing}, 27\penalty0 (3):\penalty0 599--623,
  2017.

\bibitem[Elvira et~al.(2017)Elvira, Martino, Luengo, and
  Bugallo]{elvira2017improving}
V.~Elvira, L.~Martino, D.~Luengo, and M.~F. Bugallo.
\newblock Improving {P}opulation {M}onte {C}arlo: Alternative weighting and
  resampling schemes.
\newblock \emph{Signal Processing}, 131\penalty0 (12):\penalty0 77--91, 2017.

\bibitem[Capp\'e et~al.(2008)Capp\'e, Douc, Guillin, Marin, and
  Robert]{Cappe08}
O.~Capp\'e, R.~Douc, A.~Guillin, J.~M. Marin, and C.~P. Robert.
\newblock Adaptive importance sampling in general mixture classes.
\newblock \emph{Statistics and Computing}, 18:\penalty0 447--459, 2008.

\bibitem[Martino et~al.(2015)Martino, Elvira, Luengo, and Corander]{APIS15}
L.~Martino, V.~Elvira, D.~Luengo, and J.~Corander.
\newblock An adaptive population importance sampler: Learning from the
  uncertanity.
\newblock \emph{IEEE Transactions on Signal Processing}, 63\penalty0
  (16):\penalty0 4422--4437, 2015.

\bibitem[Cornuet et~al.(2012)Cornuet, Marin, Mira, and Robert]{CORNUET12}
J.~M. Cornuet, J.~M. Marin, A.~Mira, and C.~P. Robert.
\newblock Adaptive multiple importance sampling.
\newblock \emph{Scandinavian Journal of Statistics}, 39\penalty0 (4):\penalty0
  798--812, December 2012.

\bibitem[Bugallo et~al.(2017)Bugallo, Elvira, Martino, Luengo, M\'iguez, and
  Djuric]{bugallo2017adaptive}
M.~F. Bugallo, V.~Elvira, L.~Martino, D.~Luengo, J.~M\'iguez, and P.~M. Djuric.
\newblock Adaptive importance sampling: The past, the present, and the future.
\newblock \emph{IEEE Signal Processing Magazine}, 34\penalty0 (4):\penalty0
  60--79, 2017.

\bibitem[Kong et~al.(2003)Kong, McCullagh, Meng, Nicolae, and
  Tan]{kong2003theory}
A~Kong, P~McCullagh, X-L Meng, D~Nicolae, and Z~Tan.
\newblock A theory of statistical models for monte carlo integration.
\newblock \emph{Journal of the Royal Statistical Society: Series B (Statistical
  Methodology)}, 65\penalty0 (3):\penalty0 585--604, 2003.

\bibitem[Kahn and Marshall(1953)]{kahn1953methods}
Herman Kahn and Andy~W Marshall.
\newblock Methods of reducing sample size in monte carlo computations.
\newblock \emph{Journal of the Operations Research Society of America},
  1\penalty0 (5):\penalty0 263--278, 1953.

\bibitem[Gordon et~al.(1993)Gordon, Salmond, and Smith]{Gordon93}
N.~Gordon, D.~Salmond, and A.~F.~M. Smith.
\newblock Novel approach to nonlinear and non-{G}aussian {B}ayesian state
  estimation.
\newblock \emph{IEE Proceedings-F Radar and Signal Processing}, 140:\penalty0
  107--113, 1993.

\bibitem[Geweke(1989)]{Geweke89}
J.~Geweke.
\newblock {B}ayesian inference in econometric models using {M}onte {C}arlo
  integration.
\newblock \emph{Econometrica}, 24:\penalty0 1317--1399, 1989.

\bibitem[Kong(1992)]{Kong92}
A.~Kong.
\newblock A note on importance sampling using standardized weights.
\newblock \emph{University of Chicago, Dept. of Statistics, Tech. Rep}, 348,
  1992.

\bibitem[Kong et~al.(1994)Kong, Liu, and Wong]{Kong94}
A.~Kong, J.~S. Liu, and W.~H. Wong.
\newblock Sequential imputations and {B}ayesian missing data problems.
\newblock \emph{Journal of the American Statistical Association}, 9:\penalty0
  278--288, 1994.

\bibitem[Elvira et~al.(2016)Elvira, Martino, Luengo, and
  Bugallo]{elvira2016heretical}
V.~Elvira, L.~Martino, D.~Luengo, and M.~F. Bugallo.
\newblock Heretical multiple importance sampling.
\newblock \emph{IEEE Signal Processing Letters}, 23\penalty0 (10):\penalty0
  1474--1478, 2016.

\bibitem[Douc et~al.(2007{\natexlab{a}})Douc, Marin, and Robert]{Douc07a}
G.R. Douc, J.M. Marin, and C.~Robert.
\newblock Convergence of adaptive mixtures of importance sampling schemes.
\newblock \emph{Annals of Statistics}, 35:\penalty0 420--448,
  2007{\natexlab{a}}.

\bibitem[Douc et~al.(2007{\natexlab{b}})Douc, Marin, and Robert]{Douc07b}
G.R. Douc, J.M. Marin, and C.~Robert.
\newblock Minimum variance importance sampling via population {M}onte {C}arlo.
\newblock \emph{ESAIM: Probability and Statistics}, 11:\penalty0 427--447,
  2007{\natexlab{b}}.

\bibitem[Elvira et~al.(2015{\natexlab{b}})Elvira, Martino, Luengo, and
  Corander]{elvira2015gradient}
V.~Elvira, L.~Martino, L.~Luengo, and J.~Corander.
\newblock A gradient adaptive population importance sampler.
\newblock In \emph{IEEE International Conf. on Acoustics, Speech and Signal
  Processing (ICASSP)}, pages 4075--4079, 2015{\natexlab{b}}.

\bibitem[Haario et~al.(1999)Haario, Saksman, and Tamminen]{haario1999adaptive}
H.~Haario, E.~Saksman, and J.~Tamminen.
\newblock Adaptive proposal distribution for random walk metropolis algorithm.
\newblock \emph{Computational Statistics}, 14\penalty0 (3):\penalty0 375--396,
  1999.

\bibitem[Haario et~al.(2001)Haario, Saksman, and Tamminen]{Haario2001}
H.~Haario, E.~Saksman, and J.~Tamminen.
\newblock An adaptive {M}etropolis algorithm.
\newblock \emph{Bernoulli}, 7\penalty0 (2):\penalty0 223--242, April 2001.

\bibitem[Niederreiter(1992)]{Niederreiter92}
H.~Niederreiter.
\newblock \emph{Random Number Generation and Quasi-{M}onte {C}arlo Methods}.
\newblock Society for Industrial Mathematics, 1992.

\bibitem[Douc and Capp{\'e}(2005)]{douc2005comparison}
R.~Douc and O.~Capp{\'e}.
\newblock Comparison of resampling schemes for particle filtering.
\newblock In \emph{ISPA 2005. Proceedings of the 4th International Symposium on
  Image and Signal Processing and Analysis, 2005.}, pages 64--69. IEEE, 2005.

\bibitem[Hardy et~al.(1952)Hardy, Littlewood, and
  P{\'o}lya]{hardy1952inequalities}
G.~H. Hardy, J.~E. Littlewood, and G.~P{\'o}lya.
\newblock \emph{Inequalities}.
\newblock Cambridge Univ. Press, 1952.

\bibitem[Abramowitz and Stegun(1972)]{abramowitz1972handbook}
M.~Abramowitz and I.~A. Stegun.
\newblock \emph{Handbook of mathematical functions: with formulas, graphs, and
  mathematical tables}.
\newblock Dover Pub., number 55., 1972.

\bibitem[Gwanyama(2004)]{gwanyama2004hm}
P.~W. Gwanyama.
\newblock The {HM-GM-AM-QM} inequalities.
\newblock \emph{The College Mathematics Journal}, 35\penalty0 (1):\penalty0
  47--50, Jan. 2004.

\end{thebibliography}

\appendix
%%%%%%%%%%%%%%%%%%%%%%%%%%%%%%%%%%%%%%%%%
%%%%%%%%%%%%%%%%%%%%%%%%%%%%%%%%%%%%%%%%%
\section{Further observations about the sampling $\mathcal{S}_3$}
\label{sec_further_obs}
%%%%%%%%%%%%%%%%%%%%%%%%%%%%%%%%%%%%%%%%%
%%%%%%%%%%%%%%%%%%%%%%%%%%%%%%%%%%%%%%%%%

In the sampling procedure $\mathcal{S}_3$, ${\bf X}_n \sim q_{n}({\bf x})$ for $n=1,\dots,N$, i.e., the selection of the index is deterministic.  Note that the set of samples $\{ {\bf x}_n \}_{n=1}^N$ is used in the IS estimators regardless of the order they are drawn. It can be interpreted that the $N$ samples are drawn from the mixture
$\psi({\bf x}) = \frac{1}{N} \sum_{n=1}^N q_n({\bf x})$ via Rao-Blackwellization (see \cite[Section 9.12]{Owen13} for more details).  More formally, if we define the r.v. $\X=\X_n \quad \mbox{ with } \quad  n \sim \mathcal{U}\{1,2,\ldots,N\}$, then ${\bf X} \sim \psi({\bf x})$. 
The procedure $\mathcal{S}_3$ follows a similar principle as a well-known variance reduction method, known as the stratified sampling \citep{Robert04,Liu04b}, where the domain of ${\bf X}$ is divided into different regions that, in the case of sampling $\mathcal{S}_3$, are unbounded and overlapped \cite[Section 9.12]{Owen13}. Finally, note that the approach $\mathcal{S}_3$ can also be seen as the application of a quasi-Monte Carlo technique \citep{Niederreiter92} for generating the deterministic sequence of indexes $j_1=1,j_2=2,\ldots, j_N=N$ (uniform, in the sense of low-discrepancy sequence) and then drawing $\x_{n}\sim q_{j_n}({\bf x})= q_{n}({\bf x})$ for $n=1,\ldots,N$. Note also, that $\mathcal{S}_3$ can be seen as a residual resampling step of the indexes of the proposals. Since all weights of the proposals are the same, the resampling is fully deterministic, which explains part the variance reduction of the MIS schemes with sampling $\mathcal{S}_3$.

%%%%%%%%%%%%%%%%% 
\section{{Connections with resampling methods}}
\label{appendix_resampling}

%%%%%%%%%%%%%%%%% 
Resampling methods are used in PFs to replace a set of weighted particles with another set of {equally weighted particles}. The way we address the sampling process in MIS has clear connections with the resampling step in PFs (e.g., see \citep{douc2005comparison}). An important difference of the proposed framework is that {the MIS} proposals are equally weighted in the mixture. The sampling method $\mathcal{S}_1$ is then equivalent to the multinomial resampling, whereas the sampling methods $\mathcal{S}_2$ and $\mathcal{S}_3$ correspond to residual resampling (note that, since $M=N$ and all the proposals are equally weighted, exactly one sample per proposal is drawn). In future works, it would be interesting to analyze sampling schemes related to residual, stratified and systematic resamplings, which can be incorporated quite naturally in MIS schemes, when the weights of the proposals are different (see for instance \citep{he2014optimal}).

%%%%%%%%%%%%%%%%%%%%%%%%%%%%%%%%%%%%%%%%%
%%%%%%%%%%%%%%%%%%%%%%%%%%%%%%%%%%%%%%%%%
\section{Proofs of unbiasedness of the MIS estimators}
\label{appendix_unbiasedness}
%%%%%%%%%%%%%%%%%%%%%%%%%%%%%%%%%%%%%%%%%
%%%%%%%%%%%%%%%%%%%%%%%%%%%%%%%%%%%%%%%%%

In this appendix we prove the unbiasedness of the estimator $\hat I$ of Eq. \eqref{eq_IS_estimator_unnorm} for the five weighting options described in Section \ref{sec_weighting}. We recall that the general expression for the expectation of $\hat I$ within the proposed framework is 
\begin{eqnarray}
\E[\hat I] &=&  \frac{1}{ZN}\sum_{n=1}^N \sum_{j_{1:N}} \int \frac{\pi(\x_n)g(\x_n)}{\varphi_{\mathcal{P}_n}(\x_n)} P(j_{1:N}) p(\x_n|j_n) d\x_n.
\label{general_estimator_appendix}
\end{eqnarray}

\noindent \textbf{\textsc{Option 1}} ($\mathcal{W}_1$): $\varphi_{\mathcal{P}_n}(\x_n) = \varphi_{j_{1:n-1}}(\x_n) =  p(\x_n|j_{1:n-1})$. 
We first marginalize in Eq. \eqref{general_estimator_appendix} over all indexes that do not affect the $n$-th weight {($j_{n:N}$)}:
\begin{eqnarray}
\E[\hat I] 
 &=&  \frac{1}{ZN}\sum_{n=1}^N \sum_{j_{1:n-1}} \int \frac{\pi(\x_n)g(\x_n)}{\varphi_{j_{1:n-1}}(\x_n)} p(\x_n,j_{1:n-1})d\x_n \nonumber \\
  &=&  \frac{1}{ZN}\sum_{n=1}^N \sum_{j_{1:n-1}} \int \frac{\pi(\x_n)g(\x_n)}{\varphi_{j_{1:n-1}}(\x_n)} p(\x_n|j_{1:n-1})P(j_{1:n-1})d\x_n.
\label{bias_4}
\end{eqnarray}
Then, substituting $\varphi_{j_{1:n-1}}(\x_n) = p(\x_n|j_{1:n-1})$  into Eq. \eqref{bias_4}, canceling terms and marginalizing $j_{1:n-1}$, we have:
\begin{eqnarray}
\E[\hat I]
  &=& \frac{1}{ZN}\sum_{n=1}^N \int \pi(\x_n)g(\x_n) d\x_n \nonumber  \\
   &=&  \frac{1}{Z} \int \pi(\x)g(\x) d\x = I. \nonumber \qed
\end{eqnarray}

\noindent\textbf{\textsc{Option 2}} ($\mathcal{W}_2$): $\varphi_{\mathcal{P}_n}(\x_n) = \varphi_{j_{n}}(\x_n) =  p(\x_n|j_n)$. We substitute $\varphi_{j_{n}}(\x_n)=  p(\x_n|j_n)$ into Eq. \eqref{general_estimator_appendix}, which cancels the denominator:
\begin{eqnarray}
\E[\hat I]
&=&  \frac{1}{ZN}\sum_{n=1}^N \sum_{j_{1:N}} \int \pi(\x_n)g(\x_n) P(j_{1:N})  d\x_n \nonumber \\
  &=& \frac{1}{ZN}\sum_{n=1}^N \int \pi(\x_n)g(\x_n) d\x_{n}\nonumber \\
   &=&  \frac{1}{Z} \int \pi(\x)g(\x) d\x = I. \nonumber \qed
\label{bias_1}
\end{eqnarray}

\noindent\textbf{\textsc{Option 3}} ($\mathcal{W}_3$): $\varphi_{\mathcal{P}_n}(\x_n) = \varphi_n(\x_n) =  p(\x_n)$. Since $\varphi_n$ does not depend on any index, we can first marginalize over the whole set of indexes $j_{1:N}$ in Eq. \eqref{general_estimator_appendix}:
\begin{eqnarray}
\E[\hat I] 
&=&  \frac{1}{ZN}\sum_{n=1}^N \int \frac{\pi(\x_n)g(\x_n)}{\varphi_n{(\x_n)}}  p(\x_n) d\x_n.
\label{eq_temp_opt3} 
\end{eqnarray}
Then, substituting $\varphi_{n}=  p(\x_n)$ in Eq. \eqref{eq_temp_opt3}:
\begin{eqnarray}
\E[\hat I] 
&=&  \frac{1}{ZN}\sum_{n=1}^N \int \pi(\x_n)g(\x_n)d\x_{n} \nonumber \\ &=&   \frac{1}{Z} \int \pi(\x)g(\x) d\x = I. \nonumber \qed
\end{eqnarray}

\noindent\textbf{\textsc{Option 4}} ($\mathcal{W}_4$): $\varphi_{\mathcal{P}_n}(\x) = \varphi_{j_{1:N}}(\x)  = f(\x|j_{1:N}) = \frac{1}{N}\sum_{n=1}^N q_{j_n}(\x)$. In this case, the expectation of $\hat I$ can be expressed as:
\begin{eqnarray}
\E[\hat I] 
&=&   \frac{1}{ZN} \sum_{j_{1:N}}  P(j_{1:N})  \int   \frac{\pi(\x)g(\x)}{\varphi_{j_{1:N}}(\x)} \sum_{n=1}^N q_{j_n}(\x) d\x.
\label{bias_2}
\end{eqnarray}
Substituting $\varphi_{j_{1:N}}(\x) = f(\x|j_{1:N}) = \frac{1}{N}\sum_{n=1}^N q_{j_n}(\x) $ in Eq. \eqref{bias_2}, and cancelling the denominator:
\begin{eqnarray}
\E[\hat I] 
&=&  \frac{1}{Z} \sum_{j_{1:N}}   \int  \pi(\x)g(\x) P(j_{1:N}) d\x \nonumber \\
&=&     \frac{1}{Z} \int  \pi(\x)g(\x)d\x = I. \nonumber \qed
\label{bias_2b}
\end{eqnarray}

\noindent\textbf{\textsc{Option 5}} ($\mathcal{W}_5$): $\varphi_{\mathcal{P}_n}(\x) = \varphi(\x)  = f(\x) = \frac{1}{N}\sum_{n=1}^N q_{n}(\x)= \psi(\x) $.  Now, the expectation of $\hat I$ becomes
\begin{eqnarray}
\E[\hat I] 
&=&  \frac{1}{Z} \int    \frac{\pi(\x)g(\x)}{\varphi(\x)}  \sum_{j_{1:N}} \Bigg[ \frac{1}{N} \sum_{n=1}^N q_{j_n}(\x)  \Bigg] P(j_{1:N}) d\x \nonumber  \\
&=&   \frac{1}{Z} \int  \frac{\pi(\x)g(\x)}{\varphi(\x)} \psi(\x) d\x,
\label{bias_3}
\end{eqnarray}
where, in the last step, we have used the identity 
$$\sum_{j_{1:N}} \left[ \frac{1}{N}\sum_{n=1}^N q_{j_n}(\x)  \right]P(j_{1:N}) = f(\x) = \psi(\x)$$
for any valid sampling procedure within this framework (see Remark \ref{remark_sampling} and Section \ref{sec_BeyondXn} for more details). Substituting $\varphi(\x) = \psi(\x)$  in Eq. \eqref{bias_3}
\begin{eqnarray}
\E[\hat I] &=&   \frac{1}{Z} \int  \frac{\pi(\x)g(\x)}{\psi(\x)}\psi(\x) d\x \nonumber  \nonumber  \\
&=&     \frac{1}{Z} \int  \pi(\x)g(\x)d\x = I. \qed
\label{bias_2b}
\end{eqnarray}

%%%%%%%%%%%%%%%%%%%%%%%%%%%%%%%%%%%%
%%%%%%%%%%%%%%%%%%%%%%%%%%%%%%%%%%%%
\section{Variance analysis of the MIS estimators}
\label{sec_appendix_variance}
%%%%%%%%%%%%%%%%%%%%%%%%%%%%%%%%%%%%
%%%%%%%%%%%%%%%%%%%%%%%%%%%%%%%%%%%%
Let us consider the unbiased estimator,
\begin{eqnarray}
\hat{I} = \frac{1}{ZN}\sum_{n=1}^N w_n(\x_n) g(\x_n),
\label{estimator_temp}
\end{eqnarray}
that approximates $I$. The variance of $\hat{I}$ can be expressed in the general form as
\begin{eqnarray}
\Var(\hat{I}) &=& E_{p(\x_{1:N},j_{1:N})}\left[\left(\hI - E_{p(\x_{1:N},j_{1:N})}[\hI] \right)^2 \right] \nonumber\\
&=& E_{p(\x_{1:N},j_{1:N})}[\hI^2] - E_{p(\x_{1:N},j_{1:N})}^2[\hI].%\\
\label{eq_general_variance_1}
\end{eqnarray}

In the general case of Eq. \eqref{eq_general_variance_1}, the $N$ terms of the sum of the estimator in $\hI$ are dependent. However, in the specific cases where they are independent, the variance of a sum of r.v.'s can be simplified as the sum of the variances, i.e.,
{\scriptsize
\begin{eqnarray}
\Var(\hat{I})&=& \frac{1}{Z^2N^2} \Bigg[ \sum_{n=1}^N E_{p(\x_n,j_n)}[w_n^2(\x_n)g^2(\x_n)] -   \sum_{n=1}^N E_{p(\x_n,j_n)}^2[w_n(\x_n)g(\x_n)] \Bigg] \nonumber \\
&=& \frac{1}{Z^2N^2} \Bigg[ \sum_{n=1}^N  \sum_{j_n =1}^N \int \frac{\pi^2(\x_n)g^2(\x_n)}{\varphi_{\mathcal{P}_n}^2(\x_n)}p(\x_n|j_n)P(j_n)d\x_n  \nonumber \\ 
&&- \sum_{n=1}^N \left(  \sum_{j_n =1}^N \int \frac{\pi(\x_n)g(\x_n)}{\varphi_{\mathcal{P}_n}(\x_n)}p(\x_n|j_n)P(j_n)d\x_n \right)^2\Bigg]. \nonumber \\
\label{eq_independent_variance}
\end{eqnarray}
}

In some MIS schemes, the $N$ terms are dependent (due to a sampling without replacement or because the $n$-th weight depends on several indexes $j_k$, with at least one $k \neq n$). However, conditioned to the whole set of indexes $j_{1:N}$, the terms of the sum in Eq. \eqref{estimator_temp} are always conditionally independent, so we can apply
{\scriptsize  
\begin{align}
&\Var(\hat{I}) =  \frac{1}{Z^2N^2} \sum_{j_{1:N}} \Bigg[ \sum_{n=1}^N E_{p(\x_n|j_n)}[w_n^2(\x_n)g^2(\x_n)]  -   \sum_{n=1}^N E_{p(\x_n|j_n)}^2[w_n(\x_n)g(\x_n)] \Bigg] P(j_{1:N}) \nonumber \\
&= \frac{1}{Z^2N^2} \sum_{j_{1:N}} \Bigg[ \sum_{n=1}^N  \int \frac{\pi^2(\x_n)g^2(\x_n)}{\varphi_{\mathcal{P}_n}^2(\x_n)} p(\x_n|j_n)d\x_n  - \sum_{n=1}^N \left( \int \frac{\pi(\x_n)g(\x_n)}{\varphi_{\mathcal{P}_n}(\x_n)} p(\x_n|j_n) d\x_n \right)^2 \Bigg] P(j_{1:N}). \nonumber \\
\label{eq_cond_independent_variance}
\end{align}
}

%%%%%%%%%%%%
\subsection{Variance of the estimators of the MIS schemes }
%%%%%%%%%%%%

In the following, we analyze the variance of the six MIS schemes discussed through this paper under the assumptions described in Theorem \ref{theorem_1} (see Section \ref{sec_variance} for more details). Since some schemes arise under more than one sampling/weighting combination (see Table \ref{table_sampling_weights}), here we always use the combination that facilitates the analysis. 

%%---------------------------%%
%\item  \texttt{[\Ra]} Sampling 1 / Weighting 2:
\noindent \textbf{1.} \texttt{[\Ra]} \textbf{Sampling 1 / Weighting 2:}
%%---------------------------%%
In this scheme, all the terms of the sum in Eq. \eqref{estimator_temp} are independent, so we can use Eq. \eqref{eq_independent_variance} for computing the variance of $\hI$. Substituting $\varphi_{j_n}(\x_n) = p(\x_n|j_n) = q_{j_n}(\x_n)$ in \ref{eq_independent_variance}, 
{\scriptsize
\begin{align}
& \Var ( \hI_{\Ra} )  = \frac{1}{Z^2N^2}\sum_{n=1}^N  \sum_{j_n = 1}^N  \left[  \int  \frac{\pi^2(\x_n)g^2(\x_n)}{p^2(\x_n|j_n)} p(\x_n|j_n)P(j_n) d\x_n \right] - \frac{I^2}{N}\nonumber \\
	&=   \frac{1}{Z^2N^2}\sum_{n=1}^N   \left[ \int  \sum_{j_n = 1}^N \frac{\pi^2(\x_n)g^2(\x_n)}{q_{j_n}(\x_n)} P(j_n) d\x_n  \right]  - \frac{I^2}{N}\nonumber  \\
		&=   \frac{1}{Z^2N^2}\sum_{n=1}^N    \left[  \int \frac{1}{N} \sum_{k = 1}^N \frac{\pi^2(\x_n)g^2(\x_n)}{q_k(\x_n)}d\x_n \right]   - \frac{I^2}{N}\nonumber  \\
			&=   \frac{1}{Z^2N^2}  \sum_{k = 1}^N \int  \frac{\pi^2(\x)g^2(\x)}{q_k(\x)} d\x  - \frac{I^2}{N},
\label{eq_var_r1}
\end{align}
}
were we have used that $P(j_n) = \frac{1}{N}$, $\forall j_n \in \{1,...,N\}$. 

\noindent \textbf{2.} \texttt{[\Rb]}  \textbf{Sampling 1 / Weighting 4:} The expression for the conditional independence of Eq. \eqref{eq_cond_independent_variance} is used substituting $\varphi_{j_{1:N}}(\x_n) = f(\x_n|j_{1:N}) = \frac{1}{N} \sum_{k=1}^N q_{j_k}(\x_n)$ and averaging it over the $N^N$ equiprobable sequences of indexes $j_{1:N}$:
%%---------------------------%% )
{\scriptsize
\begin{align}
	& \Var ( \hI_{\Rb} )  = \frac{1}{Z^2N^2} \Bigg[  \sum_{j_{1:N}} \bigg[ \sum_{n=1}^N  \int \frac{\pi^2(\x_n)g^2(\x_n)}{\varphi_{j_{1:N}}^2(\x_n)} p(\x_n|j_n)d\x_n - \sum_{n=1}^N \left( \int \frac{\pi(\x_n)g(\x_n)}{\varphi_{j_{1:N}}(\x_n)} p(\x_n|j_n) d\x_n \right)^2 \bigg] P(j_{1:N}) \Bigg] \nonumber  \\
	&= \frac{1}{Z^2N^2}   \frac{1}{N^N}  \Bigg[  \sum_{j_{1:N}} \sum_{n=1}^N  \int \frac{\pi^2(\x_n)g^2(\x_n)}{f^2(\x_n|j_{1:N})} q_{j_n}(\x_n)d\x_n  -  \sum_{j_{1:N}}   \sum_{n=1}^N \left( \int \frac{\pi(\x_n)g(\x_n)}{f(\x_n|j_{1:N})}q_{j_n}(\x_n) d\x_n \right)^2 \Bigg] \nonumber \\
	&= \frac{1}{Z^2N}  \frac{1}{N^N} \Bigg[ \sum_{j_{1:N}}   \int \frac{\pi^2(\x)g^2(\x)}{f^2(\x|j_{1:N})}  \left( \frac{1}{N} \sum_{n=1}^N  q_{j_n}(\x) \right) d\x  -  \frac{1}{N}  \sum_{j_{1:N}}   \sum_{n=1}^N \left( \int \frac{\pi(\x_n)g(\x_n)}{f(\x_n|j_{1:N})}q_{j_n}(\x_n) d\x_n \right)^2 \nonumber \Bigg]\\
  &=	 \frac{1}{Z^2N}  \frac{1}{N^N} \Bigg[ \sum_{j_{1:N}} \int \frac{\pi^2(\x)g^2(\x)}{f(\x|j_{1:N})}d\x  -  \frac{1}{N}  \sum_{j_{1:N}}   \sum_{n=1}^N  \left(\int \frac{\pi(\x_n)g(\x_n)}{f(\x_n|j_{1:N})} q_{j_n}(\x_n)d\x_n \right)^2 \Bigg].  \nonumber \\
\label{eq_var_r2}
\end{align}
}
where we have used the identity $f(\x|j_{1:N}) = \frac{1}{N}  \sum_{n=1}^N  q_{j_n}(\x_n) $. This expression for the variance resembles that of scheme \texttt{[\Nc]}, averaged over the $N^N$ possible mixtures (combinations) that can arise with sampling $\mathcal{S}_1$.
%%---------------------------%%

\noindent \textbf{3.}  \texttt{[\Rc]} \textbf{Sampling 1 / Weighting 3:} All the elements are independent in the sum, and the weights do not depend on any index of the set $j_{1:N}$. Therefore, we can start with  Eq. \eqref{eq_independent_variance}, marginalize over the indexes, and substitute $\varphi_n(\x_n)= p(\x_n) = \psi(\x_n)$, %%---------------------------%%
{\scriptsize
\begin{eqnarray}
	\Var ( \hI_{\Rc} )    &=& \frac{1}{Z^2N^2} \sum_{n=1}^N  \int \frac{\pi^2(\x_n)g^2(\x_n)}{\varphi_{n}^2(\x_n)}p(\x_n)d\x_n  -  \frac{1}{Z^2N^2} \sum_{n=1}^N \left(  \int \frac{\pi(\x_n)g(\x_n)}{\varphi_{n}(\x_n)}p(\x_n)d\x_n \right)^2\nonumber \\
 &=& \frac{1}{Z^2N^2} \sum_{n=1}^N  \int \frac{\pi^2(\x_n)g^2(\x_n)}{\psi^2(\x_n)}\psi(\x_n)d\x_n  -  \frac{1}{Z^2N^2} \sum_{n=1}^N \left(  \int \frac{\pi(\x_n)g(\x_n)}{\psi(\x_n)}\psi(\x_n)d\x_n \right)^2\nonumber \\
	&=& \frac{1}{Z^2N} \int  \frac{ {\pi}^2(\x)g^2(\x)}{\psi(\x)}d\x - \frac{I^2}{N}.
\label{eq_var_r3}
\end{eqnarray}
}
%%---------------------------%%
\noindent \textbf{4.}  \texttt{[\Na]} \textbf{Sampling 3 / Weighting 3:} The methods that use sampling without replacement introduce correlation at the selection of the proposals. However, under the perspective of the deterministic sampling ($\mathcal{S}_3$), the $n$-th sample $\x_n$ is a realization of the r.v. $X_n \sim q_n$ and is independent of the other samples. Marginalizing first  Eq. \eqref{eq_independent_variance} over the indexes, and substituting $\varphi_n(\x_n) = p(\x_n) = q_n(\x_n)$:
%%---------------------------%%
{ \scriptsize 
\begin{eqnarray}
	\Var ( \hI_{\Na} )  &=& \frac{1}{Z^2N^2} \sum_{n=1}^N  \int \frac{\pi^2(\x_n)g^2(\x_n)}{\varphi_{n}^2(\x_n)}p(\x_n)d\x_n -  \frac{1}{Z^2N^2} \sum_{n=1}^N \left(  \int \frac{\pi(\x_n)g(\x_n)}{\varphi_{n}(\x_n)}p(\x_n)d\x_n \right)^2\nonumber \\
&=& \frac{1}{Z^2N^2} \sum_{n=1}^N  \int \frac{\pi^2(\x_n)g^2(\x_n)}{q_{n}^2(\x_n)}q_n(\x_n)d\x_n  -  \frac{1}{Z^2N^2} \sum_{n=1}^N \left(  \int \frac{\pi(\x_n)g(\x_n)}{q_{n}(\x_n)}q_n(\x_n)d\x_n \right)^2\nonumber \\
	&=& \frac{1}{Z^2N^2} \sum_{n=1}^N  \int  \frac{ {\pi}^2(\x_n)g^2(\x_n)}{q_n(\x_n)}d\x_n - \frac{I^2}{N}.
\label{eq_var_n1}
\end{eqnarray}
}

%%---------------------------%% 
\noindent \textbf{5.}  \texttt{[\Nb]} \textbf{Sampling 2 / Weighting 1:} In this scheme, we use again the expression for conditional independence of Eq. \eqref{eq_cond_independent_variance}. Substituting  $\varphi_{j_{1:n-1}} = p(\x_n|j_{1:n-1})$,
%%---------------------------%%
{ \scriptsize 
\begin{align}
	&\Var ( \hI_{\Nb} )  = \frac{1}{Z^2N^2} \sum_{j_{1:N}} \Bigg[ \sum_{n=1}^N  \int \frac{\pi^2(\x_n)g^2(\x_n)}{\varphi_{j_{1:n-1}}^2(\x_n)} p(\x_n|j_n)d\x_n  -  \sum_{n=1}^N \left( \int \frac{\pi(\x_n)g(\x_n)}{\varphi_{j_{1:n-1}}(\x_n)} p(\x_n|j_n) d\x_n \right)^2 \Bigg] P(j_{1:N}) \nonumber \\
		&=  \frac{1}{Z^2N^2}\sum_{n=1}^N  \sum_{j_{1:n}} \Bigg[ \int  \frac{\pi^2(\x_n)g^2(\x_n)}{p^2(\x_n|j_{1:n-1})}p(\x_n|j_{n}) d\x_n  -    \left(\int \frac{\pi(\x_n)g(\x_n)}{p(\x_n|j_{1:n-1})} p(\x_n|j_n)d\x_n \right)^2 \Bigg] P(j_{1:n}) \nonumber \\
	&=  \frac{1}{Z^2N^2}\sum_{n=1}^N  \sum_{j_{1:n-1}} \int  \frac{\pi^2(\x_n)g^2(\x_n)}{p(\x_n|j_{1:n-1})}P(j_{1:n-1}) d\x_n  - \frac{1}{Z^2N^2} \sum_{n=1}^N \sum_{j_{1:n}} \left(\int \frac{\pi(\x_n)g(\x_n)}{p(\x_n|j_{1:n-1})} q_{j_n}d\x_n \right)^2  P(j_{1:n}) \nonumber \\
	\label{eq_var_n2}
\end{align}
}
Since the the integrals only depend on the set of indexes $j_{1:n}$, each term of the sum has been first marginalized over $j_{n+1:N}$. The first term in the sum can  then be further marginalized over $j_n$ to obtain the final expression. Note that the variance is the average of the variance of all the $N!$ possible sequences of indexes in the sampling without replacement. 

%%---------------------------%%
\noindent \textbf{6.}  \texttt{[\Nc]}  \textbf{Sampling 3 / Weighting 5:} We have followed the same arguments of scheme \texttt{\Na}. Marginalizing  Eq. \eqref{eq_independent_variance} over all the set of indexes $j_{1:N}$, and substituting $\varphi_{n}(\x_n) = f(\x_n) = \psi(\x_n)$:
%%---------------------------%%
{ \scriptsize 
\begin{eqnarray}
	\Var ( \hI_{\Nc} )
&=& \frac{1}{Z^2N^2} \sum_{n=1}^N  \int \frac{\pi^2(\x_n)g^2(\x_n)}{\psi^2(\x_n)}q_n(\x_n)d\x_n  -  \frac{1}{Z^2N^2} \sum_{n=1}^N \left(  \int \frac{\pi(\x_n)g(\x_n)}{\psi(\x_n)}q_n(\x_n)d\x_n \right)^2\nonumber \\
	&=& \frac{1}{Z^2N} \int \frac{ {\pi}^2(\x)g^2(\x)}{\psi^2(\x)} \left( \frac{1}{N} \sum_{n=1}^N q_n(\x) \right)d\x -  \frac{1}{Z^2N^2} \sum_{n=1}^N  \left( \int  \frac{ {\pi}(\x)g(\x)}{\psi(\x)} q_n(\x)d\x \right)^2 \nonumber \\
    &=& \frac{1}{Z^2N}\int \frac{ {\pi}^2(\x)g^2(\x)}{\psi(\x)}d\x  - \frac{1}{Z^2N^2}\sum_{n=1}^N \left(\int \frac{{\pi}(\x)g(\x)}{\psi(\x)} q_n(\x) d\x \right)^2, 
    \label{eq_var_n3}
\end{eqnarray}
}
where we have used the identity $\psi(\x) =  \frac{1}{N} \sum_{n=1}^N q_n(\x)d\x$.

%%---------------------------%%
\subsection{Proof of Theorem \ref{theorem_1}}
The proof of Theorem \ref{theorem_1} is split in the next three propositions.
\label{sec_appendix_var_t1}
\begin{proposition}
\label{prop_1a}
$\Var (\hI_{\Ra}) = \Var (\hI_{\Na})$
\end{proposition}
\noindent\emph{\textbf{Proof:}} See that Eqs. \eqref{eq_var_r1} and \eqref{eq_var_n1} are equivalent. \qed

\begin{proposition}
\label{prop_1b}
$\Var(\hI_{\Na}) \geq  \Var(\hI_{\Rc})$. 
\end{proposition}

\noindent\emph{\textbf{Proof:}}  Subtracting  Eqs. \eqref{eq_var_r3} and \eqref{eq_var_n1}, we get
\begin{align*}
& \textrm{Var}(\hI_{\Rc}) - \textrm{Var}(\hI_{\Na})=\\ &= \frac{1}{Z^2N^2} \int{\left(\frac{N}{\frac{1}{N}\sum_{j=1}^{N}{q_j(\x)}} - \sum_{i=1}^{N}{\frac{1}{q_i(\x)}} \right)
	 	g^2(\x) {\pi}^2(\x) d\x}.
\end{align*}
Since $g^2(\x) {\pi}^2(\x) \ge 0 \ \forall \x \in \mathbb{R}^{d_x}$, it is sufficient to show that
\begin{equation}
	\frac{1}{\frac{1}{N}\sum_{j=1}^{N}{q_j(\x)}} \le \frac{1}{N} \sum_{i=1}^{N}{\frac{1}{q_i(\x)}}.
\label{eq_varIneq2}
\end{equation}
Now, let us note that the left-hand side of Eq. \eqref{eq_varIneq2} is the inverse of the arithmetic mean of $q_1(\x),\ \ldots,\ q_N(\x)$,
\begin{equation*}
	A_N = \frac{1}{N}\sum_{j=1}^{N}{q_j(\x)},
\end{equation*}
whereas the right hand side of Eq. \eqref{eq_varIneq2} is the inverse of the harmonic mean of $q_1(\x),\ \ldots,\ q_N(\x)$,
\begin{equation*}
	\frac{1}{H_N} = \frac{1}{N} \sum_{i=1}^{N}{\frac{1}{q_i(\x)}}.
\end{equation*}
Therefore, the inequality in Eq. \eqref{eq_varIneq2} is equivalent to stating that $\frac{1}{A_N} \le \frac{1}{H_N}$, or equivalently $A_N \ge H_N$, which is the well-known arithmetic mean--harmonic mean inequality for positive real numbers \citep{hardy1952inequalities,abramowitz1972handbook,gwanyama2004hm}. \qed
\begin{proposition}
\label{prop_1c}
$\Var(\hI_{\Rc}) \geq  \Var(\hI_{\Nc})$. 
\end{proposition}

\noindent\emph{\textbf{Proof:}} Subtracting \eqref{eq_var_r3} and \eqref{eq_var_n3}, we get
\begin{align*}
&	\Var(\hat{I}_{\Nc}) - \Var(\hat{I}_{\Rc}) 
=  - \frac{I^2}{N}+  \frac{1}{Z^2N^2}\sum_{n=1}^N \left(\int \frac{{\pi}(\x)g(\x)}{\psi(\x)} q_n(\x) d\x \right)^2
\end{align*}
Therefore, the proposition is proved if
\begin{eqnarray}
 \frac{1}{Z^2}\sum_{n=1}^N \left(\int \frac{{\pi}(\x)g(\x)}{\psi(\x)} q_n(\x) d\x \right)^2 &\geq& NI^2 \nonumber
\end{eqnarray}
If we substitute $I = \int g(\x) \normalized \pi(\x) d\x$, multiplying both numerator and denominator by $\psi(\x)$ in the integral of the right-hand side,
{\scriptsize 
\begin{eqnarray}
\frac{1}{Z^2}\sum_{n=1}^N \left(\int \frac{{\pi}(\x)g(\x)}{\psi(\x)}q_n(\x) d\x \right)^2 &\geq& N \left(\frac{1}{Z} \int \frac{{\pi}(\x)g(\x)}{\psi(\x)}\psi(\x) d\x \right)^2 \nonumber \\
\sum_{n=1}^N \left(\int \frac{{\pi}(\x)g(\x) }{\psi(\x)}q_n(\x) d\x \right)^2 &\geq& N \left( \int \frac{{\pi}(\x)g(\x)}{\psi(\x)}\left( \frac{1}{N} \sum_{n=1}^N q_n(\x) \right) d\x \right)^2\nonumber \\
\sum_{n=1}^N \left(\int \frac{{\pi}(\x)g(\x)}{\psi(\x)}q_n(\x) d\x \right)^2 &\geq&  \frac{1}{N} \left( \sum_{n=1}^N\int \frac{{\pi}(\x)g(\x)}{\psi(\x)}q_n(\x)d\x \right)^2\nonumber \\
N \sum_{n=1}^N a_n^2 &\geq& \left( \sum_{n=1}^N a_n\right)^2 \label{eq_cs_proof}
\end{eqnarray}}
with $a_n=\int \frac{{\pi}(\x)g(\x)}{\psi(\x)}q_n(\x)d\x$. The inequality of Eq. (\ref{eq_cs_proof}) holds, since it is the definition of the Cauchy-Schwarz inequality \citep{hardy1952inequalities}, 
 \begin{eqnarray}
\left( \sum_{n=1}^N a_n ^2\right) \left( \sum_{n=1}^N b_n ^2\right) \geq\left( \sum_{n=1}^N a_n b_n \right)^2,
 \end{eqnarray}
\noindent with $b_n=1$ for $n=1,...,N$. $\hfill\Box$

\noindent\emph{\textbf{Proof of Theorem \ref{theorem_1}}.} The proof is obtained by applying Propositions \ref{prop_1a}, \ref{prop_1b}, and \ref{prop_1c}. \qed
\subsection{Proof of Theorem \ref{theorem_2}}
\label{sec_appendix_var_t2}
Let us first particularize the variance expression for $N=2$. From Eq. \eqref{eq_var_n1},
\begin{align}
	&\Var ( \hI_{\Na} ) = \Var ( \hI_{\Ra} ) \nonumber \\
	&= \frac{1}{4Z^2} \left( \int  \frac{\pi^2(\x)g^2(\x)}{q_1(\x)}d\x +  \int  \frac{\pi^2(\x)g^2(\x)}{q_2(\x)}d\x \right)  - \frac{I^2}{2}. \nonumber \\
	\label{eq_var_n1_N2} 
\end{align}
From Eq. \eqref{eq_var_r3},
\begin{equation}
\Var ( \hI_{\Rc} )  = \frac{1}{2Z^2}\int \frac{\pi^2(\x)g^2(\x)}{\frac{q_1(\x)+q_2(\x)}{2}}d\x - \frac{I^2}{2}.
	\label{eq_var_r3_N2} 
\end{equation} 
From Eq. \eqref{eq_var_n3},
{\scriptsize
\begin{eqnarray}
\Var ( \hI_{\Nc} )  &=& \frac{1}{2Z^2}\int \frac{\pi^2(\x)g^2(\x)}{\frac{q_1(\x)+q_2(\x)}{2}}d\x  - \frac{1}{4Z^2}\left(\int \frac{\pi(\x)g(\x)}{\frac{q_1(\x)+q_2(\x)}{2}} q_1(\x)d\x  \right)^2  - \frac{1}{4Z^2}\left(\int \frac{\pi(\x)g(\x)}{\frac{q_1(\x)+q_2(\x)}{2}}q_2(\x)d\x  \right)^2. \nonumber \\
	\label{eq_var_n3_N2} 
\end{eqnarray}
} 
From Eq. \eqref{eq_var_r2},
{\scriptsize
\begin{eqnarray}
\Var ( \hI_{\Rb} ) 
&=& \frac{1}{8Z^2} \left( \int \frac{\pi^2(\x)g^2(\x)}{q_1(\x)}d\x  +  \int \frac{\pi^2(\x)g^2(\x)}{q_2(\x)}d\x \right)  - \frac{I^2}{4} + \frac{1}{4Z^2} \int \frac{\pi^2(\x)g^2(\x)}{\frac{q_1(\x)+q_2(\x)}{2}}d\x \nonumber \\
&&-  \frac{1}{8Z^2}\left(\int \frac{\pi(\x)g(\x)}{\frac{q_1(\x)+q_2(\x)}{2}} q_1(\x)d\x  \right)^2  - \frac{1}{8Z^2} \left(\int \frac{\pi(\x)g(\x)}{\frac{q_1(\x)+q_2(\x)}{2}} q_2(\x)d\x  \right)^2.
	\label{eq_var_r2_N2} 
\end{eqnarray} 
}
From Eq. \eqref{eq_var_n2}, 
{\scriptsize
\begin{eqnarray}
\Var ( \hI_{\Nb} )  
	&=& \frac{1}{4Z^2}  \int \frac{\pi^2(\x)g^2(\x)}{\frac{q_1(\x) + q_2(\x)}{2}}d\x + \frac{1}{8Z^2}\int \frac{\pi^2(\x)g^2(\x)}{q_{1}(\x)}d\x  +  \frac{1}{8Z^2}\int \frac{\pi^2(\x)g^2(\x)}{q_{2}(\x)}d\x  \nonumber \\  && -\frac{1}{8Z^2} \left( \int \frac{\pi(\x)g(\x)}{\frac{q_1(\x) + q_2(\x)}{2}} q_{1}(\x)d\x \right)^2  -\frac{1}{8Z^2} \left( \int \frac{\pi(\x)g(\x)}{\frac{q_1(\x) + q_2(\x)}{2}} q_{2}(\x)dx \right)^2 -\frac{I^2}{4}. 
	\label{eq_var_n2_N2} 
		\end{eqnarray}
}	
\begin{proposition}
\label{prop_2a}
For $N=2$, $\Var ( \hI_{\Rb} ) = \Var ( \hI_{\Nb} )$
\end{proposition}
\noindent\emph{\textbf{Proof:}} See that Eqs. \eqref{eq_var_r2_N2} and \eqref{eq_var_n2_N2} are equivalent. \qed

\begin{proposition}
\label{prop_2b}
For $N=2$, $\Var ( \hI_{\Na} )  \geq \Var ( \hI_{\Rb} )  \geq \Var ( \hI_{\Nc} ) $
\end{proposition}
\noindent\emph{\textbf{Proof:}} Analyzing Eqs. \eqref{eq_var_n1_N2} 
 and 	\eqref{eq_var_n3_N2}, we see that Eq. \eqref{eq_var_r2_N2} can be rewritten as
\begin{eqnarray}
\Var ( \hI_{\Rb} ) &=& \frac{1}{2}\Var ( \hI_{\Na} )   + \frac{1}{2} \Var ( \hI_{\Nc} ).
	\label{eq_var_r2_N2_ext1} 
\end{eqnarray}

Since in Theorem \ref{theorem_1} it is proved that $\Var ( \hI_{\Na} ) \geq \Var ( \hI_{\Nc} )$ for any $N$, the proposition holds at least for $N=2$. \qed

\noindent\emph{\textbf{Proof of Theorem \ref{theorem_2}}.} The proof is obtained by applying Propositions \ref{prop_2a} and \ref{prop_2b}. \qed

\begin{remark}
We hypothesize that Theorem \ref{theorem_2} might also hold for $N>2$. The MIS schemes $\Rb$ and $\Nb$ seem to average estimators with variance reduction (related to $\Nc$) with estimators with worse variance (related to $\Na$).
\end{remark}

\begin{remark}
Note that the scheme $\Rc$ does not appear in Theorem \ref{theorem_2}.  Eq. \eqref{eq_var_r2_N2} can be rewritten as
{\scriptsize
\begin{eqnarray}
\Var ( \hI_{\Rb} )
 &=&  \frac{1}{2}\Var ( \hI_{\Rc} )  +  \frac{1}{8Z^2} \left( \int \frac{\pi^2(\x)g^2(\x)}{q_1(\x)}d\x  +  \int \frac{\pi^2(\x)g^2(\x)}{q_2(\x)}d\x \right) \nonumber \\ 
&&  -  \frac{1}{8Z^2}\left(\int \frac{\pi(\x)g(\x)}{\frac{q_1(\x)+q_2(\x)}{2}} q_1(\x)d\x  \right)^2  - \frac{1}{8Z^2} \left(\int \frac{\pi(\x)g(\x)}{\frac{q_1(\x)+q_2(\x)}{2}} q_2(\x)d\x  \right)^2. \nonumber   
	\label{eq_var_r2_N2_ext2} 
\end{eqnarray} 
}
The question is then whether the last four terms are larger than $ \frac{1}{2}\Var ( \hI_{\Rc} )$. We hypothesize that no inequality can be established in a general case, i.e., whether the scheme $\Rc$ would outperform $\Rb$ or not for a given $\pi(\x)$ and $g(\x)$, might depend on the proposals $q_1(\x)$ and $q_2(\x)$.
\end{remark}

%%%%%%%%%%%%%%%%%%%%%%%%%%%%%%%%%%%%%%%%%
%%%%%%%%%%%%%%%%%%%%%%%%%%%%%%%%%%%%%%%%%
\subsection{Example with closed-form variances}
\label{sec_ex1_a}
%%%%%%%%%%%%%%%%%%%%%%%%%%%%%%%%%%%%%%%%%
%%%%%%%%%%%%%%%%%%%%%%%%%%%%%%%%%%%%%%%%%

Let us derive the expressions of the example of Section \ref{running_example_variances} by considering the targeted distribution
\begin{eqnarray}
\pi(\x) &=& \frac{1}{2}\Bigg[ \gauss \left( \x|-\mu,\sigma^2 \right)  + \gauss \left(\x|\mu,\sigma^2 \right) \Bigg]. 
\end{eqnarray}
We consider $N=2$ proposal densities, $q_1(\x) =  \gauss(\x|-\mu,\sigma^2)$ and $q_2(\x) = \gauss(\x|\mu,\sigma^2)$. Note that the mixture of proposals is exactly the targeted distribution, i.e. $\psi(\x) = \pi(\x)$. We address the case where we want to estimate a specific moment $g$ of $\pi$ with the $M=2$ samples. In the following, we provide explicit variances of the unnormalized estimator of Eq. \eqref{eq_IS_estimator_unnorm} for the six MIS schemes. From Eq. \eqref{eq_var_r1},
\begin{eqnarray}
\Var(\hat I_{\Na}) &=& \frac{1}{4}\left[ \int\frac{\pi^2(\x)g^2(\x)}{q_1(\x)}d\x + \int\frac{\pi^2(\x)g^2(\x)}{q_2(\x)}d\x\right] - \frac{I}{2} \nonumber \\
&=& \frac{1}{4}\left[S_1 + S_2 \right] - \frac{I}{2}. \nonumber
\label{eq_closed_var_r1_a}
\end{eqnarray}
Let us first compute
{\scriptsize
\begin{eqnarray}
S_1 &=& \int \frac{g^2(\x)\frac{1}{2}\left(q_1(\x) + q_2(\x) \right)}{q_1(\x)}\pi(\x)d\x \nonumber \\
&=& \frac{1}{2} \left[ \int g^2(\x) \pi(\x) d\x  + \int \frac{q_2(\x)}{q_1(\x)} g^2(\x) \pi(\x) d\x \right] \nonumber \\
&=& \frac{1}{4} \left[ \int g^2(\x) q_1(\x) d\x  + \int g^2(\x) q_2(\x) d\x  + \int g^2(\x) \frac{q_1(\x) + q_2(\x)}{q_1(\x)} q_2(\x)  d\x \right] \nonumber \\
&=& \frac{1}{4} \left[ \int g^2(\x) q_1(\x) d\x  +2 \int g^2(\x) q_2(\x) d\x  + \int g^2(\x) \frac{q_2(\x)}{q_1(\x)} q_2(\x)  d\x \right]. \nonumber
\end{eqnarray}
}
Since the proposals are Gaussian,
{\scriptsize
\begin{eqnarray}
\frac{q_2(\x)}{q_1(\x)}q_2(\x)  &=&  \frac{\frac{1}{\sqrt{2\pi \sigma^2}} \exp{\left( - \frac{\left(\x - \mu \right)^2}{2 \sigma^2} \right)}}{\frac{1}{\sqrt{2\pi \sigma^2}} \exp{\left( - \frac{\left(\x + \mu \right)^2}{2 \sigma^2} \right)}}\frac{1}{\sqrt{2\pi \sigma^2}} \exp{\left( - \frac{\left(\x - \mu \right)^2}{2 \sigma^2} \right)} \nonumber \\
&=& \exp{\left(\frac{4\mu \x}{2 \sigma^2}\right)} \frac{1}{\sqrt{2\pi \sigma^2}}  \exp{-\frac{\left(\x-\mu\right)^2}{2\sigma^2}}  \nonumber \\
&=& \frac{1}{\sqrt{2\pi \sigma^2}}  \exp{\left(-  \frac{\x^2+\mu^2 - 2\mu \x - 4\mu \x}{2\sigma^2} \right) }  \nonumber \\
&=& \frac{1}{\sqrt{2\pi \sigma^2}}  \exp{\left(-  \frac{\left(\x - 3\mu \right)}{2\sigma^2} \right) } \exp{\left(-  \frac{8 \mu^2}{2\sigma^2} \right). }  \nonumber
\end{eqnarray}
}
Then,
{\scriptsize
\begin{eqnarray}
S_1 &=&  \frac{1}{4} \left[ \int g^2(\x) q_1(\x) d\x  + 2 \int g^2(\x) q_2(\x) d\x  +  \exp{\left(-  \frac{8 \mu^2}{2\sigma^2} \right)} \int g^2(\x)  \frac{1}{\sqrt{2\pi \sigma^2}}  \exp{\left(-  \frac{\left(\x - 3\mu \right)}{2\sigma^2} \right) }  d\x \right] \nonumber \\
&=&  \frac{1}{4} \left[ \int g^2(\x) q_1(\x) d\x  + 2 \int g^2(\x) q_2(\x) d\x  +  \exp{\left(-  \frac{8 \mu^2}{2\sigma^2} \right)} \int g^2(\x)  \mathcal{N}(3\mu,\sigma^2)  d\x \right]. \nonumber 
\end{eqnarray}
}
Similarly, 
{\scriptsize
\begin{eqnarray}
S_2 &=& \frac{1}{4} \left[ \int g^2(\x) q_2(\x) d\x  +2 \int g^2(\x) q_1(\x) d\x  + \int g^2(\x) \frac{q_1(\x)}{q_2(\x)} q_1(\x)  d\x \right], \nonumber
\end{eqnarray}
}
where
{\scriptsize
\begin{eqnarray}
\frac{q_1(\x)}{q_2(\x)}q_1(\x)  &=&  \frac{1}{\sqrt{2\pi \sigma^2}} \exp{\left(-  \frac{\x^2+\mu^2 + 2\mu \x + 4\mu \x}{2\sigma^2} \right) }  \nonumber \\
&=& \frac{1}{\sqrt{2\pi \sigma^2}}  \exp{\left(-  \frac{\left(\x + 3\mu \right)}{2\sigma^2} \right) } \exp{\left(\frac{8 \mu^2}{2\sigma^2} \right) }.  \nonumber
\end{eqnarray}
}
Finally, from Eq. \eqref{eq_closed_var_r1_a},
{\scriptsize
\begin{eqnarray}
\Var(\hat I_{\Na}) &=& \frac{1}{16}\Bigg[3 \int g^2(\x) q_1(\x)d\x + 3\int g^2(\x)q_2(\x)d\x \nonumber \\
&+& \left( \int g^2(\x) \gauss(\x|3\mu,\sigma^2)d\x + \int g^2(\x) \gauss(\x|-3\mu,\sigma^2)d\x\right)\exp{\left( \frac{4\mu^2}{\sigma^2} \right)} \Bigg] -\frac{I}{2}.  \nonumber
\end{eqnarray}
}
Note that $\Var(\hat I_{\Ra}) = \Var(\hat I_{\Na})$. From Eq. \eqref{eq_var_r3},
{\scriptsize
\begin{eqnarray}
\Var(\hat I_{\Rc}) &=& \frac{1}{2} \int \frac{g^2(\x)\pi(\x)}{\pi(x)} \pi(\x) d\x - \frac{I}{2} \nonumber \\
&=& \frac{1}{2} \int g^2(\x) \pi(\x)d\x - \frac{1}{2} \int g(\x)\pi(\x)d\x \nonumber \\
&=& \frac{1}{2} \int g(\x)(g(\x) - 1) \pi(x)d\x.
\end{eqnarray}
}

From Eq. \eqref{eq_var_n3},
{\scriptsize
\begin{eqnarray}
\Var(\hat I_{\Nc}) &=& \frac{1}{2} \int g^2(\x)\pi(\x) d\x - \frac{1}{4} \Bigg[ \left(\int g(\x) q_1(\x) d\x \right)^2 + \left(\int g(\x) q_2(\x) d\x \right)^2 \Bigg]. \nonumber 
\end{eqnarray}
}

From Eq. \eqref{eq_var_r2_N2_ext1}, $\Var(\hat I_{\Rb}) = \frac{\Var(\hat I_{\Na})+ \Var(\hat I_{\Nc})}{2}$. Therefore,
{\scriptsize
\begin{eqnarray}
\Var(\hat I_{\Rb}) &=&  \frac{1}{32}\Bigg[3 \int g^2(\x) q_1(\x)d\x + 3\int g^2(\x)q_2(\x)d\x\nonumber \\
&+& \left( \int g^2(\x) \gauss(\x|3\mu,\sigma^2)d\x + \int g^2(\x) \gauss(\x|-3\mu,\sigma^2)d\x\right)\exp{\left( \frac{4\mu^2}{\sigma^2} \right)} \Bigg] - \frac{I}{4}  \nonumber \\
&+& \frac{1}{4} \int g^2(\x) \pi(\x) d\x - \frac{1}{8} \Bigg[ \left(\int g(\x) q_1(\x) d\x \right)^2 + \left(\int g(\x) q_2(\x) d\x \right)^2\Bigg]. \nonumber
\end{eqnarray}

Moreover, from Proposition \ref{prop_2a}, $\hat I_{\Nb} = \hat I_{\Rb}$. }

%%%%%%%%%%%%%%
\section{Multidimensional mixture of generalized Gaussian distributions}
\label{sec_ex3}
%%%%%%%%%%%%%%

Let us consider a mixture of multivariate generalized Gaussian distributions (GGD) as a target pdf. In particular
\begin{equation}
{\pi}({\bf x})  = \frac{1}{3}\sum_{k=1}^{3} \mathcal{GG}({\bf x}; {\bm \mu}_k, {\bm \alpha}_k, {\bm \beta}_k), \quad {\bf x}\in \mathbb{R}^{d_x},
\label{eq_example3_mixture}
\end{equation}  
where ${\bm \mu}_k=[\mu_{k,1},...,\mu_{k,d_x}]^{\top}$, ${\bm \alpha}_k=[\alpha_{k,1},...,\alpha_{k,d_x}]^{\top}$, and ${\bm \beta}_k=[\beta_{k,1},...,\beta_{k,d_x}]^{\top}$ are respectively the mean, scale, and shape parameters of each component of the mixture.
Each component of the mixture factorizes in all dimensions, i.e., the multivariate GGD pdf is the product of $N$ unidimensional GGD pdfs. Namely,
{\small
\begin{equation}
\nonumber
\mathcal{GG}({\bf x}; {\bm \mu}_k, {\bm \alpha}_k, {\bm \beta}_k)=\prod_{d=1}^{d_x} \kappa_{k,d} \exp{\left(-\left(\frac{|x_d - \mu_{k,d}|}{\alpha_{k,d}}\right)^{\beta_{k,d}} \right)},
\end{equation} 
} 
where $\kappa_{k,d}= \frac{\beta_{k,d}}{2\alpha_{k,d}\Gamma \left(\frac{1}{\beta_{k,d}}\right)}$, $\Gamma(\cdot)$ is the gamma function, and $x_d$ is the $d$-th dimension of $\x$. This family of distributions includes both Gaussian and Laplace distributions with $\beta=2$ and $\beta=1$, respectively. 
In this example, $\mu_{1,d}=-3$, $\mu_{2,d}=1$, $\mu_{3,d}=5$, $\beta_{1,d}=1.1$, $\beta_{2,d}=1.8$, $\beta_{3,d}=5$, $\alpha_{1,d}=\alpha_{2,d}=\alpha_{3,d}=1$  for all $d = 1,...,d_x$. %, and $\alpha_k=1$ for all $k \in \{1,2,3\}$.
The expected value of the target ${\pi}({\bf x})$ is then $E_{\pi}[{X_d}]=1$ for $d=1,...,d_x$. In order to study the performance of the different MIS schemes, we vary the dimension of the state space in Eq. \eqref{eq_example3_mixture} testing different values of $d_x$ (with $2\leq d_x \leq 10$). 
We consider the problem of approximating via Monte Carlo the expected value of the target density, and we compare the performance of all MIS schemes. In this example, we use $N=500$ non-standardized t-student densities as proposal functions, where each location parameter has been selected uniformly within the $[-6,6]^{d_x}$ square, and the scale parameters and the degree of freedom parameters have been selected as $\sigma_{n,d} =5$ and $\nu_{n,d} =5$, respectively, for $n=1,...,N$ and $d=1,...,d_x$. For each method, we draw $M=kN$ samples, with $k=32$, and we average all the results over $200$ runs.

Fig. \ref{fig_ex3_mse_norm} shows the MSE in the estimation of the mean of the target (averaged over all dimensions) when we increase the dimension $d_x$. Note that the hierarchy established in Section \ref{sec_variance} also holds in this example regardless the dimension. In this case, methods $\Ra$ and $\Na$ behave poorly even at lower dimensions, while the other MIS schemes have a similar behavior. When we increase the dimension, all the methods degrade, and, at certain point ($d_x\geq6$), the performance of all of them is similar. Note that the proposal pdfs are fixed in random locations of the space, which is well covered at low dimensions (since we are using $N=500$ pdfs), but this coverage becomes worse as the dimension increases. This can probably explain the similar performance of all the methods in higher dimensions. 

%%%%%%%%%%%%%%%%%%%%%%%%%%
%%%%%%%%%%%%%%%%%%%%%%%%%%
%%%%%%%%%%%%%%%%%%%%%%%%%%

\begin{landscape}

\begin{table*}[!t]
\caption{Summary of the distributions of the r.v.'s $J_n$, $\X_n$ and $\X$, for the three different sampling procedures.}
\begin{center}
\begin{tabular}{|c|c|c|c|c|}
\cline{2-4}
\multicolumn{1}{c|}{} & \multicolumn{3}{c|}{{\bf Selection of the indexes}} & \multicolumn{1}{c}{}  \\
\hline
\multirow{3}{*}{{\bf Distributions}}  &  \multirow{ 3}{*}{{\bf With replacement}}   & \multicolumn{2}{c|}{{\bf Without Replacement}} & \multirow{ 3}{*}{{\bf Text references}} \\
\cline{3-4}
 & & {\bf random selection} & {\bf deterministic selection}  & \\
 &{$\mathcal{S}_1$} & {$\mathcal{S}_2$} & {$\mathcal{S}_3$} &\\
\hline
\hline
$J_n\sim P(j_n)$ &  $\frac{1}{N}$ & $\frac{1}{N}$  &  $\mathbbm{1}_{j_n=n}$ & Eqs. \eqref{eq_index_conditional_1} and \eqref{eq_index_conditional_3}\\ 
\hline
$J_n|J_{1:n-1}\sim P(j_n|j_{1:n-1})$ &  $\frac{1}{N}$ & $\frac{1}{|\mathcal{I}_n|} \mathbbm{1}_{j_n\in \mathcal{I}_n}$  &  $\mathbbm{1}_{j_n=n}$ & Eqs.  \eqref{eq_index_conditional_1}-\eqref{EqMonica}-\eqref{eq_index_conditional_3}  \\ % 
\hline
\hline
$\X_n|J_{1:n-1} \sim p(\x_n|j_{1:n-1})$ & $\psi({\x_n})$  & $\frac{1}{|\mathcal{I}_n|} \sum_{\forall k\in \mathcal{I}_n} q_k(\x_n)$  & $q_n(\x_n)$ & Sect. \ref{DistInXn} \\
\hline
$\X_n|J_n \sim p(\x_n|j_n)$ & $q_{j_n}(\x_n)$ & $q_{j_n}(\x_n)$ &  $q_{j_n}(\x_n) = q_n(\x_n)$ & Sect. \ref{sec:samplingMixture}\\
\hline
$\X_n\sim p(\x_n)$ &  $\psi({\x_n})$ & $\psi({\x_n})$  &  $q_n(\x_n)$ & Eq. \eqref{Eqmarginal}\\
\hline
\hline
$\X|J_{1:N} \sim f(\x|j_{1:N})$ &   $\frac{1}{N}\sum_{n=1}^N q_{j_n}(\x)$  & $\psi({\x})$ & $\psi({\x})$ & Eq. \eqref{eq_exotic_mixture}\\
\hline
$\X \sim f(\x)$ & $\psi({\x})$  & $\psi({\x})$ & $\psi({\x})$ & Eq. \eqref{eq_events_union} \\
\hline
\hline
$\X_{1:N} \sim p(\x_{1:N})$ & $\prod_{n=1}^N\psi({\x_n})$ & $\psi(\x_1) \prod_{n=2}^N \frac{1}{|\mathcal{I}_{n}|} \sum_{\ell \in {I}_{n}} q_{\ell}(\x_n)$ &  $\prod_{n=1}^N q_n({\x_n})$ & Sect. \ref{sec_BeyondXn}; Eq. \eqref{eq_joint_monica}\\
\hline
\end{tabular}
\end{center}
\label{table_distributions}
\end{table*}%

\begin{table*}[!t]
\caption{Specific function, {$\varphi_{\mathcal{P}_n}$}, at the denominator of weight, $w_n = \frac{\pi(\x_n)}{\varphi_{\mathcal{P}_n}(\x_n)}$, resulting from the combination of the different sampling schemes (Section \ref{sec_BeyondXn}) and weighting functions (Section \ref{sec_wfunctions}).}
%\vspace{-0.5cm}
\begin{center}
\begin{tabular}{|c|c|c|c|c|c|}
\hline
\multirow{2}{*}{$\varphi_{\mathcal{P}_n}$}   & $\mathcal{W}_1$  &  $\mathcal{W}_2$  & $\mathcal{W}_3$  & $\mathcal{W}_4$ & $\mathcal{W}_5$ \\
& $p(\x_n|j_{1:n-1})$ &$p(\x_n|j_n)$ & $p(\x_n)$ & $f(\x|j_{1:N})$ & $f(\x)$  \\ 
\hline \hline
 $\mathcal{S}_1$:  {\bf with replacement} &   $\psi(\x_n)$  {[R3]}  & $q_{j_n}(\x_n)$  {[R1]} &  $\psi(\x_n)$  {[R3]} &    $\frac{1}{N}\sum_{k=1}^N q_{j_k}(\x_n)$  {[R2]}  & $\psi(\x_n)$  {[R3]} \\ 
\hline
$\mathcal{S}_2$: {\bf w/o (random)} &   $\frac{1}{|\mathcal{I}_n|} \sum_{\forall k\in \mathcal{I}_n} q_k(\x_n)$  {[N2]} & $q_{j_n}(\x_n)$  {[N1]} &  $\psi(\x_n)$  {[N3]} &  $\psi(\x_n)$ {[N3]}  & $\psi(\x_n)$ {[N3]} \\ 
\hline
$\mathcal{S}_3$: {\bf w/o (deterministic)} &  $q_n(\x_n)$ {[N1]} & $q_n(\x_n)$ {[N1]} & $q_n(\x_n)$ {[N1]} &  $\psi(\x_n)$ {[N3]} & $\psi(\x_n)$ {[N3]} \\ 
\hline
\end{tabular}
\end{center}
\label{table_sampling_weights}
\end{table*}%

\begin{figure}[!htb]
\centering
\includegraphics[width=0.5\textwidth]{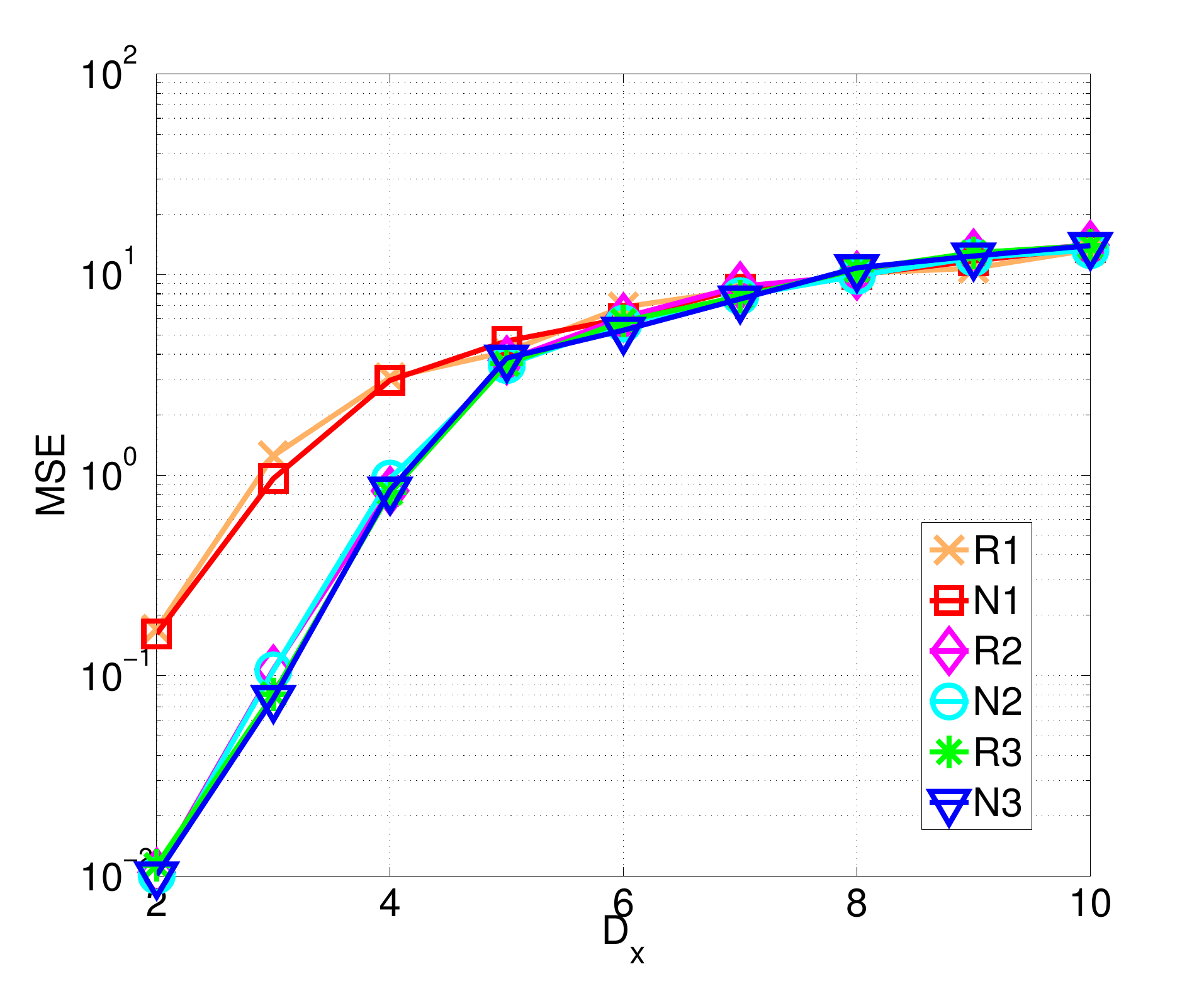}
\caption{\textbf{(Ex. of Section. \ref{sec_ex3})} MSE of the self-normalized estimator $\tilde I$ for all MIS schemes when we increase the dimension $d_x$ of the state space.}
\label{fig_ex3_mse_norm}
\end{figure}

\begin{table*}[!htb]
\caption{ Summary of possible MIS strategies in an adaptive framework.}
\begin{center}
\begin{tabular}{|c||c|c|c|c||c|}
\hline
 \multirow{2}{*}{{\bf MIS scheme}} &  \multirow{2}{*}{{\bf Function $\varphi_{j,t}({\bf x})$}}    &   \multirow{2}{*}{${N}$}  & $P$ & $L$ & \multirow{2}{*}{{\bf Corresponding Algorithm} }    \\
\cline{4-5}
 && & \multicolumn{2}{c||}{$LP=N$} &  \\
\hline
\hline
$\Na$ & $q_{j,t}({\bf x})$ & & $JT$ & $1$ & PMC \citep{Cappe04} \\
\cline{1-2} \cline{4-6} 
\emph{Full} $\Nc$ & $\psi({\bf x})= \frac{1}{JT}\sum_{j=1}^J \sum_{t=1}^T q_{j,t}({\bf x})$  &  & $1$ & $JT$ & suggested in \citep{elvira2015efficient}  \\
\cline{1-2} \cline{4-6}  
\emph{Partial} (temporal) $\Nc$ & $\xi_{j}({\bf x})= \frac{1}{T}\sum_{t=1}^T q_{j,t}({\bf x})$ & $JT$ & $J$ & $T$ &   AMIS \citep{CORNUET12}, with $J=1$  \\
\cline{1-2} \cline{4-6} 
\emph{Partial} (spatial) $\Nc$ & $\phi_{t}({\bf x})= \frac{1}{J}\sum_{j=1}^J q_{j,t}({\bf x})$&  & $T$ & $J$ &  APIS \citep{APIS15} \\
\cline{1-2} \cline{4-6} 
\emph{Partial} (spatial) $\Rc$ & $\phi_{t}({\bf x})= \frac{1}{J}\sum_{j=1}^J q_{j,t}({\bf x})$&  & $T$ & $J$ &   \citep{Cappe08,Douc07a,Douc07b} \\
\cline{1-2} \cline{4-6} 
\emph{Partial} (generic) $\Nc$ & generic $\varphi_{j,t}({\bf x})$ in Eq. \eqref{PartialMixProposal2}  &  & $P$ & $L$ & suggested in \citep{elvira2015efficient}\\ 
\hline
\end{tabular}
\end{center}
\label{FantasticTable}
\end{table*}%  

\end{landscape}

\end{document}